\documentclass[11pt]{article}
\pdfoutput=1
\usepackage[T1]{fontenc}
\usepackage[latin1]{inputenc}
\usepackage[nosort]{cite}
\usepackage{color}
\usepackage[bulletsep]{collref}
\usepackage{graphicx}
\usepackage{bbm}
\usepackage{amsmath}
\usepackage{amssymb}
\usepackage{subfig}
\usepackage{multirow}
\usepackage{tikz}
\usepackage{hyperref}

\makeatletter \@addtoreset{equation}{section} \makeatother

\makeatletter
\let\old@startsection=\@startsection
\let\oldl@section=\l@section
\renewcommand{\@startsection}[6]{\old@startsection{#1}{#2}{#3}{#4}{#5}{#6\mathversion{bold}}}
\renewcommand{\l@section}[2]{\oldl@section{\mathversion{bold}#1}{#2}}
\makeatother

\makeatletter
\let\old@makecaption=\@makecaption
\def\@makecaption{\small\old@makecaption}
\makeatother

\let\oldPhi=\Phi
\let\oldPsi=\Psi
\let\oldGamma=\Gamma
\let\oldDelta=\Delta
\let\oldSigma=\Sigma
\let\oldTheta=\Theta
\let\oldPi=\Pi
\let\oldUpsilon=\Upsilon
\renewcommand{\Phi}{\mathnormal{\oldPhi}}
\renewcommand{\Psi}{\mathnormal{\oldPsi}}
\renewcommand{\Gamma}{\mathnormal{\oldGamma}}
\renewcommand{\Sigma}{\mathnormal{\oldSigma}}
\renewcommand{\Delta}{\mathnormal{\oldDelta}}
\renewcommand{\Theta}{\mathnormal{\oldTheta}}
\renewcommand{\Pi}{\mathnormal{\oldPi}}
\renewcommand{\Upsilon}{\mathnormal{\oldUpsilon}}

\newcommand{\tr}{\mathop{\mathrm{tr}}}

\renewcommand{\Re}{\mathop{\mathrm{Re}}}
\renewcommand{\Im}{\mathop{\mathrm{Im}}}

\newcommand{\Sphere}{\mathrm{S}}  
\newcommand{\AdS}{\mathrm{AdS}}

\ifx\genfrac\sdflkaj

\else

\fi

\newcommand{\p}{\partial}
\newcommand{\X}{\mathbb{X}}

\newcommand{\A}{\mathbb{A}}
\newcommand{\HH}{\mathbb{H}}

\def\[{\begin{equation}}
\def\]{\end{equation}}

\makeatletter
\def\mr@ignsp#1 {\ifx\:#1\@empty\else #1\expandafter\mr@ignsp\fi}%
\newcommand{\multiref}[1]{\begingroup
\xdef\mr@no@sparg{\expandafter\mr@ignsp#1 \: }%
\def\mr@comma{}%
\@for\mr@refs:=\mr@no@sparg\do{\mr@comma\def\mr@comma{,}\ref{\mr@refs}}%
\endgroup}
\makeatother

\newcommand{\hypref}[2]{\ifx\href\asklfhas #2\else\href{#1}{#2}\fi}

\renewcommand{\eqref}[1]{(\multiref{#1})}

\ifx\href\asklfhas\newcommand{\href}[2]{#2}\fi

\newcommand{\be}{\begin{eqnarray}}
\newcommand{\ee}{\end{eqnarray}}

\begin{document}

\thispagestyle{empty}
\begin{flushright}\footnotesize
\texttt{NORDITA-2015-008}\\
\end{flushright}
\vspace{1cm}

\begin{center}%
{\Large\textbf{\mathversion{bold}%
Wilson Loops and Minimal Surfaces\\ Beyond the Wavy Approximation
}\par}

\vspace{1.5cm}

\textrm{Amit Dekel} \vspace{8mm} \\
\textit{%
Nordita, KTH Royal Institute of Technology and Stockholm University, \\
Roslagstullsbacken 23, SE-106 91 Stockholm, Sweden
} \\

\texttt{\\ amit.dekel@nordita.org}

\par\vspace{14mm}

\textbf{Abstract} \vspace{5mm}

\begin{minipage}{14cm}
We study Euclidean Wilson loops at strong coupling using the AdS/CFT correspondence,
where the problem is mapped to finding the area of minimal surfaces in Hyperbolic space.
We use a formalism introduced recently by Kruczenski to perturbatively compute the area corresponding to boundary contours which are deformations of the circle.
Our perturbative expansion is carried to high orders compared with the wavy approximation and yields new analytic results.
The regularized area is invariant under a one parameter family of continuous deformations of the boundary contour which are not related to the global symmetry of the problem.
We show that this symmetry of the Wilson loops breaks at weak coupling at an a priori unexpected order in the perturbative expansion.
We also study the corresponding Lax operator and algebraic curve for these solutions.

\end{minipage}

\end{center}

\newpage

\tableofcontents

\bigskip
\noindent\hrulefill
\bigskip

\section{Introduction}\label{sec:introduction}

According to the AdS/CFT correspondence string theory on $\AdS_5\times \Sphere^5$ background is dual to $\mathcal{N}=4$ super Yang-Mills theory in four space-time dimensions \cite{Maldacena:1997re}.
This theory is considered to be integrable in the planar limit, and over the years integrability based techniques where used and developed for making progress in solving the spectrum of the theory, see \cite{Beisert:2010jr} for review.
Most of the study of integrability in this theory is related to the spectral problem. Much progress was also made in the context of the scattering amplitudes/lightlike Wilson loops duality.
On the other hand, much less is known about integrability properties of Wilson loops, and in particular Euclidean Wilson loops.
As is well known, at strong coupling the problem of computing the expectation value of a Wilson loop is translated to finding the area of a minimal surface in AdS space, ending on the AdS boundary, where the contour is defined by the Wilson loop \cite{Maldacena:1998im,Rey:1998ik}.
In this paper we focus on the problem of finding this area in Euclidean $\AdS_3$ or equivalently the three dimensional hyperbolic space, $\HH_3$.

Over the years several approaches for studying the problem were proposed.
The corresponding bosonic sigma model can be classically reduced to the cosh-Gordon model using Pohlmeyer reduction \cite{Pohlmeyer:1975nb,DeVega:1992xc}.
The solution to the cosh-Gordon equation is known in terms of theta-functions \cite{babich1993,Ishizeki:2011bf,Kruczenski:2013bsa}.
Thus, for any solution there corresponds a Riemann surface, or equivalently an (odd genus) algebraic curve which satisfies some special properties in order for the solution to be real. This algebraic curve was recently shown to be the same curve which is extracted from the Lax operator \cite{Cooke:2014uga}.
Unfortunately, given a Wilson loop contour, it is not known what is the corresponding algebraic curve.
Similarly, given an algebraic curve, the properties of the Wilson loop, e.g. the area, contour etc.
are a priori not known, moreover it is not even clear that the solution corresponds to a closed Wilson loop (for example, the solution may correspond to some minimal surface ending on an infinite self intersecting line, or ending on more than one contour, which might be unstable).
These issues make it hard to use this formal solution for practical purposes as well as studying other properties of these objects.

Another interesting approach was recently introduced by Kruczenski \cite{Kruczenski:2014bla}.
In this approach the worldsheet boundary is fixed to be the unit circle, and the problem is mapped to finding the correct parametrization of the contour in the conformal gauge. Once the correct parametrization is found, one has all the required ingredients to compute the area in terms of a contour integral around the unit circle on the worldsheet.
The main drawback is that it is not known how to find the correct parametrization either analytically or numerically.

A very different approach to the same problem in Minkowskian $\AdS_3$ space was recently proposed in \cite{Toledo:2014koa}.
This approach uses the known solution to the problem where the Wilson loop contour is given by a set of null lines found in \cite{Alday:2010vh}, where the nontrivial limit of infinite cusped segments is taken, so that the contour becomes smooth. The resulting area is given in terms of an integral TBA equation, which one should generally solve numerically.

As can be understood form the discussion above, the solution to the problem is not expected to be simple by any means.
More precisely, the only "simple" examples of known solutions are the circular Wilson loop and the infinite straight Wilson line.
Beyond that, the simplest solutions are the "genus-one"\footnote{Here we refer to the genus of the corresponding Riemann surface.} solutions which correspond for example to the correlation function of two circular Wilson loops, the $q\bar q$-potential and the infinite cusp, which are all given in terms of Jacobi elliptic functions.
Other solutions should correspond to higher genus Riemann theta-functions.
In order to understand better the properties of these object it is desirable to have other simple examples at hand which can be studied analytically.
One useful approach is the so called "wavy expansion" \cite{Semenoff:2004qr,Polyakov:2000ti}, which is an expansion in a deformation parameter around the circular or infinite line contours, keeping terms up to second order.
However, this is not always satisfactory and some interesting features can be missed since to this order all the properties are fixed by conformal symmetry \cite{Semenoff:2004qr}.

In this paper we use Kruczenski's formalism introduced in \cite{Kruczenski:2014bla} to study the problem perturbatively around the circular contour.
We also adapt this approach for studying perturbations around the infinite straight line contour.
We show that quite easily we can go far beyond the wavy approximation, and apply this approach to several simple perturbative deformations.
Among these Wilson loop contours are some families of symmetric contours which include the ellipse and some less symmetric contours which include the lima\c{c}on.
Besides for providing us analytic expansion for the area to very high orders in the expansion parameter,
having this analytic data for these Wilson loops allows us to study some nontrivial properties of these objects.

One aspect we shall study is the expansion of the Lax operator and the resulting algebraic curve, which hopefully can shed some light on the relation to the Riemann theta-functions.
Another aspect is related to a curious property which was pointed out in \cite{Ishizeki:2011bf}.
There it was shown that there exists a one parameter family of deformations of the target space contour which leaves the area invariant.
The deformation depends on the spectral parameter $\lambda$ which must be a phase in this case, and is not related to any obvious global symmetry of the problem (i.e. global conformal symmetry). 
We shall call such deformations \textit{$\lambda$-deformations} throughout the paper.
Because of the universality of the wavy correction due to conformal invariance \cite{Semenoff:2004qr}, it is expected that this deformation leaves the Wilson loop expectation value invariant to third order in the expansion parameter for any value of the coupling.
However, this universality breaks at the fourth order in the perturbation of the circular Wilson loop \cite{Galakhov:2008ax} and it is not obvious whether this invariance at strong coupling should survive quantum corrections, and in particular what happens at weak coupling.
Using our approach we check how this deformation affects the one-loop weak coupling expectation value.
We find a dependence on the deformation parameter in all of the examples we consider.
Interestingly in all of the examples except one the dependence shows up starting at the 8th order in the expansion parameter (in the lima\c{c}on example it starts to show only at the 16th order), and the effect is very mild.
Moreover, given an arbitrary contour, it is not trivial at all to find its $\lambda$-deformed contour.

Since there are no other analytic results available for the area of minimal surfaces ending on simple contours as we study in this paper (e.g. the ellipse), we can only compare our results to numerical data, as we do in the case of the ellipse where such numerical data in available \cite{Fonda:2014cca}, and find agreement.
Furthermore, our analytic results should be useful for testing new approaches to come for solving the general problem.

Finally, although our main interest is the study of Wilson loops and their properties, let us mention that the same mathematical problem is of great interest also in the study of entanglement entropy of conformal field theories in $2+1$ dimensions \cite{Ryu:2006bv}, where our approach and results can be also applied.

The paper is organized as follows,
in section \ref{sec:general_setup} we start by giving a brief introduction to the formalism presented by Kruczenski in \cite{Kruczenski:2014bla}.
We provide only the relevant results which we shall use throughout the paper, and omit the derivation which can be found in \cite{Kruczenski:2014bla}.
In section \ref{sec:general_pert} we perform a general wavy perturbation around the circular Wilson loop solution, to low orders in the wavy expansion, which should set the ground for the later analysis of concrete examples where the same procedure is applied to much higher orders.
We then study some specific examples in detail in section \ref{sec:applications}, where the ellipse is set as our prime example for which we provide a more explicit analysis.
Later, in section \ref{sec:infinite_line} we slightly modify the formalism such that we map the upper half plane to the minimal surface instead of mapping the unit disk.
This is of course equivalent in principle, and we can study the same deformations in this gauge, however, it is more convenient for some deformations of the infinite straight line solution as we shall demonstrate.
Finally, in section \ref{sec:discussion} we summarize and discuss our results.
In appendix \ref{app:general_higher_orders} we provide details relevant for the general solution of the area to fourth order in the expansion parameter.

\section{General setup}\label{sec:general_setup}

In this section we briefly repeat the analysis presented by Kruczenski in \cite{Kruczenski:2014bla}, where the reader is referred to for more details.
We start with the bosonic string sigma model on $\HH_3$, and the first step is to perform a Pohlmeyer reduction.
We use the embedding coordinates $X_{\mu=0,..,3}$ in $\mathbb{R}^{3,1}$ subject to $X^2 = X_0^2-X_1^2-X_2^2-X_3^2=1$, where the $SO(3,1)\approx SL(2,\mathbb{C})$ symmetry acts linearly. The Poincar\'{e} coordinates are given by
\begin{equation}
\mathrm{X}+i \mathrm{Y} = \frac{X_1 + i X_2}{X_0 - X_3},\quad \mathrm{Z} = \frac{1}{X_0 - X_3}.
\end{equation}
The worldsheet metric is taken to be Euclidean and the coordinates are given by $z = \sigma + i \tau$, $\bar z = \sigma -i \tau$, and the corresponding derivatives are given by $\partial = \frac{1}{2}\left(\frac{\partial}{\partial \sigma} - i \frac{\partial}{\partial \tau} \right)$ and $\bar \partial = \frac{1}{2}\left(\frac{\partial}{\partial \sigma} + i \frac{\partial}{\partial \tau} \right)$.
The action in conformal gauge is given by
\begin{equation}
S = \frac{1}{2}\int d\sigma d\tau \left(\partial X_\mu \bar \partial X^\mu - \Lambda (X_\mu X^\mu - 1)\right),
\end{equation}
and it should be supplemented by the Virasoro constraints
\begin{equation}
\partial X_\mu \partial X^\mu  = 0 = \bar \partial X_\mu \bar \partial X^\mu.
\end{equation}
The equations of motion are given by
\begin{equation}
\p\bar \p X_\mu + \Lambda X_\mu = 0,
\end{equation}
where consistency with the constraint $X^2 = 1$ yields $\Lambda = \p X_\mu  \bar\p X^\mu$.
Following \cite{Ishizeki:2011bf} we rewrite the embedding coordinates in terms of a matrix $\X = X_\mu \sigma^\mu$ where $\sigma^\mu = \{1,\sigma^i\}$ and $\sigma^i$ are the Pauli matrices. In this way the constraints and equations of motion are given by
\begin{equation}
\X^\dag = \X,\quad
\det \left(\X\right)=1,\quad
\det \left(\p \X\right)=0,\quad
\det \left(\bar \p \X\right)=0,\quad
\p \bar \p\X - \frac{1}{4}\tr\left(\p \X \sigma_\mu \bar \p \X\sigma^\mu\right)\X = 0,
\end{equation}
where we used the relation $A_\mu B^\mu = -\frac{1}{4}\tr\left(\mathbb{A} \sigma_\mu \mathbb{B} \sigma^\mu\right)$.
The first constraint is solved by introducing a new matrix $\A$ by $\X=\A\A^\dag$. The second constraint implies that $\det \A = e^{i\phi}$, but the phase can be cancelled by gauge fixing as we shall see immediately so we can assume $\det \A = 1$, so $\A\in SL(2,\mathbb{C})$.
The equations are invariant under the global symmetry transformation $\X\to U \X U^\dag$ and $\A\to U\A$ with $U\in SL(2,\mathbb{C})$.
Furthermore, there is a gauge symmetry $\A\to \A H(z,\bar z)$ which leaves $\X$ unchanged if $H(z,\bar z)\in U(2)$.

Next we define the traceless currents $J=\A^{-1} \p \A$ and $\bar J=\A^{-1} \bar \p \A$ and rewrite the constraints and equations of motion in terms of the new variables.
Using the equations of motion and constraints one can construct a flat connection given by
\begin{equation}
J(\lambda) = \left(
      \begin{array}{cc}
        -\frac{1}{2}\p \alpha & f e^{-\alpha} \\
        \lambda e^{\alpha} & \frac{1}{2}\p\alpha \\
      \end{array}
    \right),\quad
\bar J(\lambda) = \left(
      \begin{array}{cc}
        \frac{1}{2}\bar \p \alpha & \frac{1}{\lambda} e^{\alpha} \\
        -\bar f e^{-\alpha} & -\frac{1}{2}\bar \p\alpha \\
      \end{array}
    \right),
\end{equation}
where $\alpha(z,\bar z)$ and $f(z)$ satisfy the generalized cosh-Gordon equation
\begin{equation}\label{eq:SGlikeEOM}
\p\bar\p \alpha(z,\bar z) = e^{2\alpha(z,\bar z)}+|f(z)|^2e^{-2\alpha(z,\bar z)},
\end{equation}
and where $\lambda \in \mathbb{C}$ is the spectral parameter.
$\alpha(z,\bar z)$ is a real function and $f(z)$ is holomorphic where the solution is defined.
For $\lambda=1$ we get back the currents defined by the string solution.
Interestingly, in case where $\lambda$ is a phase this still corresponds to a string solution, generally defined by a different contour, with the same area for the minimal surface ending on the original contour.
This deformation of the original contour is not related to any obvious global symmetry of the problem and we shall call such deformations \textit{$\lambda$-deformations} throughout the paper.
Notice also that defining $j(\lambda) = J(\lambda) dz + \bar J(\lambda) d\bar z$ we have the relation
\begin{equation}
j\left(\lambda\right)^\dag = -j\left(-\lambda\right) = -j\left(-\frac{1}{\bar \lambda}\right).
\end{equation}

Next, following \cite{Kruczenski:2014bla} we consider a minimal surface solution parameterized such that its contour in target space is mapped from the unit circle on the worldsheet.
Using polar coordinates on the worldsheet $z=r e^{i\theta}$, let us denote the contour by in target space by $X(\theta)= \mathrm{X}(\theta)+i \mathrm{Y}(\theta)$ where $\mathrm{X}(\theta)$ and $\mathrm{Y}(\theta)$ are the Poincar\'{e} coordinates at $\mathrm{Z}=0$, or equivalently at $r=1$.
Defining $r = \sqrt{1-\xi}$, close to the boundary the solution for $\alpha(z,\bar z)$ is given by
\begin{align}
\alpha(\xi,\theta) \simeq -\ln \xi + \xi(1+\xi^2)\beta_2(\theta) + \xi^4 \beta_4(\theta) + \mathcal{O}(\xi^5),
\end{align}
where $\beta_n(\theta)$ are real functions, fixed by $\beta_2(\theta)$ and $f(z)$.

Very interestingly, assuming one knows the correct parametrization function $F(\theta)$ for the boundary contour $X(\theta)$, using the linear problem
\begin{equation}\label{eq:linearproblem}
\p\psi = \psi J,\quad
\bar \p\psi = \psi \bar J,
\end{equation}
it is possible to derive the following relation
\begin{align}\label{eq:schwaderrelation}
\{X^{\lambda}(F(\theta)),\theta\} = \frac{1}{2} - 12\beta_2(\theta) - 2 \lambda f(\theta)e^{2 i \theta} + \frac{2}{\lambda}\bar f(\theta)e^{-2 i \theta},
\end{align}
where $\{,\}$ stands for the Schwarzian derivative.
$X^{\lambda=0}(\theta) = X(\theta)$ is contour we start with, and for $\lambda$ being a phase, $X^{\lambda}(\theta)$ is a deformed contour giving rise to a minimal surface with the same regularized area.
Throughout the paper we shall parameterize $\lambda=e^{i\varphi}$ with $\varphi\in [0,2\pi]$.
Equation (\ref{eq:schwaderrelation}) will be of prime importance in the current paper.
The expression implies that if we know the correct parametrization in the gauge we are using, then in principal we have all the required information in order to reconstruct $f(z)$ and $\alpha(z,\bar z)$.
Moreover, using further manipulation (which can be found in \cite{Kruczenski:2014bla}), Kruczenski presents the following expression for the regularized area
\begin{align}\label{eq:area}
A_{\text{reg}}
= -2\pi -4\int d\tau d\sigma |f|^2 e^{-2 \alpha}
= -2\pi - \left|\frac{i}{2}\oint d\theta \frac{\Re\{X,\theta\}-\{w,\theta\}}{\partial_\theta \ln w}\right|,
\end{align}
where $w(z) = \int^z \sqrt{f} dz$, and the integral is performed around the unit disk.
Although this construction is very appealing, at present it is not known how to find the correct parametrization either analytically or numerically.
In the following we shall approach the problem perturbatively around the simplest available solutions, namely the circle and the infinite line,
and show that this formalism enables us to perform such a perturbative expansion very efficiently without finding the minimal surface explicitly, and to reach very high orders in the perturbation expansion compared to the wavy approximation.

\section{Perturbations around the circular Wilson loop solution}\label{sec:general_pert}
Using the formalism introduced in \cite{Kruczenski:2014bla}, we study perturbations of minimal surfaces around the circular Wilson loop solution.
We start with a general analysis which yields a simple expression for the area to third order in the wavy expansion.
This result was already obtained in \cite{Galakhov:2008ax} using different techniques, however we believe it is useful to re-derive it here in order to set the ground for higher order perturbations which we will compute later.
We compare the result with the weak coupling computation and observe the expected universality up to this order.
General higher order analysis can be carried in principle, however it is far more complicated since conformal invariance is less restrictive at higher orders.
We give the general solution to the next order in appendix \ref{app:general_higher_orders}, however we do not find a simple general expression for the area in terms of a contour integral.
In the next section we shall turn to some explicit examples and demonstrate the power of applying the formalism to obtain results to very high order in the perturbative expansion.

We start by collecting the required properties of the circular contour solution which is the starting point of the later analysis.

\subsection{The circular Wilson loop properties}\label{sec:The_circular_Wilson_loop_properties}
The circular Wilson loop is the simplest example which can be solved analytically, where the minimal surface is the semi-sphere.
The (non-unique) $\A$ matrix which we will use is given by
\begin{equation}
\A_c = \frac{1}{\sqrt{1-z\bar z}}\left(
                                   \begin{array}{cc}
                                     1 & \bar z \\
                                     z & 1 \\
                                   \end{array}
                                 \right).
\end{equation}
From this we can extract $\alpha_c = \ln\frac{1}{1-z\bar z}$, $f_c(z) = 0$ and $\beta_2^c(\theta) = 0$, where the generalized cosh-Gordon equation is reduced to the simpler Liouville equation.
The boundary contour in the conformal gauge is given by $X(\theta)=e^{i \theta}$ and we have $\{X(\theta),\theta\}=\frac{1}{2}$.
We can also solve explicitly for $\A_c(\lambda)$
\begin{equation}
\A_c(\lambda) = \frac{1}{\sqrt{1-z\bar z}}\left(
                                   \begin{array}{cc}
                                      a(\lambda)z+c(\lambda) & \frac{1}{\lambda} (a(\lambda)+c(\lambda)\bar z) \\
                                      \tilde a(\lambda)z+\tilde c(\lambda) & \frac{1}{\lambda} (\tilde a(\lambda)+\tilde c(\lambda)\bar z) \\
                                   \end{array}
                                 \right)\to
                                 \frac{1}{\sqrt{1-z\bar z}}\left(
                                   \begin{array}{cc}
                                     1 & \frac{1}{\lambda} \bar z \\
                                     \lambda z & 1 \\
                                   \end{array}
                                 \right),
\end{equation}
where we picked $a(1)=0,c(1)=1$ and $\tilde a(1)=1,\tilde c(1)=0$ for simplicity and concreteness.
Thus, obviously in case where $\lambda$ is a phase, this corresponds to a rotation of the surface\footnote{This should be expected since the deformation leaves the regularized area invariant, and there is only one contour with the regularized area equal to $-2\pi$, namely the circle}.
Finally, the Lax operator is given by\footnote{Here we refer to the Lax operator of the Pohlmeyer reduced model which is related to the sigma model's Lax operator by a simple gauge transformation \cite{Cooke:2014uga}.}
\begin{align}
L_0(z,\bar z) = \frac{\lambda}{1-z \bar z}\left(
                                      \begin{array}{cc}
                                        1 + z \bar z & 2\frac{\bar z}{\lambda} \\
                                        -2 \lambda z & -1-z\bar z \\
                                      \end{array}
                                    \right).
\end{align}
Notice that the Lax operator is not unique, we choose it such that the resulting algebraic curve is given by
\begin{align}
y^2 = -\det L_0 = \lambda^2,
\end{align}
so that it coincides with the limit of the genus one solution up to a constant \cite{Kruczenski:2013bsa}, namely
\begin{align}
y^2 = \lim_{a\to\infty} = \lambda(\lambda - a)(\lambda + \frac{1}{a}).
\end{align}
This is equivalent to the trivial curve $y^2 = 1$ found in \cite{Dekel:2013dy}.

\subsection{General perturbation of the circular Wilson loop minimal surface}

In this section we find a simple general expression for the area of the wavy circle to third order, by adding a small perturbation to the circular contour. 
We compute $\alpha(z,\bar z)$ and $F(\theta)$ to first order where $F(\theta)$ is the parametrization function which will be properly defined below, and $f(z)$ to second order, which fixes the area to third order.
As we shall see, the reason that the third order area formula is fixed by the lower order functions is that $f(z)$ vanishes to leading order for the circle.
The fourth order result which is not fixed by conformal symmetry is much more complicated.
In appendix \ref{app:general_higher_orders} we given the general solution to $\alpha(z,\bar z)$ and $f(z)$ to the second and third orders which can be used to compute the area integral to fourth order, however we do not find a simple expression for this integral in terms of a contour integral.

\subsubsection{The general setup}

Our starting point is a given contour on the $\HH_3$ boundary which depends continuously on a parameter $\epsilon$, and reduces to the circle, $X(\theta) = e^{i \theta}$, when $\epsilon=0$.
Let us denote the contour by $X(\theta) = e^{i F(\theta) + \sum_{n=1}^{\infty} \epsilon^n G_n(F(\theta))}$ where the function $F(\theta)$ is the correct parametrization function which is unknown. 
Notice that the given functions $G_n(\theta)$ are arbitrary complex periodic functions, $G_n(s) = \sum_{k\in \mathbb{Z}} G_n^{(k)}e^{i k s}$.
Obviously, by these definitions $F(\theta)$ should also depend on $\epsilon$ and should reduce to $F(\theta) = \theta$ when $\epsilon=0$.
Thus, we expand the parametrization function in powers of $\epsilon$ as follows, $F(\theta) = \theta + \sum_{n=1}^{\infty} \epsilon^n F_n(\theta)$.

Next, we wish to compute the $\epsilon$ expansion of the minimal surface area ending on $X(\theta)$, which to leading order is given by $-2\pi$, the area of the semi-sphere.
In principle we need to find $\alpha(z,\bar z)$ and $f(z)$ which enter the area integral (\ref{eq:area}).
We expand these functions as follows,
\begin{align}
\alpha(z,\bar z) = & \ln\frac{1}{1-z \bar z}
= \ln \frac{1}{1-r^2}
+ \sum_{n=2}^{\infty} \alpha_n(z,\bar z) \epsilon^n \nonumber\\
f(z) = & \sum_{n=1}^\infty f_n(z) \epsilon^n.
\end{align}
Notice that there is no term linear in $\epsilon$ in the $\alpha(z,\bar z)$ expansion since it vanishes due to the boundary conditions.
We decompose the perturbation to the real and imaginary parts as $G_n(\theta) = G^r_n(\theta) + i G^i_n(\theta)$, where $G^r_n(\theta), G^i_n(\theta) \in \mathbb{R}$.
Our starting point is the Schwarzian derivative of the contour, satisfying
\begin{align}\label{eq:SDrelation}
\Re \{X(\theta),\theta\} & = \frac{1}{2}-12 \beta_2(\theta),\nonumber\\
\Im \{X(\theta),\theta\} & = -4\Im (e^{2 i \theta} f(e^{i \theta})).
\end{align}
The second relation implies that
\begin{align}\label{eq:f_intermsof_ImXs}
f(e^{i \theta}) &  = -\frac{i}{2} e^{-2 i \theta} \mathcal{P}(\Im\{X(\theta),\theta\}),
\end{align}
where $\mathcal{P}$ projects on positive frequencies, that is $\mathcal{P} = \frac{1}{2}(1+\mathcal{H})$ with $\mathcal{H}$ being the Hilbert transform on the unit circle, normalized such that
\begin{align}
\mathcal{H}\left[G(t)\right](s) = -\frac{1}{2 \pi i}\mathrm{p.v.}\int_{-\pi}^{\pi}dt G(t)\cot \left(\frac{t-s}{2}\right),
\end{align}
for which we have $\mathcal{H}\left[e^{i n t}\right](s) = \text{sign}(n) e^{i n s}$.
Finally, $\alpha(z,\bar z)$ and $f(z)$ are related by the generalized cosh-Gordon equation (\ref{eq:SGlikeEOM}).

\subsubsection{Finding $f(z)$ to second order}

Next, we are going to use these relations to find $f(z)$ to second order, which is enough to fix the area to third order in $\epsilon$.
To leading order we have
\begin{align}\label{eq:lowestSDrelation}
\Re \{X(\theta),\theta\} & = \frac{1}{2} +\epsilon \mathcal{L}_3 (F_1(\theta) + G_1^i(\theta))+\mathcal{O}(\epsilon^2) = \frac{1}{2}-12 \beta_2(\theta),\nonumber\\
\Im \{X(\theta),\theta\} & = \quad - \epsilon \mathcal{L}_3  G_1^r(\theta) +\mathcal{O}(\epsilon^2) = -4\Im (e^{2 i \theta} f(e^{i \theta})),\nonumber\\
\end{align}
where we defined the operator $\mathcal{L}_3 = \p_\theta^3 + \p_\theta$.
Since $\beta_2(\theta) = \mathcal{O}(\epsilon^2)$, we have
\begin{align}
F_1(\theta) = -G_1^i(\theta) + F^h(\theta),
\end{align}
where $F^h(\theta) = a + b \sin \theta + c \cos \theta$ is the homogenous solution, which we shall drop\footnote{One might want to keep the homogenous solution in some cases so that $F(0) = 0$ and $F(2\pi) = 2\pi$. This can always be fixed by choosing the constant piece of $F^h$ appropriately, however $\mathcal{L}_3$ does not "see" this term so it doesn't change the current analysis.}, so $F_1(\theta) = -G_1^i(\theta)$.
From the second equation in (\ref{eq:lowestSDrelation}) we read $f_1(z)$ using (\ref{eq:f_intermsof_ImXs}), so
\begin{align}
f_1(z) &  = \frac{i}{2} e^{-2 i \theta} \mathcal{P}(\mathcal{L}_3  G_1^r(\theta))|_{\theta \to -i \ln z},
\end{align}
Using $F_1(\theta)$ given above, the $\epsilon^2$ term in $\Im \{X(\theta),\theta\}$ follows,
\begin{align}
\Im \{X(\theta),\theta\} = -\epsilon \mathcal{L}_3  G_1^r(\theta) - \epsilon^2 \mathcal{L}_3  \left(G_2^r(\theta) - G_1^i(\theta) \p_\theta G_1^r(\theta)\right) +\mathcal{O}(\epsilon^3),
\end{align}
so $f_2(z)$ is given by
\begin{align}
f_2(z) &  = \frac{i}{2} e^{-2 i \theta} \mathcal{P}\left(\mathcal{L}_3  \left(G_2^r(\theta) - G_1^i(\theta) \p_\theta G_1^r(\theta)\right)\right)|_{\theta \to -i \ln z}.
\end{align}
This is all we need in order to fix the area to third order.

\subsubsection{The area}
The regularized area expansion is given by
\begin{align}
\mathcal{A}_{\text{reg}} =
-2\pi
& -2\epsilon^2 \int d^2 z f_1(z)\bar f_1(\bar z) e^{-2\alpha_0(z,\bar z)}
 -2\epsilon^3 \int d^2 z \left( f_1(z)\bar f_2(\bar z) + \bar f_1(\bar z)f_2(z)\right) e^{-2\alpha_0(z,\bar z)}\nonumber\\
& -2\epsilon^4 \int d^2 z \left( f_1(z)\bar f_3(\bar z) + \bar f_1(\bar z)f_3(z) + \bar f_2(\bar z)f_2(z) - 2 \bar f_1(\bar z)f_1(z)\alpha_2(z,\bar z)\right) e^{-2\alpha_0(z,\bar z)}\nonumber\\
& +\mathcal{O}(\epsilon^5).
\end{align}
Let us define
\begin{align}
\Im\{X(\theta),\theta\} = \sum_{n=0}^{\infty} \epsilon^n \mathcal{L}_3 S_n(\theta),
\end{align}
so that
\begin{align}\label{eq:defOfGtilde}
f_n(z) &  = \frac{i}{2} e^{-2 i \theta} \mathcal{P}(\mathcal{L}_3  S_n(\theta))|_{\theta \to -i \ln z}  = -\frac{1}{2}\left(z \p^3 + 3 \p^2\right)(\mathcal{P} S_n(\theta)|_{\theta \to -i \ln z})
\equiv -\frac{1}{2}\left(z \p^3 + 3 \p^2\right) \tilde G_n(z),
\end{align}
where 
\begin{align}
\tilde G_1(z) \equiv & \mathcal{P} G_1^r(\theta)|_{\theta \to -i \ln z},\nonumber\\
\tilde G_2(z) \equiv &
\mathcal{P}  \left(G_2^r(\theta) - G_1^i(\theta) \p_\theta G_1^r(\theta)\right)|_{\theta \to -i \ln z},\nonumber\\
\tilde G_3(z) \equiv &
\mathcal{P}  \Bigg(\left(G_3^r - G_1^i G_2^{r'} - G_2^i G_1^{r'}-\frac{1}{8}G_1^{r'}\p (G_1^r)^2 + \frac{1}{2}((G_1^i)^2 G_1^{r'})\right)\nonumber\\
&+\mathcal{L}_3^{-1}\bigg[2 K^{'} \mathcal{L}_3 G_1^{r}  +K \mathcal{L}_3 G_1^{r'}
-\frac{1}{4}\left((G_1^{r'})^2 -  2 G_1^r G_1^{r''}\right) \mathcal{L}_3 G_1^{r}\nonumber\\
&\qquad\qquad-\frac{1}{2}\left((G_1^{r'})^3 + 3 G_1^r G_1^{r'} G_1^{r''}\right)\bigg]\Bigg)\bigg|_{\theta \to -i \ln z}.
\end{align}
(See appendix \ref{app:general_higher_orders} for definition of $\tilde G_3(z)$).
Now, as was shown to first order in \cite{Kruczenski:2014bla}, with these definition we generally have
\begin{align}
f_i(z) \bar f_j(\bar z) e^{-2 \alpha_0(z,\bar z)} = &
-\frac{1}{2}\p \left(\bar f_j(\bar z)\left(2\bar z \tilde{G}_i(z) + (1-z\bar z)(2 \p \tilde{G}_i(z) + z(1-z \bar z)\p^2 \tilde{G}_i(z))\right)\right)
,\nonumber\\
\end{align}
so we can use Stokes theorem to convert the surface integrals over the disk into line integrals over the disk's boundary, that is
\begin{align}
\int d^2 z f_i(z) \bar f_j(\bar z) e^{-2 \alpha_0(z,\bar z)} = &
-\oint d\theta e^{- 2 i \theta} \bar f_j(e^{- i \theta}) \tilde{G}_i(e^{i \theta})
=-\frac{i}{2} \oint d\theta \mathcal{L}_3 \bar{\tilde{G}}_j \tilde{G}_i,
\end{align}
or
\begin{align}
\int d^2 z (f_i(z) \bar f_j(\bar z) + f_j(z) \bar f_i(\bar z))e^{-2 \alpha_0(z,\bar z)} = &
-\frac{i}{2}\oint d\theta \mathcal{H}(S_i) \mathcal{L}_3 S_j
=-\frac{i}{2}\oint d\theta \mathcal{H}(S_j) \mathcal{L}_3 S_i.
\end{align}
Before writing the area, let us mention that these integrals can be expressed also as
\begin{align}
\int d^2 z (f_i(z) \bar f_j(\bar z) + f_j(z) \bar f_i(\bar z))e^{-2 \alpha_0(z,\bar z)}
& = \frac{1}{4 \pi}\oint d s \oint d t \cot\left(\frac{s-t}{2}\right) S_i(t) \mathcal{L}_{3,s} S_j(s)\nonumber\\
& = -\frac{3}{16 \pi}\oint d s \oint d t \csc^4\left(\frac{s-t}{2}\right) S_i(t)  S_j(s).
\end{align}
where we wrote explicitly the Hilbert transform and integrated by parts to act on the kernel, instead of on the $S_i$ functions\footnote{This form is more similar to the expressions one gets when computing the Wilson loop at one-loop at weak coupling.}.
Using the expressions above, we can write the regularized area as
\begin{align}
\mathcal{A}_{\text{reg}} & = -2\pi
+\epsilon^2 \frac{i}{2}\oint d\theta \mathcal{H}(G_1^r) \mathcal{L}_3 G_1^r
+\epsilon^3 i \oint d\theta \mathcal{H}(G_1^r) \mathcal{L}_3 (G_2^r - G_1^i \p_\theta G_1^r)
+\mathcal{O}(\epsilon^4).
\end{align}
We did not find such a nice expression for the fourth order contribution, see appendix \ref{app:general_higher_orders}.

Finally, let us comment on the universality of this result. The weak coupling one-loop integral is proportional to
\begin{align}
W_1 =  \oint ds \oint dt \frac{\dot x(t)\cdot \dot x(s)-|\dot x(s)||\dot x(t)|}{(x(s)-x(t))^2}.
\end{align}
Throughout the paper we ignore the $\frac{\lambda}{16 \pi^2}$ prefactor\footnote{here $\lambda$ corresponds to the 't Hooft coupling and not the spectral parameter.} which should appear in front of $W_1$.
It is straightforward to expand $W_1$ to third order in $\epsilon$, and after some algebraic manipulations to arrive at
\begin{align}
W_1 & = -2\pi^2
+2\pi \epsilon^2 \frac{i}{2}\oint d\theta \mathcal{H}(G_1^r) \mathcal{L}_3 G_1^r
+2\pi \epsilon^3 i \oint d\theta \mathcal{H}(G_1^r) \mathcal{L}_3 (G_2^r - G_1^i \p_\theta G_1^r)
+\mathcal{O}(\epsilon^4).
\end{align}
Thus, the second and third order terms have the same form at weak and strong coupling so these terms are universal.
The reason for that is conformal symmetry \cite{Semenoff:2004qr,Galakhov:2008ax}.
It was shown explicitly in \cite{Galakhov:2008ax} that the form of the fourth order terms is different.

\section{Applications}\label{sec:applications}
In this section we study some specific examples in detail.
We will examine several deformations of the circle with different symmetry properties.
We will consider a family of contours which interpolate between the circle and hypocycloids.
These hypocycloids have cusps, and we do not expect the expansion to be valid for values where the cusps become significant (and logarithmic divergences are expected), instead, the expansion should be valid when the contour is relatively smooth.
One special case of these contours is the ellipse which we will consider in more details first.
These contours have both reflection and discrete rotational symmetries, according to the number of cusps.
We will also consider a different family of contours with the same symmetries as the previous family.
After that we will consider a contour with less symmetries, the lima\c{c}on which has one reflectional symmetry.
At last, we will consider a contour with no discrete symmetries.

Since the analysis for different contours is conceptually very much the same, we chose to present a more detailed derivation only for the ellipse.
In the other case we only quote the final results.
In all cases we present results to very high order in the expansion parameter relative to the ordinary wavy approximation.

\subsection{Ellipse}

We express the ellipse curve by
\begin{align}
X(\theta)=e^{i F(\theta)} + i \epsilon \sin F(\theta),
\end{align}
where the parametrization function $F(\theta)$ is a real monotonic function satisfying $F(0)=0$ and $F(2\pi) = 2\pi$ (for definiteness, although there is some freedom which allows more general boundary conditions), and $\epsilon$ is a real small number, which will serve as an expansion parameter.
$\epsilon = 0 $ corresponds to the circle where all the details of the solution for the minimal surface are known and provided in subsection \ref{sec:The_circular_Wilson_loop_properties}.
Next we are going to solve the problem order by order in $\epsilon$.
First we take the parameterizations function to have the form
\begin{equation}\label{eq:ellipse_param}
F(\theta) = \theta + \sum_{n=1}^{\infty} F_n(\theta)\epsilon^n,
\end{equation}
with $F_n(\theta)$ satisfying $F_n(0) = F_n(2\pi)  =0$.
We shall rely heavily on the Schwarzain derivative relations (\ref{eq:SDrelation}), where in this case the Schwarzian derivative is given by
\begin{equation}
\{X(\theta),\theta\} =
\frac{\left(1-\frac{3 \epsilon  (2+\epsilon )}{((1+\epsilon ) \cos(F(\theta))+i \sin(F(\theta)))^2}\right) F'(\theta)^4-3 F''(\theta)^2+2 F'(\theta) F'''(\theta)}{2 F'(\theta)^2},
\end{equation}
where the primes represent a derivative with respect to $\theta$.
In the following we will need the real and imaginary parts of the Schwarzian derivative which are given by
\begin{equation}
\Re\{X(\theta),\theta\} =
\frac{\left(1+\frac{3 \epsilon  (2+\epsilon ) \left(-(1+\epsilon )^2 \cos(F(\theta))^2+\sin(F(\theta))^2\right)}{\left((1+\epsilon )^2 \cos(F(\theta))^2+\sin(F(\theta))^2\right)^2}\right) F'(\theta)^4-3 F''(\theta)^2+2 F'(\theta) F'''(\theta)}{2 F'(\theta)^2},
\end{equation}
\begin{equation}
\Im\{X(\theta),\theta\} =
\frac{6 \epsilon  (1+\epsilon ) (2+\epsilon ) \sin(2 F(\theta)) F'(\theta)^2}{(2+\epsilon  (2+\epsilon )+\epsilon  (2+\epsilon ) \cos(2 F(\theta)))^2}.
\end{equation}
From these equations we can read $f(z)$ and $\beta_2(\theta)$,
\begin{align}\label{eq:beta_and_f_SD}
\beta_2(\theta) = & \frac{1}{12}\left(\frac{1}{2}-\Re\{X(\theta),\theta\}\right),\nonumber\\
f(e^{i\theta}) = & -\frac{i}{2}e^{- 2 i \theta}\mathcal{P}\left(\Im\{X(\theta),\theta\}\right).
\end{align}
Next we expand the real function $\alpha(r,\theta)$ and the holomorphic function $f(z)$ as,
\begin{align}
\alpha(r,\theta) = \sum_{n=0}^{\infty} \alpha_n(r,\theta)\epsilon^n,\quad
f(z) = \sum_{n=1}^{\infty} f_n(z)\epsilon^n,
\end{align}
where $\alpha_0 = \ln\frac{1}{1-z\bar z} = \ln\frac{1}{1-r^2}$.

As for the general analysis, the procedure goes as follows.
First we use the fact that given a solution to the problem to order $\epsilon^n$, $\Im\{X,\theta\}$ is known to order $\epsilon^{n+1}$.
This allows us to (easily) extract $f(z)$ to order $\epsilon^{n+1}$.
Thus, we start with the circular Wilson loop solution and immediately read $f_1 = -\frac{3}{4}$.
Next we plug $f(z)$ in the generalized cosh-Gordon equation and solve for $\alpha(r,\theta)$.
We use the two implicit boundary conditions, $\alpha_{n>0}(r=1) = 0$ and $\alpha_{n}(r=0) = \text{finite}$.
Then we expand $\alpha(r,\theta)$ around $r=1$, extract $\beta_2(\theta)$ and plug in (\ref{eq:beta_and_f_SD}) to solve for $F_n(\theta)$.

\begin{figure}
    \centering
    \includegraphics[trim = 0mm 0mm 0mm 0mm,clip,width = 0.7\textwidth]{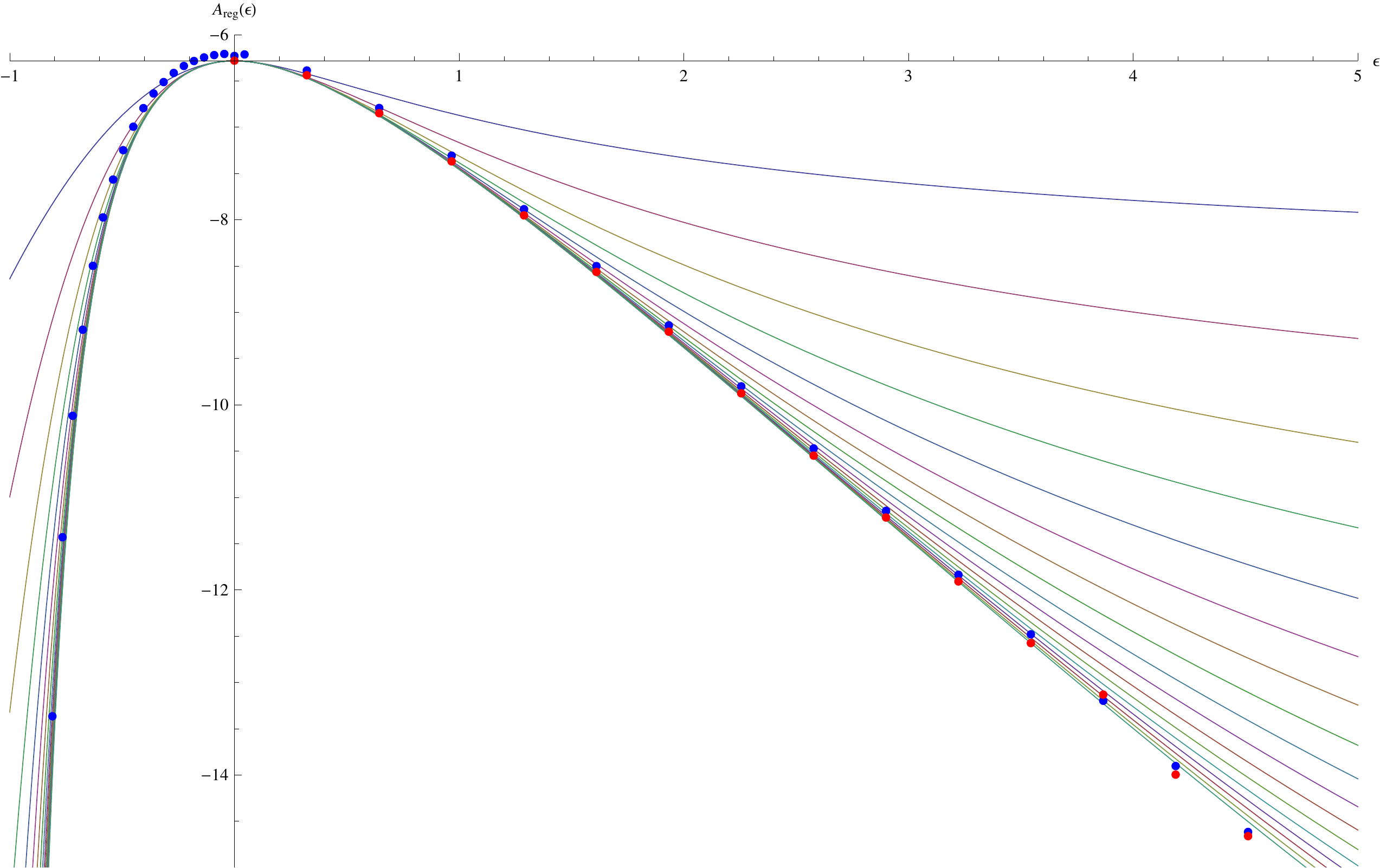}
    \caption{The area of the minimal surface ending on an ellipse contour parameterized by $(\cos(s),(1+\epsilon)\sin(s))$.
    The different lines correspond to different orders in the $\epsilon$ expansion (from order $\epsilon^2$ up to order $\epsilon^{18}$), where it is seen how the lines converge. For $\epsilon>0$ we plot the map of the solution for $-1<\epsilon<0$ which behaves better at this region as explained. On top of the graph we added the points generated numerically in \cite{Fonda:2014cca}. In the simulation in \cite{Fonda:2014cca}, the authors parameterized the ellipse as $\frac{x^2}{R_1^2}+\frac{y^2}{R_2^2}=1$, and the blue points correspond to taking $R_1=1$ and so our $x$-axis is $\epsilon = (R_2/R_1)-1$. The red dots correspond to taking $R_1=2$. As one can see the accuracy in these two cases in slightly different. Moreover, the numerics are limited by the IR cutoff (which was 0.03 in this case), so we do not expect a perfect agreement, although it is easy to see how our analytic prediction agrees with their data. We see that our approximation begins to break for $\epsilon>4$, which correspond to $R_2/R_1 = 5$ where the ellipse is already very different from the circle we started with.}
    \label{fig:ellipse_area}
\end{figure}

\begin{figure}
\begin{center}
\subfloat[][]{
  \includegraphics[width=70mm]{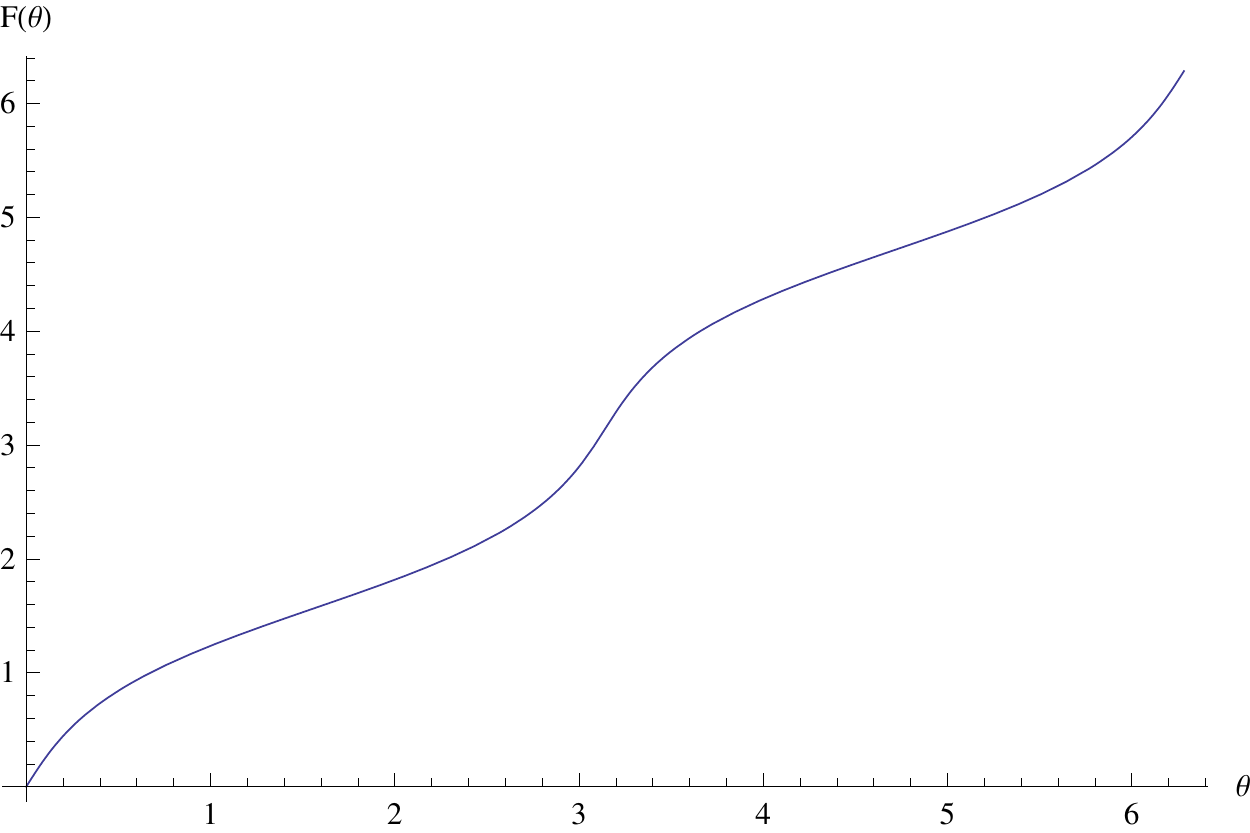}} \hspace{3mm}
\subfloat[][]{
  \includegraphics[width=70mm]{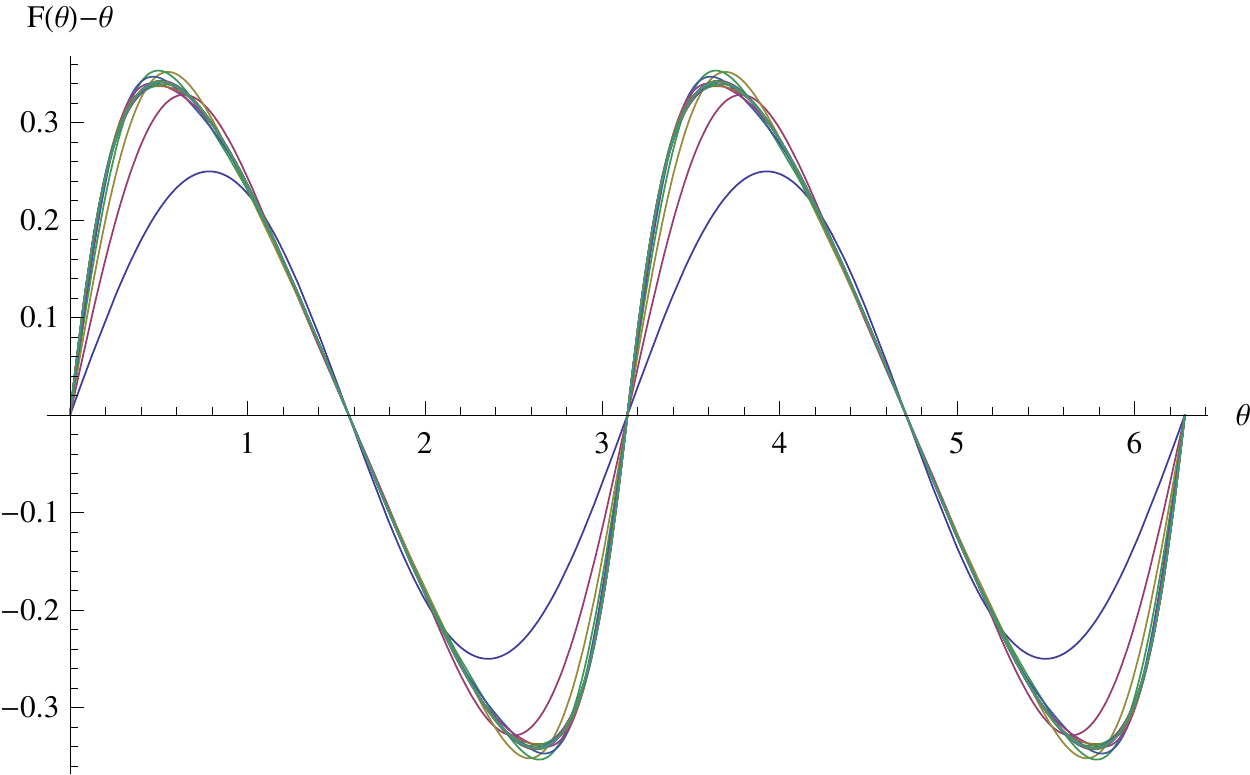}} \hspace{3mm}
\caption{(a) The parametrization as a function of $s$ for $\epsilon = -1/2$ (which is equivalent to $\epsilon  = 1$) up to order $\epsilon^{17}$
(b) $F(\theta)-\theta$ as a function of $\theta$ for $\epsilon = -1/2$ (which is equivalent to $\epsilon  = 1$) for all orders up to order $\epsilon^{17}$.}
\label{fig:ellipse_param}
\end{center}
\end{figure}

Due to the symmetries of the ellipse the expansions take the form 
\begin{align}
F(\theta) = \theta+\sum_{n=1}^\infty F_n(\theta)\epsilon^n,\quad
F_n(\theta) = \sum_{k=1}^n F_{n,k}\sin 2 k \theta,
\end{align}
\begin{align}
f(z) = \sum_{n=1}^\infty f_n(z)\epsilon^n,\quad
f_n(z) = \sum_{k=0}^{n-1} f_{n,k} z^{2 k},
\end{align}
\begin{align}
\alpha(z,\bar z) & = \sum_{n=0}^\infty \alpha_n(z,\bar z)\epsilon^n,\quad
\alpha_n(z,\bar z) = \sum_{k=0}^{n-2} \alpha_{n,k}(r) \cos 2 k \theta,
\end{align}
and it turns out that the expansion procedure is quite simple and can be carried to high orders.
Moreover, we observe a pattern in the functional form of $\alpha(r,\theta)$,
\begin{align}
\alpha_{n,k}(r) & = r^{2 k}(1-r^2)^2\sum_{l=0}^{\frac{1}{4}\left(6(n-k)-11+3(-1)^{n-k}\right)}\alpha_{n,k,l}r^{2 l},
\end{align}
which holds at least to the 18th order in $\epsilon$, and we suspect it might hold to any order.
This reduces the problem to an algebraic one of finding constant coefficients.
Finally, we find the following expression for the regularized area
\begin{align}\label{eq:ellipse_area}
A_{\text{reg}}(\epsilon) = & -2\pi -\frac{3 \pi  \epsilon ^2}{4}+\frac{3 \pi  \epsilon ^3}{4}-\frac{237 \pi  \epsilon ^4}{320}+\frac{117 \pi  \epsilon ^5}{160}-\frac{64881 \pi  \epsilon ^6}{89600}+\frac{64443 \pi  \epsilon ^7}{89600}-\frac{14373577 \pi  \epsilon ^8}{20070400}\nonumber\\
&+\frac{3584953 \pi  \epsilon ^9}{5017600}-\frac{110314688219 \pi  \epsilon ^{10}}{154542080000}+\frac{22064732579 \pi  \epsilon ^{11}}{30908416000}-\frac{6630907488364381 \pi  \epsilon ^{12}}{9281797324800000}\nonumber\\
&+\frac{1106373532973931 \pi  \epsilon ^{13}}{1546966220800000}-\frac{40943000996733445243 \pi  \epsilon ^{14}}{57175871520768000000}\nonumber\\
&+\frac{1952095942839819321 \pi  \epsilon ^{15}}{2722660548608000000}
-\frac{157750690929831538029244697 \pi  \epsilon ^{16}}{219774901986388869120000000}\nonumber\\
&+\frac{19736906966190071806502297 \pi  \epsilon ^{17}}{27471862748298608640000000}
-\frac{801650044535506237372382994066703 \pi  \epsilon ^{18}}{1115068403809909423032238080000000}\nonumber\\
&
+\mathcal{O}(\epsilon^{19}).
\end{align}
Notice the alternating sign in the expression which implies that $A_{\text{reg}}(\epsilon)$ gives a better approximation for $-1<\epsilon<0$ rather than $\epsilon > 0$.
Fortunately, due to conformal symmetry it is enough to consider $-1<\epsilon<0$ only, where the $\epsilon>0$ values are related by  $\epsilon' = \frac{-\epsilon}{1+\epsilon}$ where $-1<\epsilon<0$.
In figure \ref{fig:ellipse_area} we plot the area as a power of $\epsilon$ using different $\epsilon$ power approximations from order $\epsilon^2$ to $\epsilon^{18}$. For $\epsilon>0$ we plot the image value using (\ref{eq:ellipse_area}) for $\epsilon<0$.
We also plot numerical data points on top of the functions using the data from \cite{Fonda:2014cca}\footnote{I am grateful to P. Fonda and E. Tonni for sharing this numerical data.}.
It is also interesting to see how the parametrization function looks like, keeping in mind that it is enough to find this function in order to solve the problem. We plot the function in figure (\ref{fig:ellipse_param}).

\begin{figure}
\begin{center}
\subfloat[][]{
  \includegraphics[width=28mm]{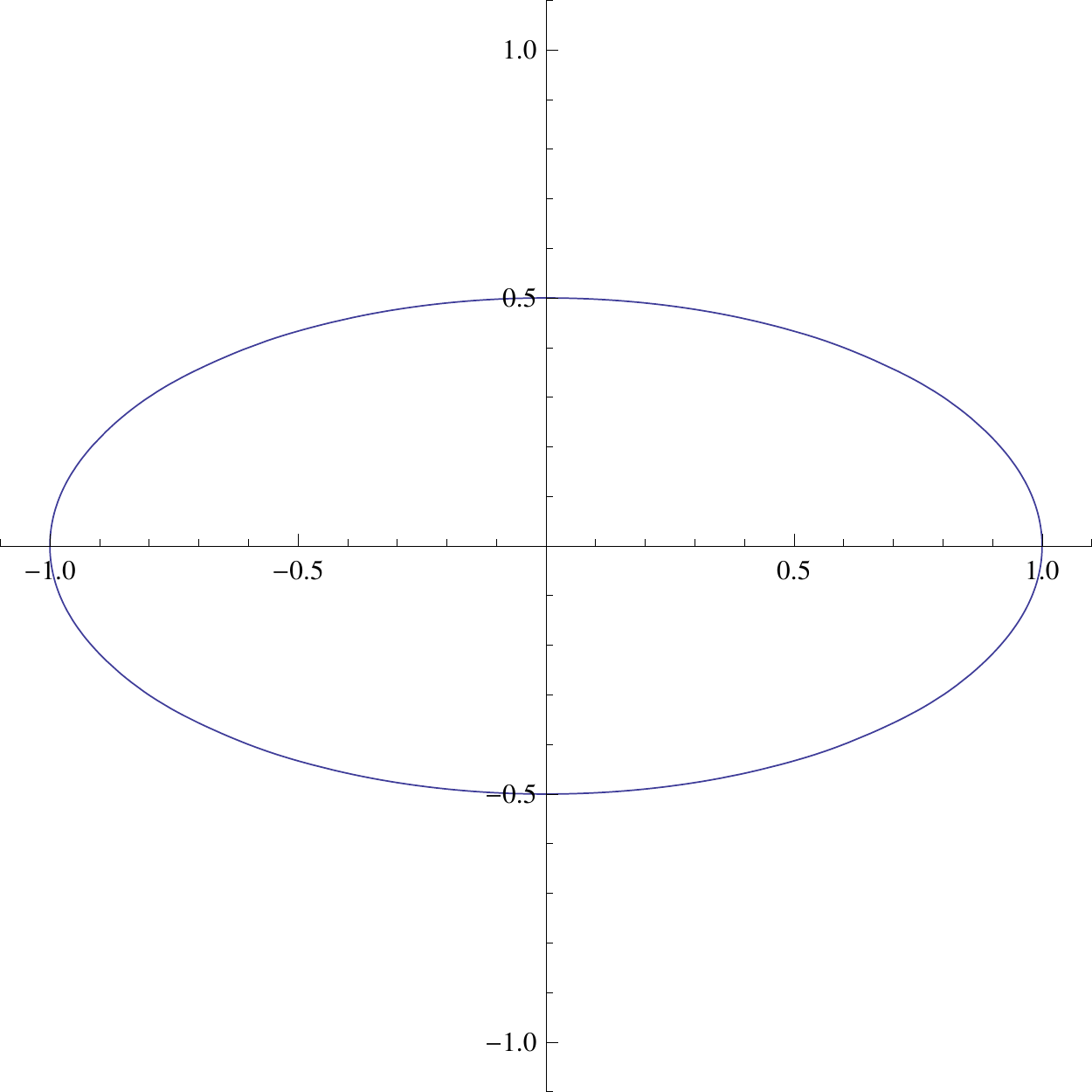}} \hspace{3mm}
\subfloat[][]{
  \includegraphics[width=28mm]{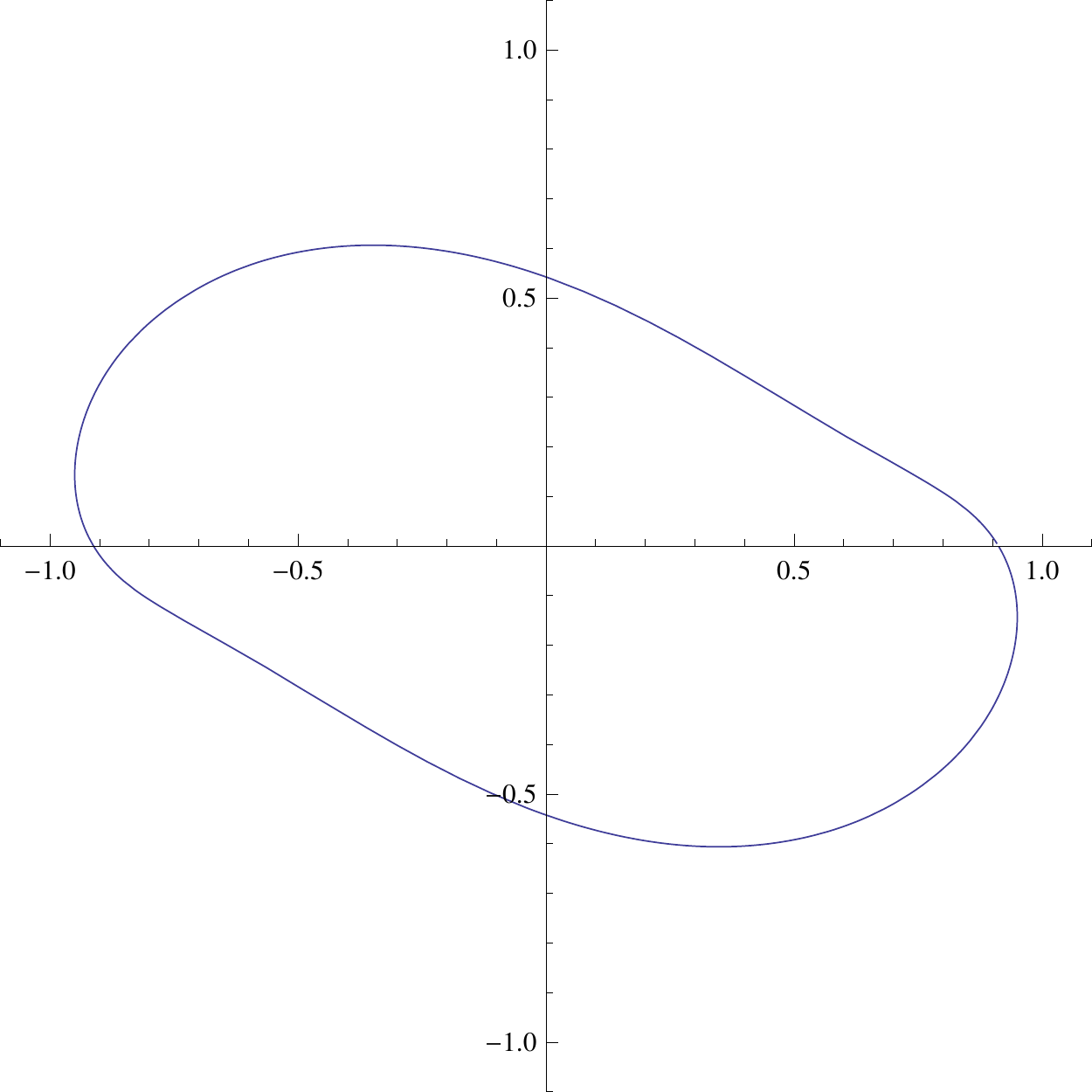}} \hspace{3mm}
\subfloat[][]{
  \includegraphics[width=28mm]{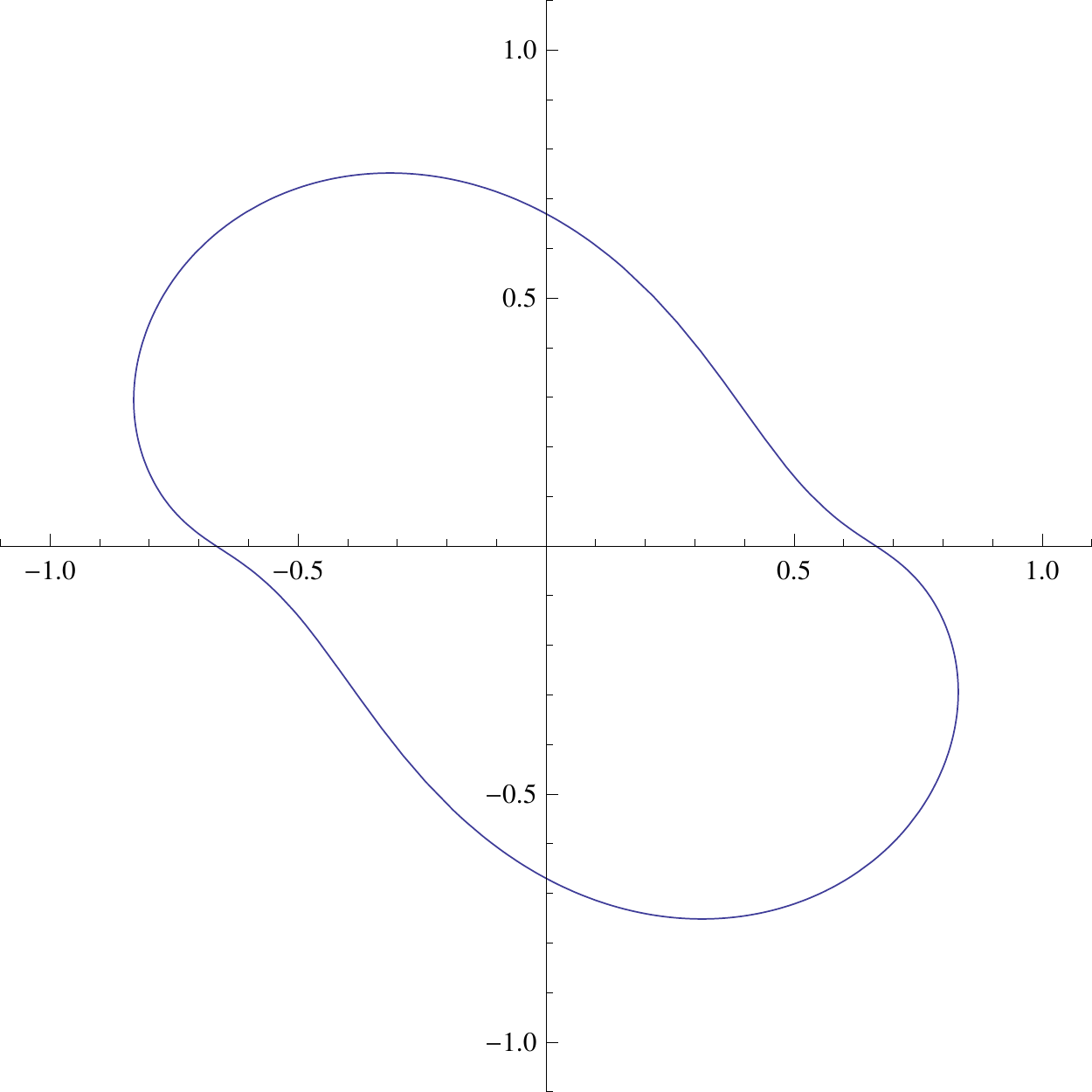}} \hspace{3mm}
\subfloat[][]{
  \includegraphics[width=28mm]{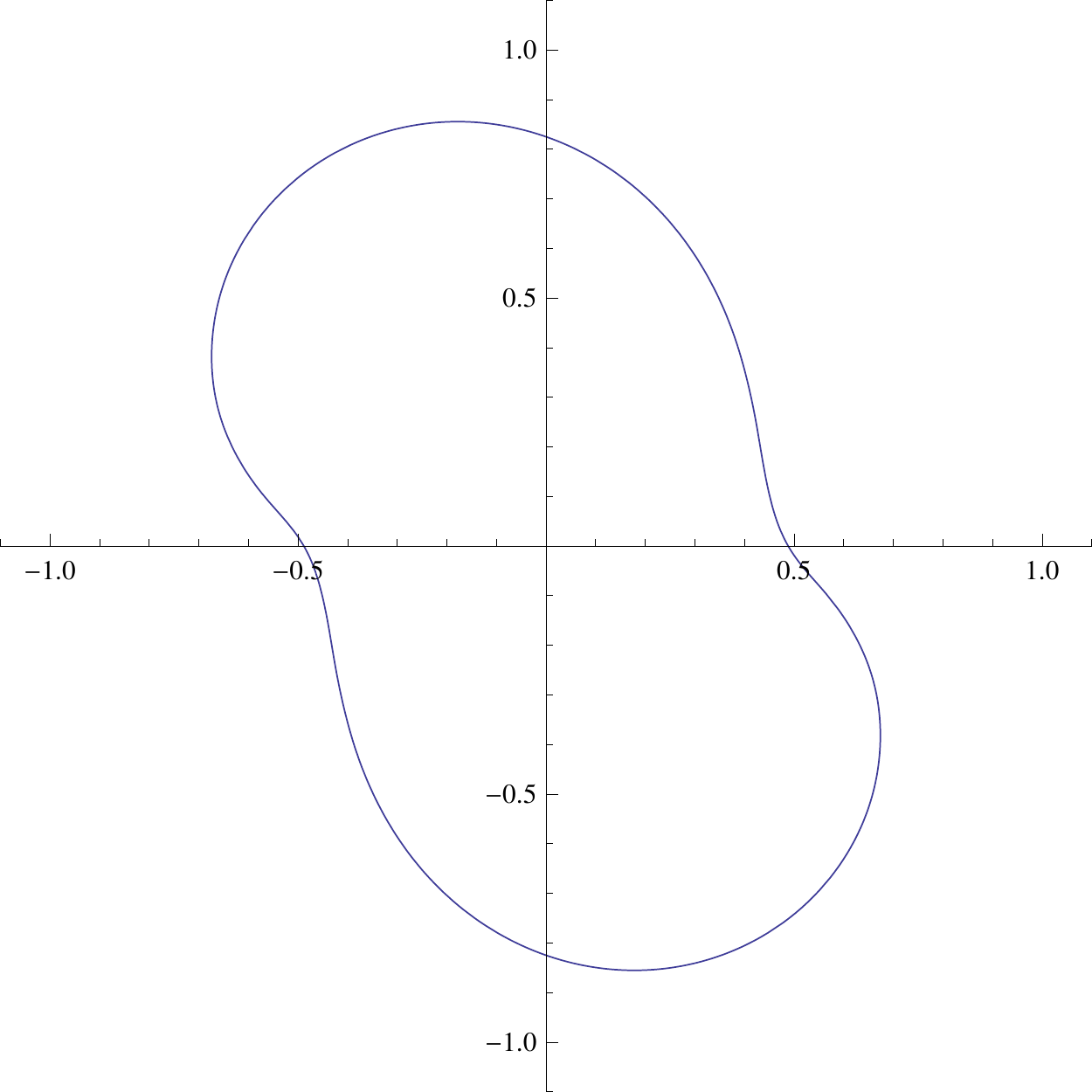}} \hspace{3mm}  
\subfloat[][]{
  \includegraphics[width=28mm]{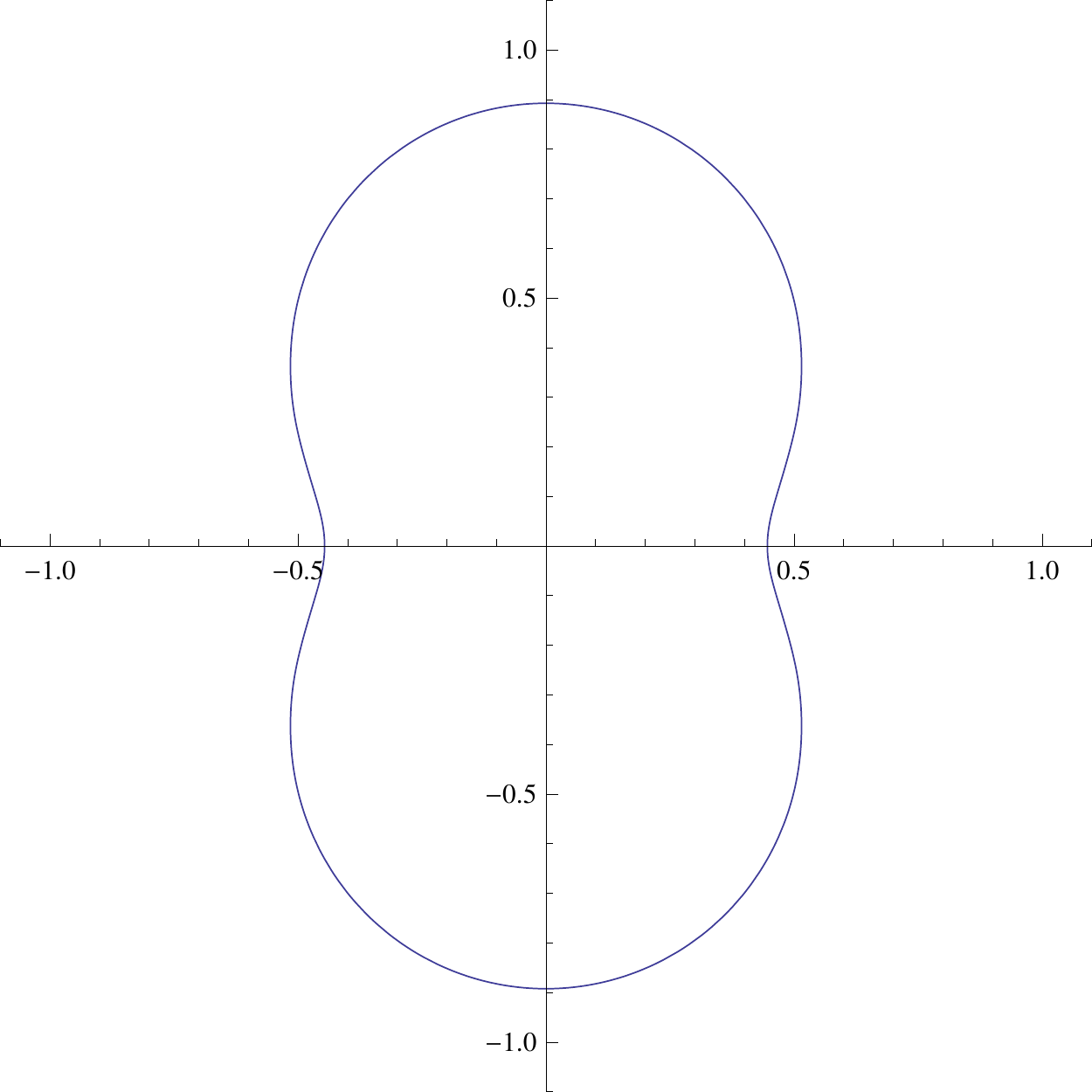}} \hspace{3mm}  
\caption{$\lambda$ deformations for the $\epsilon=-1/2$ (or equivalently $\epsilon=1$) ellipse. We use $\lambda = e^{i \varphi}$ and in 
(a) $\varphi = 0$, 
(b) $\varphi = \frac{\pi}{4}$ 
(c) $\varphi = \frac{\pi}{2}$ 
(d) $\varphi = \frac{3\pi}{4}$ 
and in (e) $\varphi = \pi$. 
The computation is carried to order $\mathcal{O}(\epsilon^{13})$.}
\label{fig:ellipse_deform}
\end{center}
\end{figure}

As discussed in the previous sections, there exists a one parameter family of deformations of the ellipse contour, $\lambda$-deformations, for which the area does is left invariant.
Given the analysis above it is relatively easy to find these contours $X^{\lambda}$ to the same order in $\epsilon$, by solving
\begin{align}
\{X^{\lambda}(\theta),\theta\} = \Re\{X^{\lambda=1}(\theta),\theta\} - 2 \lambda f(\theta)e^{2 i \theta} + \frac{2}{\lambda}\bar f(\theta)e^{-2 i \theta},
\end{align}
where $f(\theta) = f(r=1,\theta)$, with the "initial" condition $X^{\lambda = 0}(\theta) = X(\theta)$.
Since $\lambda$ must be a phase, we define $\lambda = e^{i \varphi}$ with $\varphi\in [0,2\pi]$. We plot the deformed contour for $\varphi = 0, \pi/4 \pi/2, 3\pi/4, \pi$ in figure (\ref{fig:ellipse_deform}).
Notice that for the circle this deformation amounts to rotations and does not give new solutions.

\subsubsection{Weak coupling and $\lambda$-deformation}
\begin{figure}
    \centering
    \includegraphics[trim = 0mm 0mm 0mm 0mm,clip,width = 0.6\textwidth]{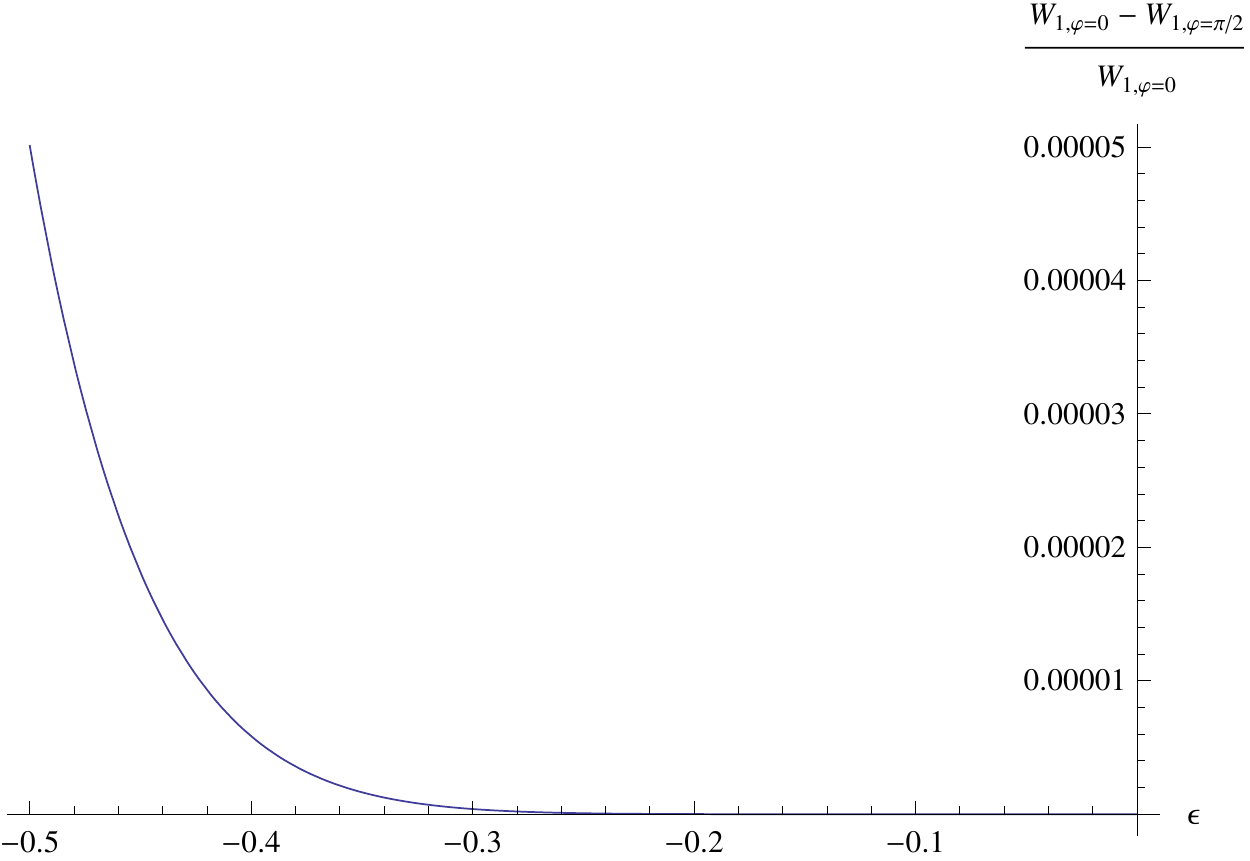}
    \caption{The relative deviation of the one-loop weak coupling expectation value of the $\lambda$-deformed ellipse with $\varphi=\pi/2$ compared to the undeformed ellipse, using equation (\ref{eq:ellipse_weak_coupling}).
    As can be seen, the difference is very small. 
    Notice that the expectation value of the deformed Wilson loop is smaller than the one of the ellipse for any $\varphi$.}
    \label{fig:ellipseAreaLambdaDef}
\end{figure}

One can speculate whether the $\lambda$-deformation of the Wilson loop contour leaves it invariant beyond the strong coupling limit.
$\alpha'$ quantum corrections are hard to compute, however, the one loop weak coupling quantum correction is easier to evaluate.
Next, we compute the one loop weak coupling correction to the for the $\lambda$-deformed ellipse $X^\lambda(\theta)$, by expanding in powers of $\epsilon$.
We take $\lambda$ to be a phase, $\lambda = e^{i \varphi}$, and denote by a subscript the $\varphi$  dependence.
We find
\begin{align}\label{eq:ellipse_weak_coupling}
W_{1,\varphi}   = &
 \oint ds \oint dt \frac{\dot x(t)\cdot \dot x(s)-|\dot x(s)||\dot x(t)|}{(x(s)-x(t))^2} \nonumber\\
= & -2 \pi ^2-\frac{3 \pi ^2 \epsilon ^2}{2}+\frac{3 \pi ^2 \epsilon ^3}{2}-\frac{101 \pi ^2 \epsilon ^4}{64}  + \frac{53 \pi ^2 \epsilon^5}{32} -\frac{1759 \pi ^2 \epsilon^6}{1024} + \frac{1805 \pi ^2 \epsilon^7}{1024} \nonumber\\
 &-\frac{\pi ^2 (8227199+7776 \cos (2 \varphi)) \epsilon^8}{4587520} + \frac{\pi ^2 (2078399+7776 \cos (2 \varphi)) \epsilon^9}{1146880}\nonumber\\
 & -\frac{\pi ^2 (4632279678637+40879629488 \cos( 2 \varphi )) \epsilon ^{10}}{2543321088000}\nonumber\\
& +\frac{\pi^2 (927152406637 +
    15013543088 \cos(2 \varphi)) \epsilon^{11}}{508664217600}\nonumber\\
& -\frac{\pi ^2 (740574919262795417+18893380267080208 \cos(2 \varphi )+30047328000000 \cos(4 \varphi )\epsilon ^{12}}{407338305454080000}\nonumber\\
& +\frac{\pi ^2 (368445596660191451+13339409812485424 \cos(2 \varphi )+90141984000000 \cos(4 \varphi )) \epsilon ^{13}}{203669152727040000}\nonumber\\
& +\mathcal{O}(\epsilon^{14}),
\end{align}
The invariance to third order is expected since the coefficients to that order are universal due to conformal symmetry.
However starting from order $\epsilon^4$ the coefficients should not be fixed by conformal symmetry alone.
Quite surprisingly, the first appearance of $\lambda$ dependence shows at the 8th order.
Thus, eventually the symmetry is broken by quantum corrections.
Interestingly, the relative difference is extremely small as can be seen is figure \ref{fig:ellipseAreaLambdaDef} where we show the relative difference between $W_{1,\varphi=0}$ and $W_{1,\varphi=\pi/2}$, where it is maximal.

\subsubsection{Lax operator}

Finally, it is interesting to study the Lax operator and the resulting algebraic curve.
It is known that minimal surface solutions in $\HH_3$ are related to theta functions defined on a hyperelliptic Riemann surface of odd genus \cite{babich1993,Ishizeki:2011bf,Kruczenski:2013bsa,Cooke:2014uga}.
However, the relation between a given Riemann surface and a contour on the AdS boundary is poorly understood.
Given a generic algebraic curve, it is not known a priori what is the corresponding extremal surface (or even if a closed smooth nonintersecting extremal surface exists) and vise verse.
For example, it is not known what is the Riemann surface corresponding to the minimal surface ending on the ellipse, what is the genus and even if it is finite.

Next we construct a Lax operator for the ellipse by solving the equation
\begin{align}
d L + [J , L]  = 0,
\end{align}
order by order in $\epsilon$, starting with the circular solution, where the Lax operator is given by
\begin{align}
L_0(z,\bar z) = \frac{\lambda}{1-z \bar z}\left(
                                      \begin{array}{cc}
                                        1 + z \bar z & 2\frac{\bar z}{\lambda} \\
                                        -2 \lambda z & -1-z\bar z \\
                                      \end{array}
                                    \right).
\end{align}
The Lax operator is not unique\footnote{For example, any power of the Lax operator yields another Lax operator as well as any of their linear combinations. In the case of the circle $L_0^2 = 1$ which is a "trivial" Lax operator, more generally we could have started with $a L_0 + b$.}, but the resulting algebraic curves for different Lax operators should be related by a bi-rational transformation.
At each order in $\epsilon$ we find some integration constants which we can set as we wish.
We choose these constants such that we arrive at the simplest result.

The explicit form of the Lax operator for higher orders in $\epsilon$ is not very illuminating, however, simple algebraic curve is found by taking its determinant.
Each time when we increase the power of $\epsilon$ the resulting Lax operator develops an extra $\lambda^{-1}$ factor in the $L_{12}$ component (similar to the one in $L_0$ if we ignore the overall $\lambda$ pre-factor),
which increases the order of the pole by $1$ (a similar observation was made in \cite{Cagnazzo:2013dqa} where the general wavy approximation was considered.).
However, the Lax operator remains a rational function of the spectral parameter which implies it is a good operator \cite{Janik:2012ws}.
Moreover, we can set the integration constants and an overall constant factor such that
\begin{align}
y^2 = -\det L = 1 + \mathcal{O}(\epsilon^{16}).
\end{align}
Thus, we do not observe any interesting structure which can point to a relation with the Riemann surface corresponding to the exact solution.
We comment more on this issue in the discussion section.
We found the same result, namely a trivial curve, in all the examples we have looked into, thus we shall not repeat this fact in the following subsections.

\subsection{Hypocycloids}
\begin{figure}
\begin{center}
\subfloat[][]{
  \includegraphics[width=29mm]{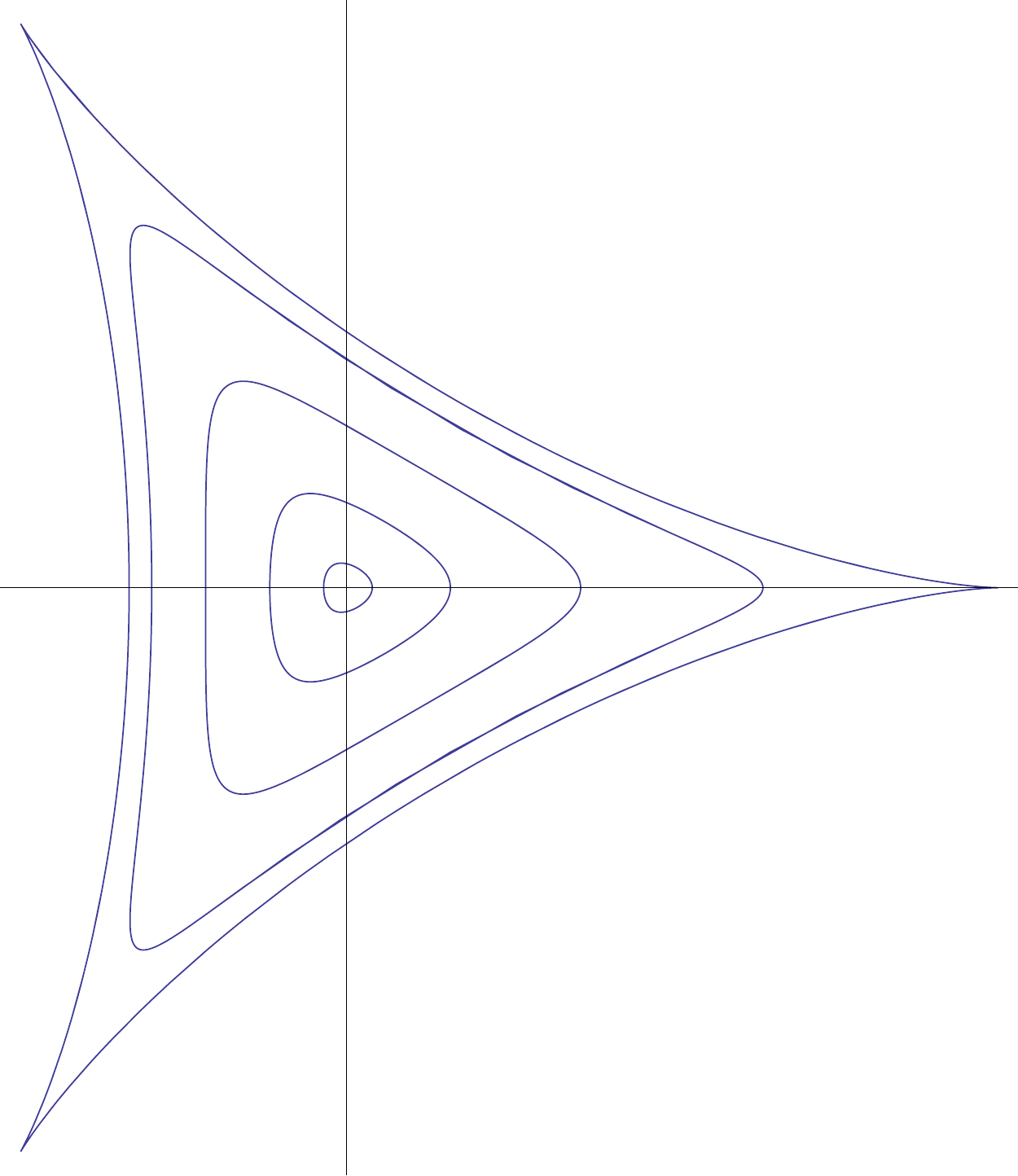}} \hspace{3mm}
\subfloat[][]{
  \includegraphics[width=29mm]{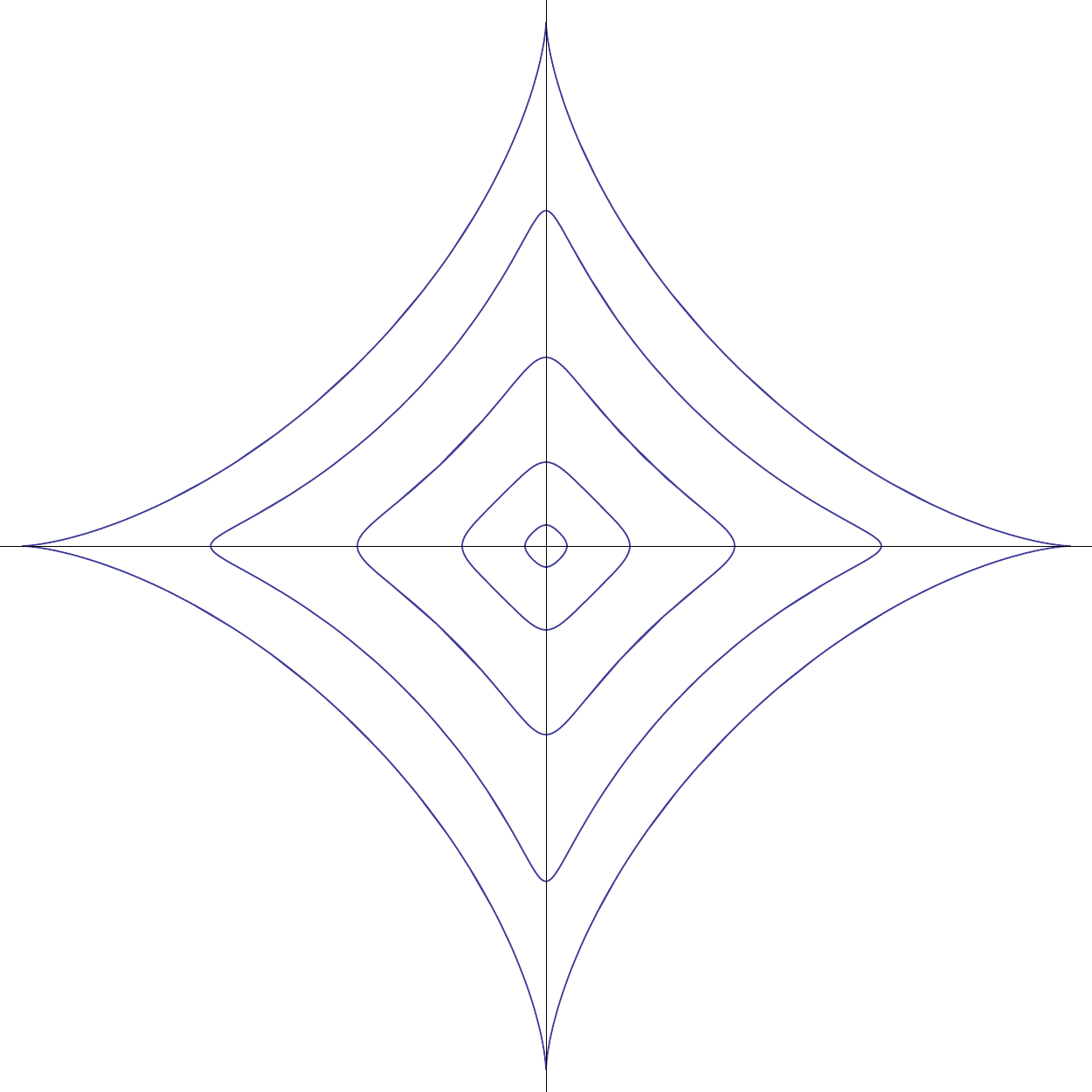}} \hspace{3mm}
\subfloat[][]{
  \includegraphics[width=29mm]{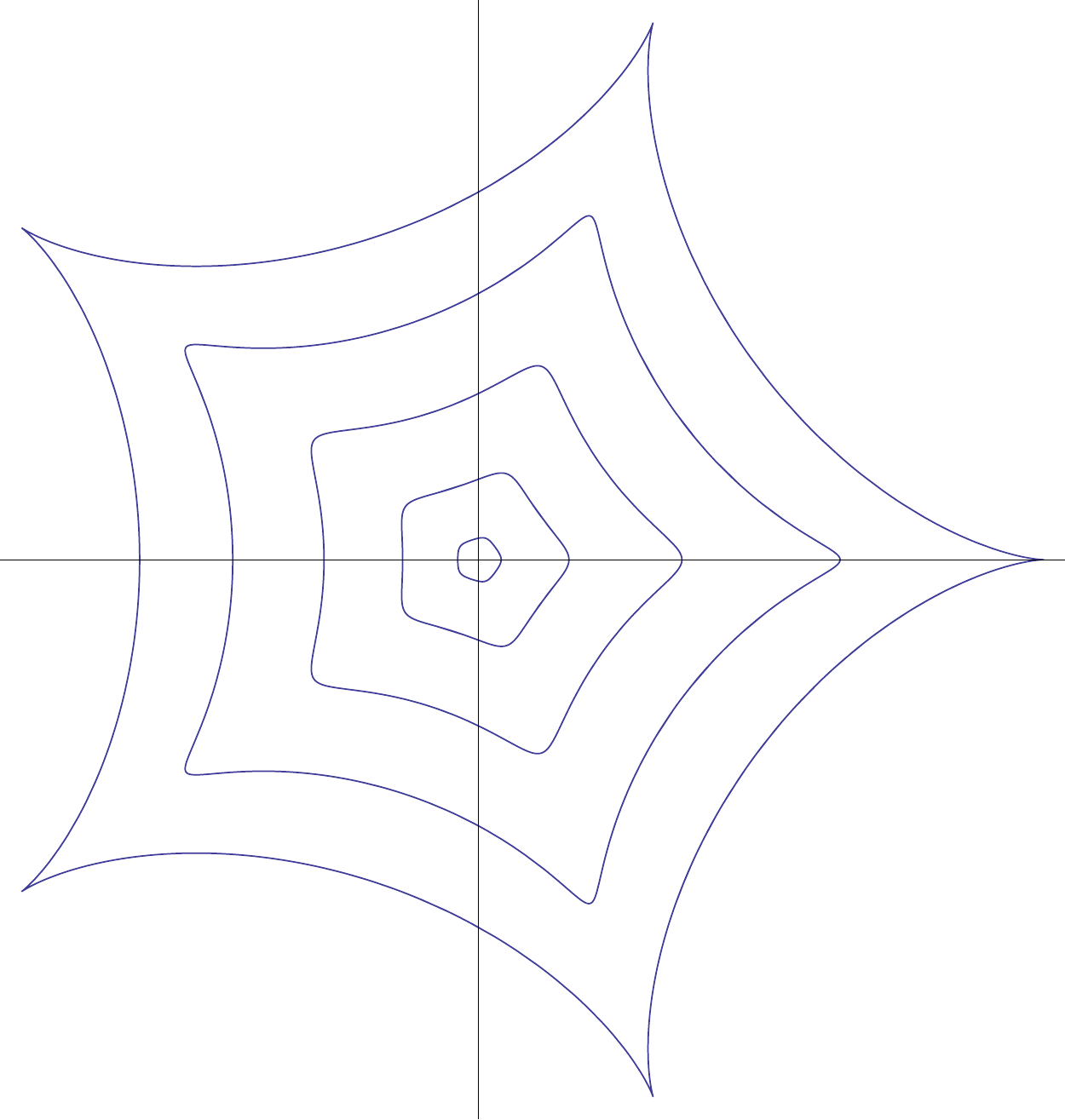}} \hspace{3mm}
\subfloat[][]{
  \includegraphics[width=29mm]{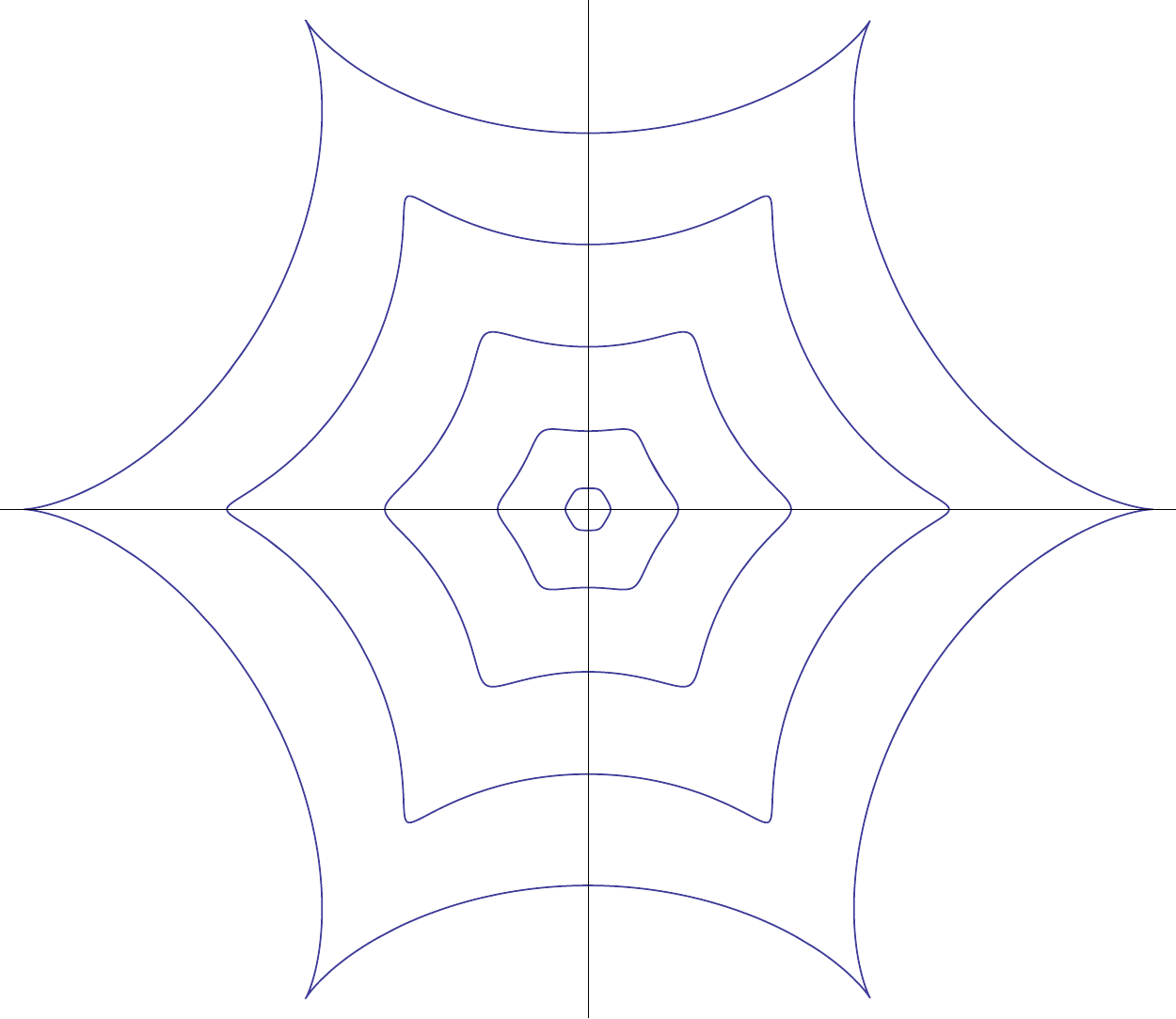}} \hspace{3mm}
\caption{\textbf{Hypocycloids to circle deformation.}
Here we plot some examples of (\ref{eq:hypo}) for $n=3,4,5,6$, for different values of $-\frac{2}{n}<\epsilon<0$.
For clarity, the figures are scaled to avoid intersections.
}
\label{fig:hypocycliods}
\end{center}
\end{figure}

In this section we apply the above procedure to a family of symmetric contours which interpolates between the circle and hypocycloids, which are n-cusped symmetric contours (see figure \ref{fig:hypocycliods}), given by
\begin{align}\label{eq:hypo}
X(\theta) = e^{i F(\theta)}+i \epsilon e^{i (1-n/2) F(\theta)}  \sin(p F(\theta)),
\end{align}
where we recover the ellipse for $n=2$.
Accordingly, the contours have a symmetry under $n-1$ discrete rotations and $n$ reflections.
The cusps appear when $\epsilon = -2/n$, and due to conformal symmetry if is enough to consider $-2/n<\epsilon<0$ (for non-intersecting contours).

\begin{figure}
\begin{center}
\subfloat{
  \includegraphics[width=25mm]{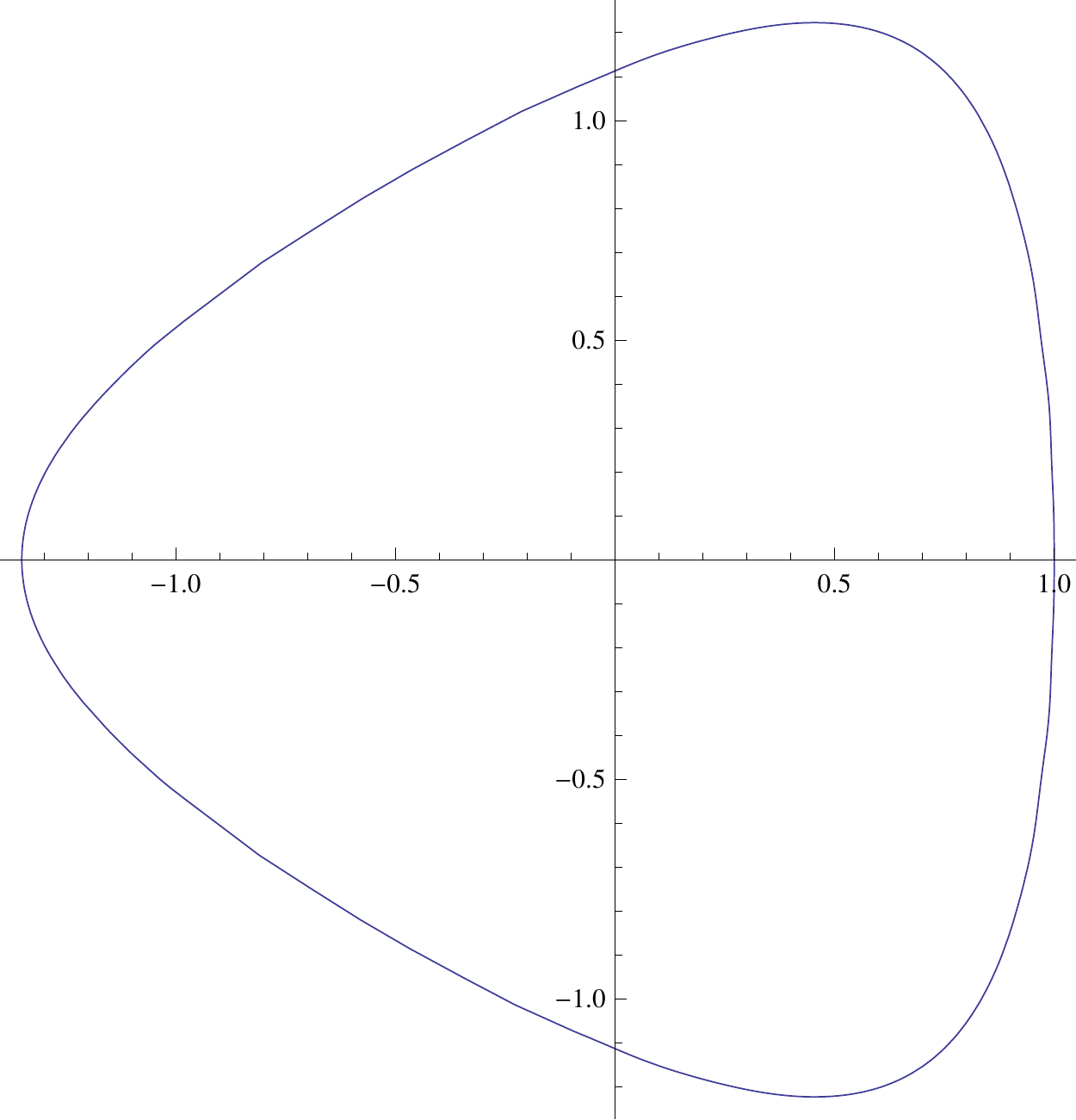}} \hspace{3mm}
\subfloat{
  \includegraphics[width=25mm]{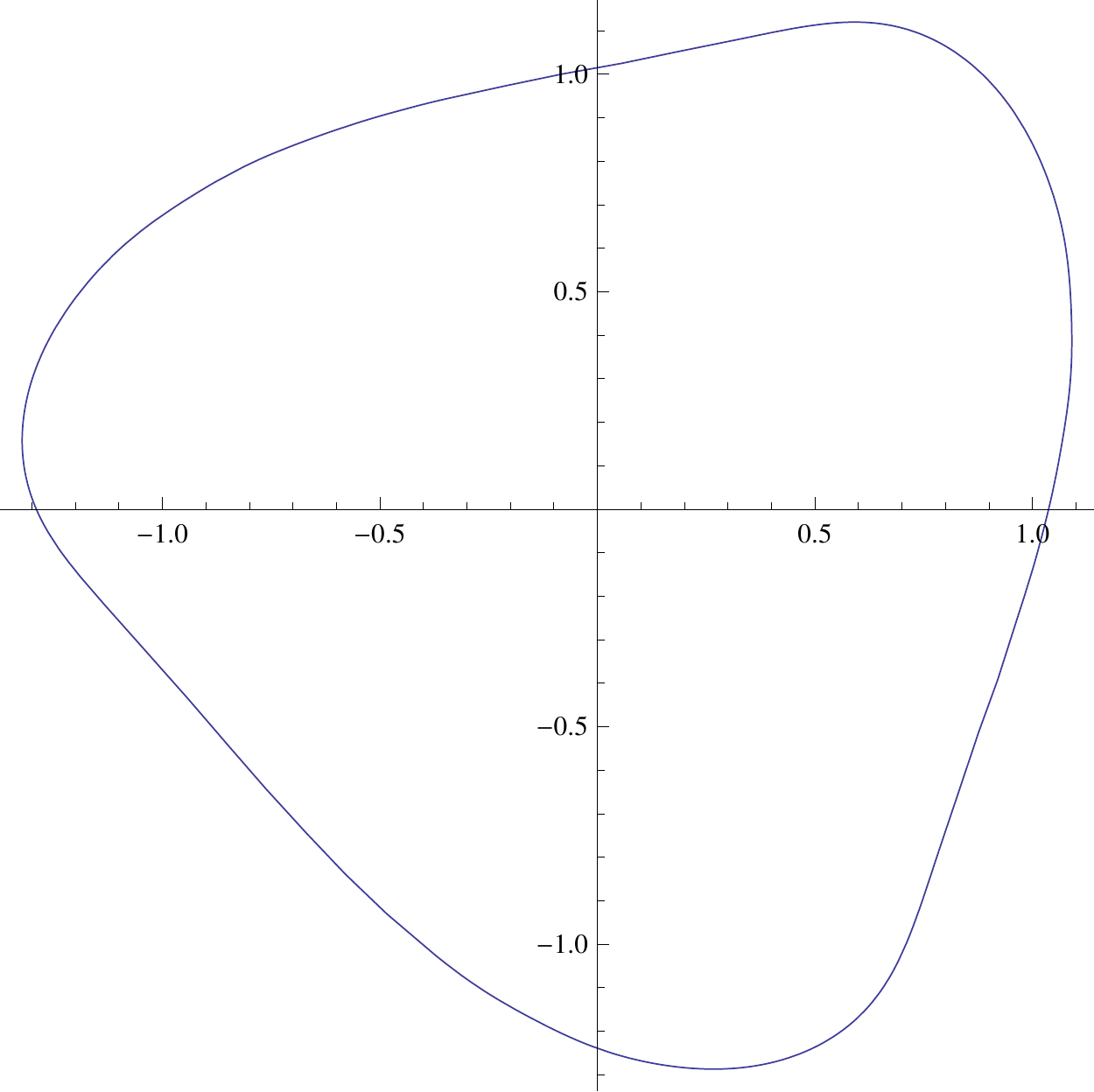}} \hspace{3mm}
\subfloat{
  \includegraphics[width=25mm]{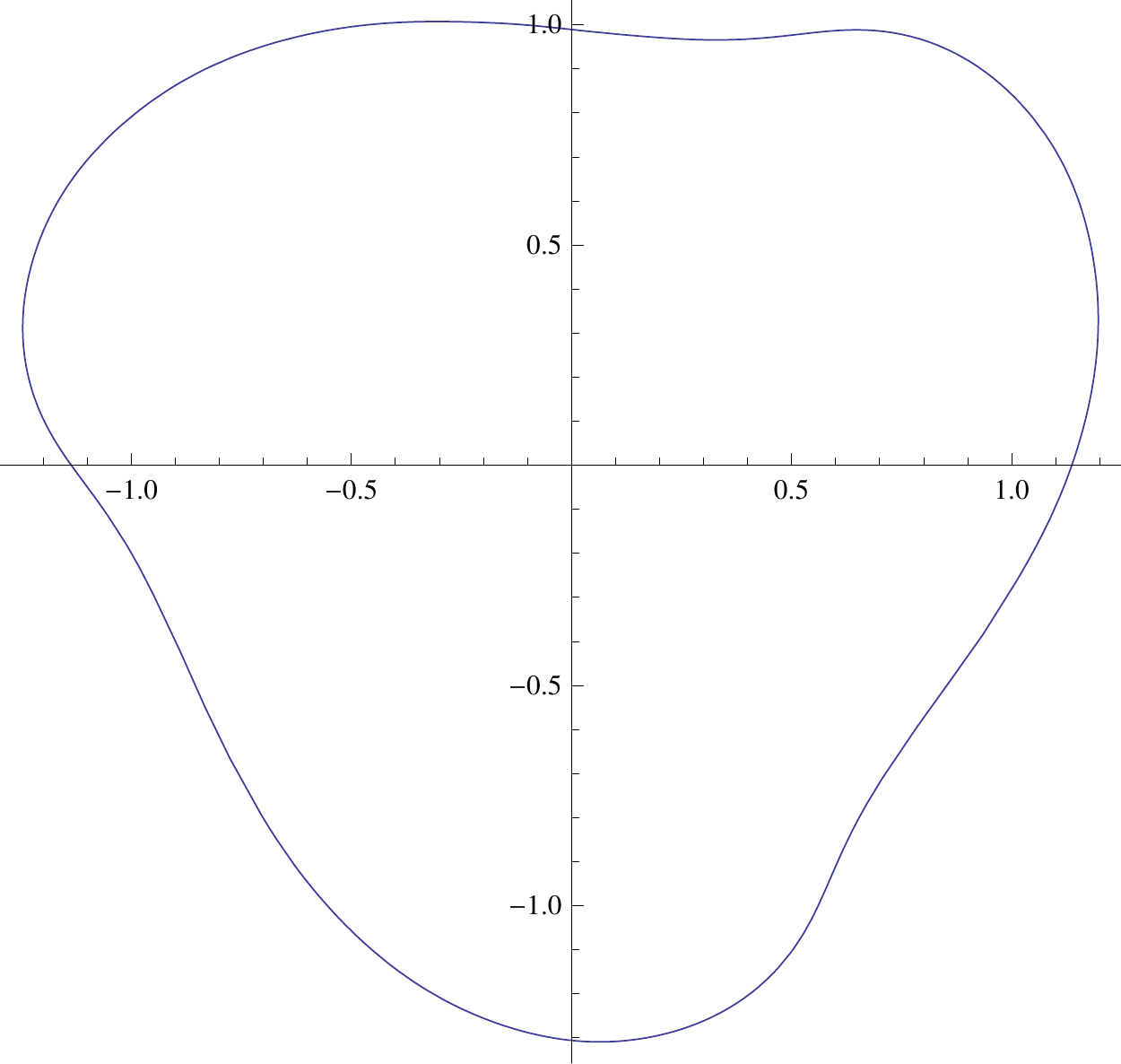}} \hspace{3mm}
\subfloat{
  \includegraphics[width=25mm]{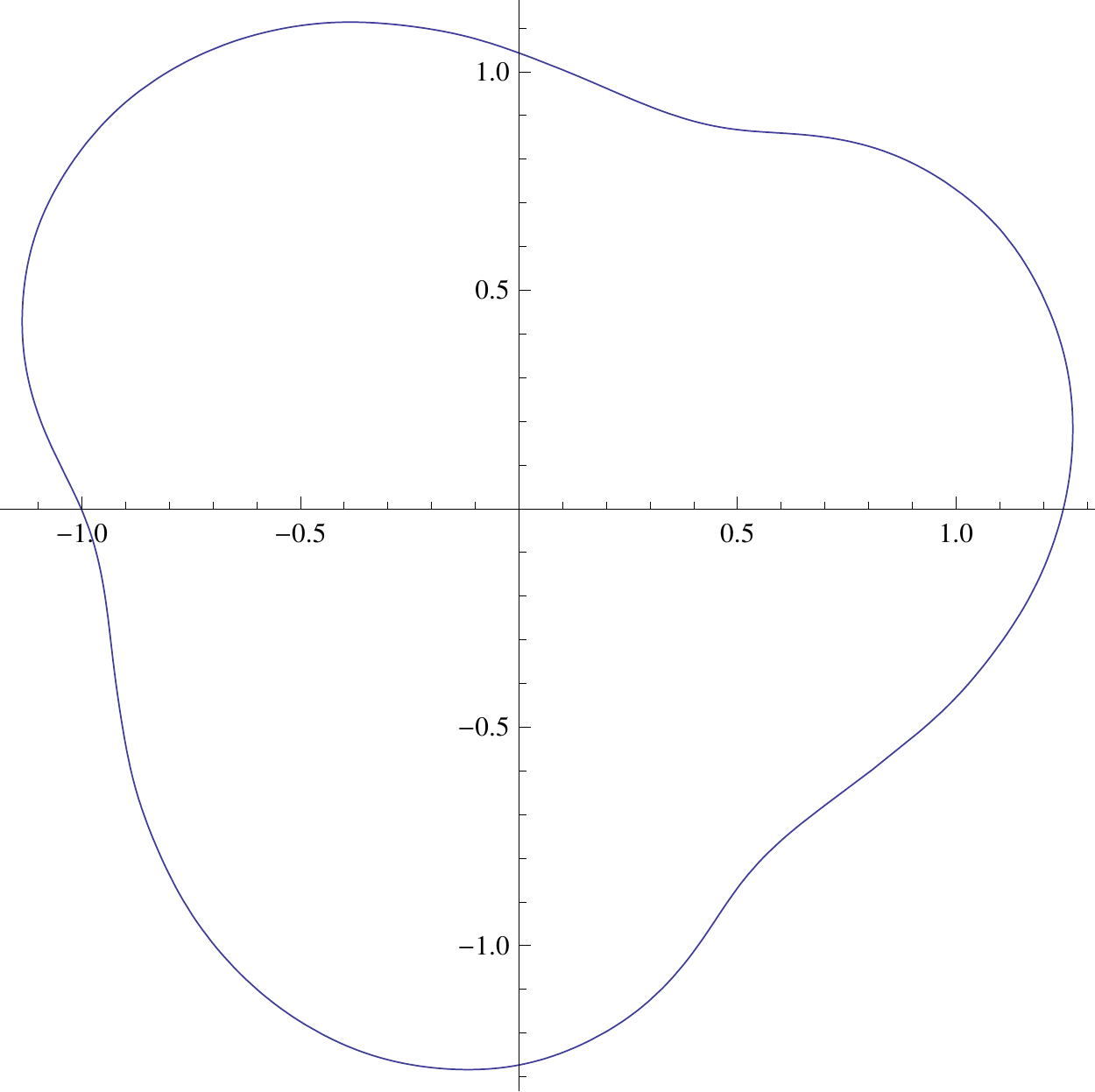}} \hspace{3mm}
\subfloat{
  \includegraphics[width=25mm]{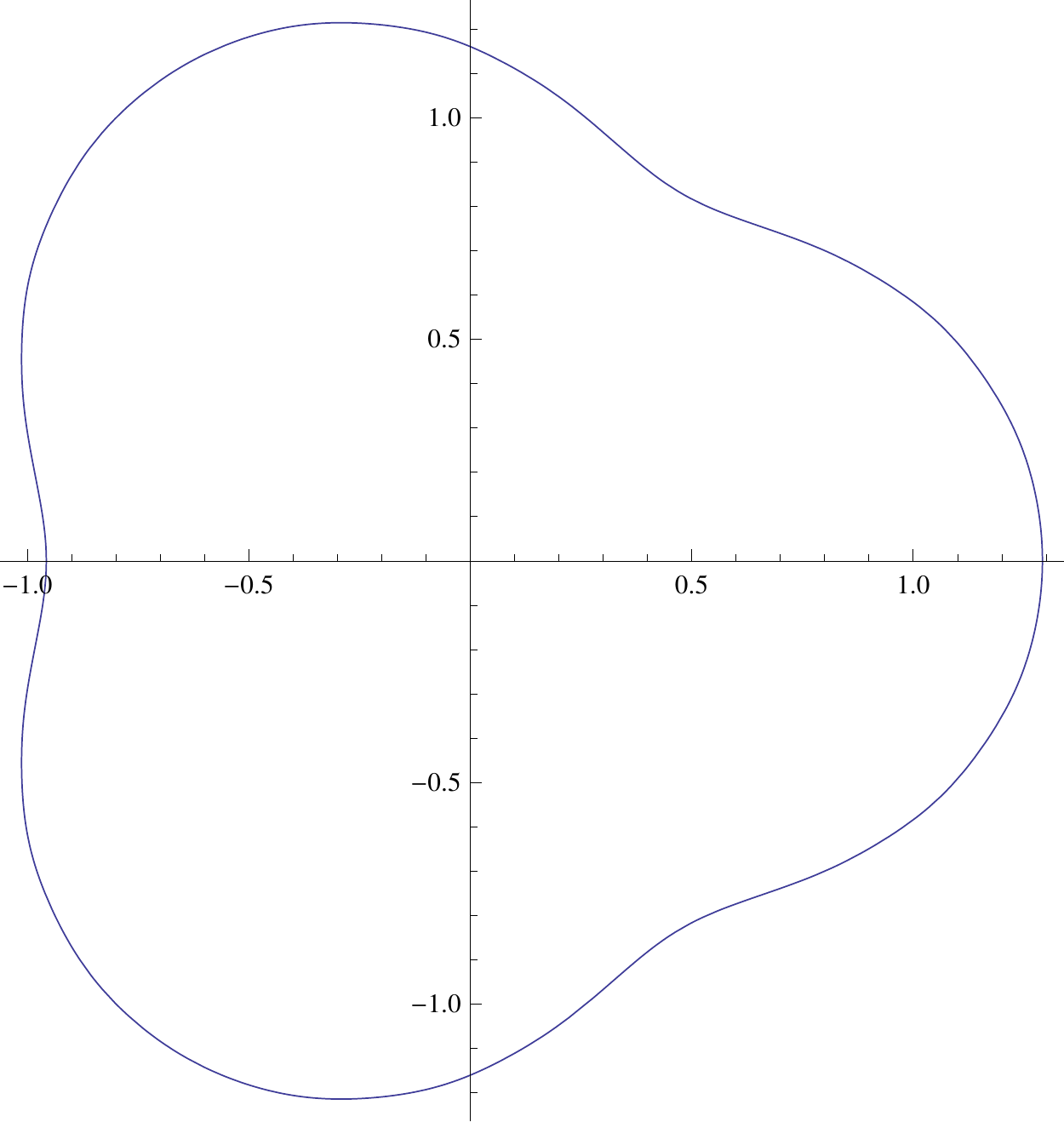}} \hspace{3mm}
\subfloat{
  \includegraphics[width=25mm]{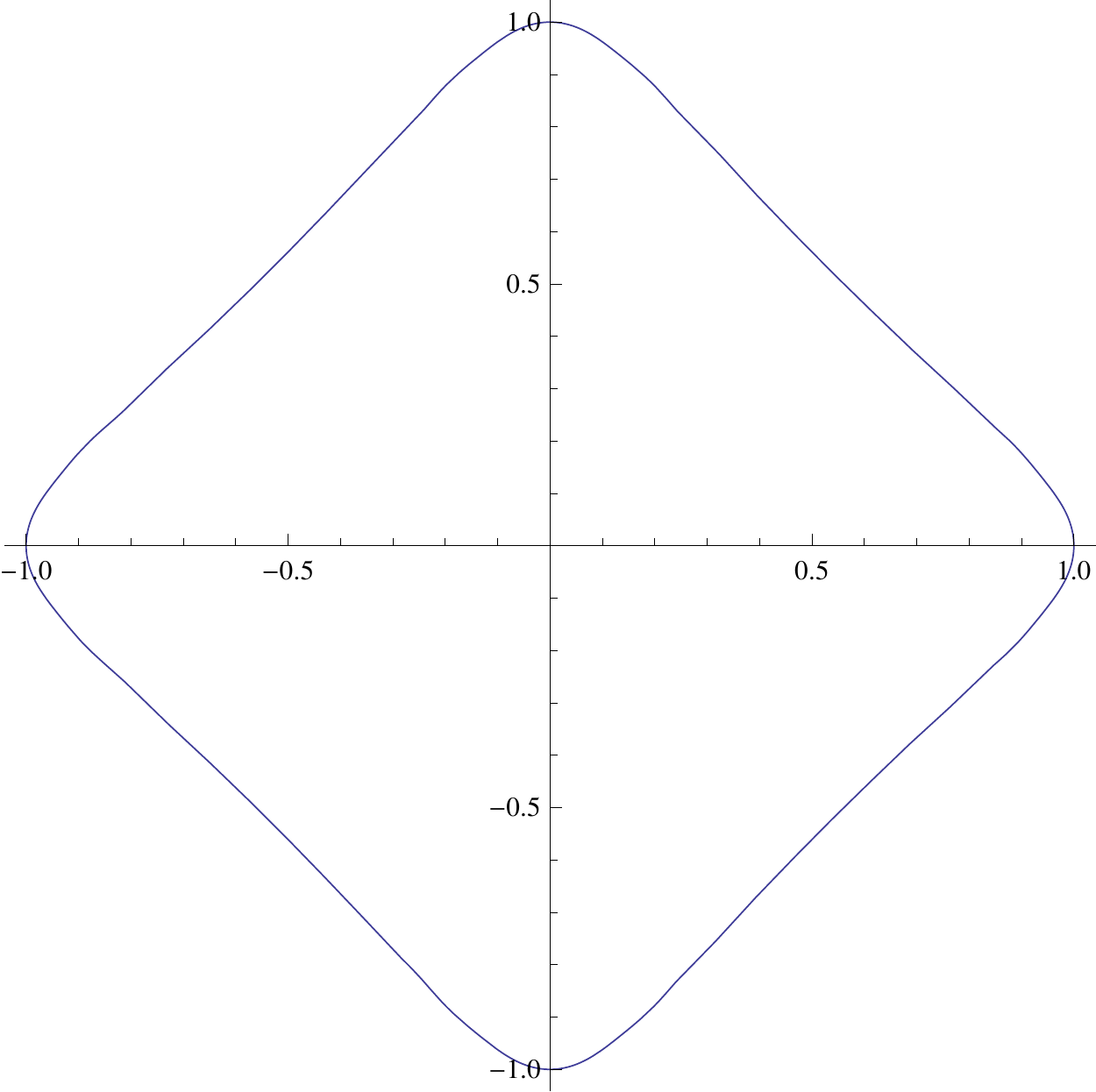}} \hspace{3mm}
\subfloat{
  \includegraphics[width=25mm]{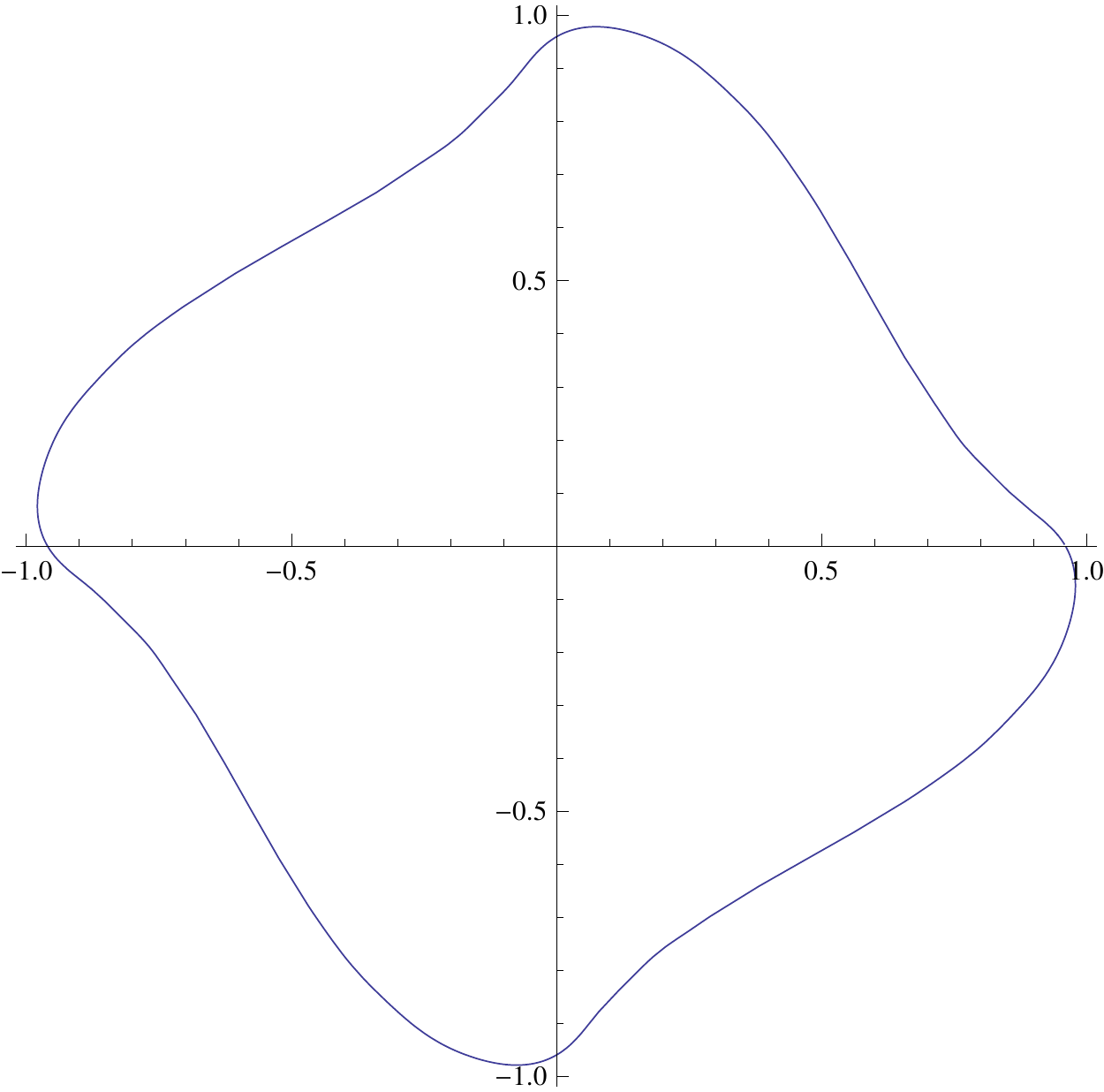}} \hspace{3mm}
\subfloat{
  \includegraphics[width=25mm]{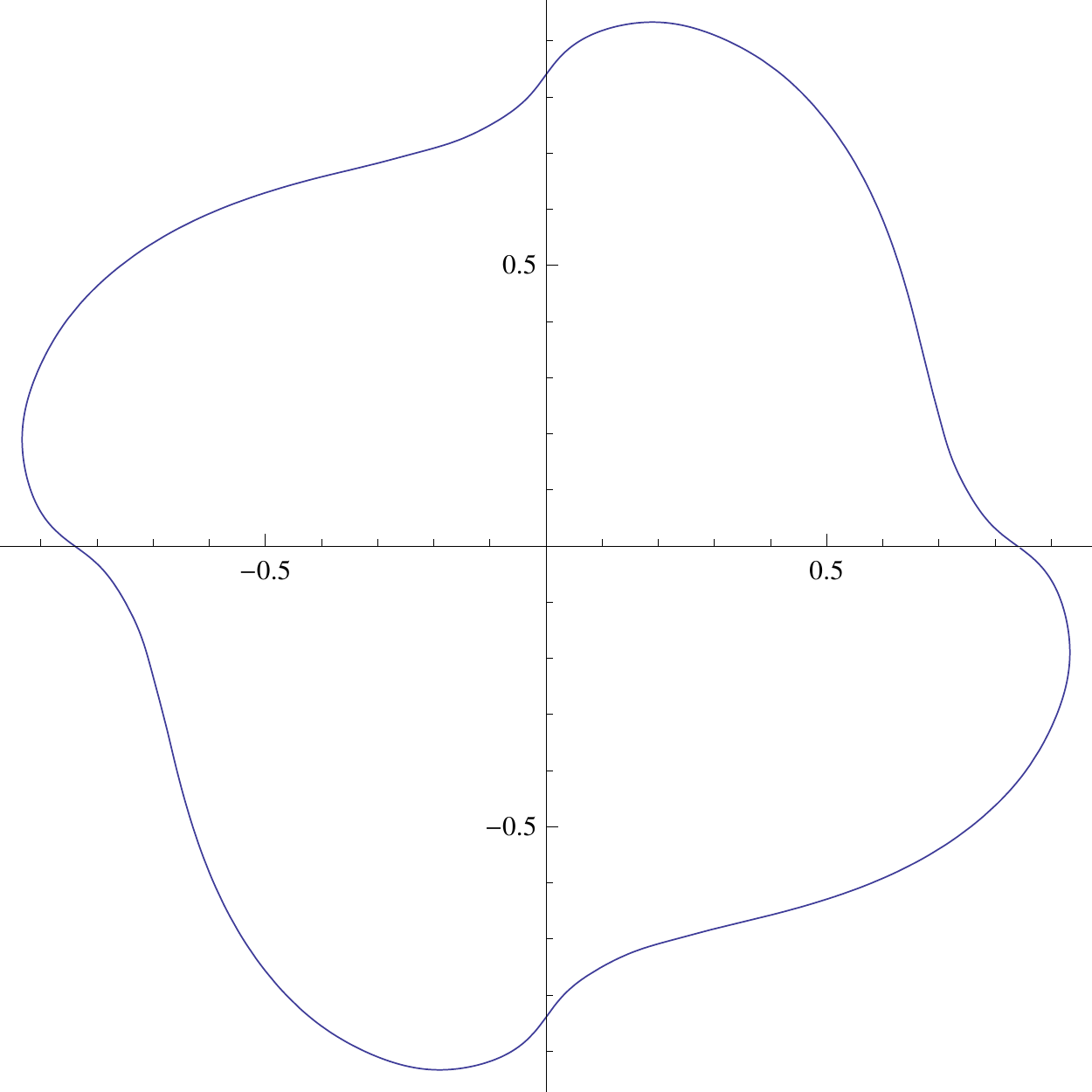}} \hspace{3mm}
\subfloat{
  \includegraphics[width=25mm]{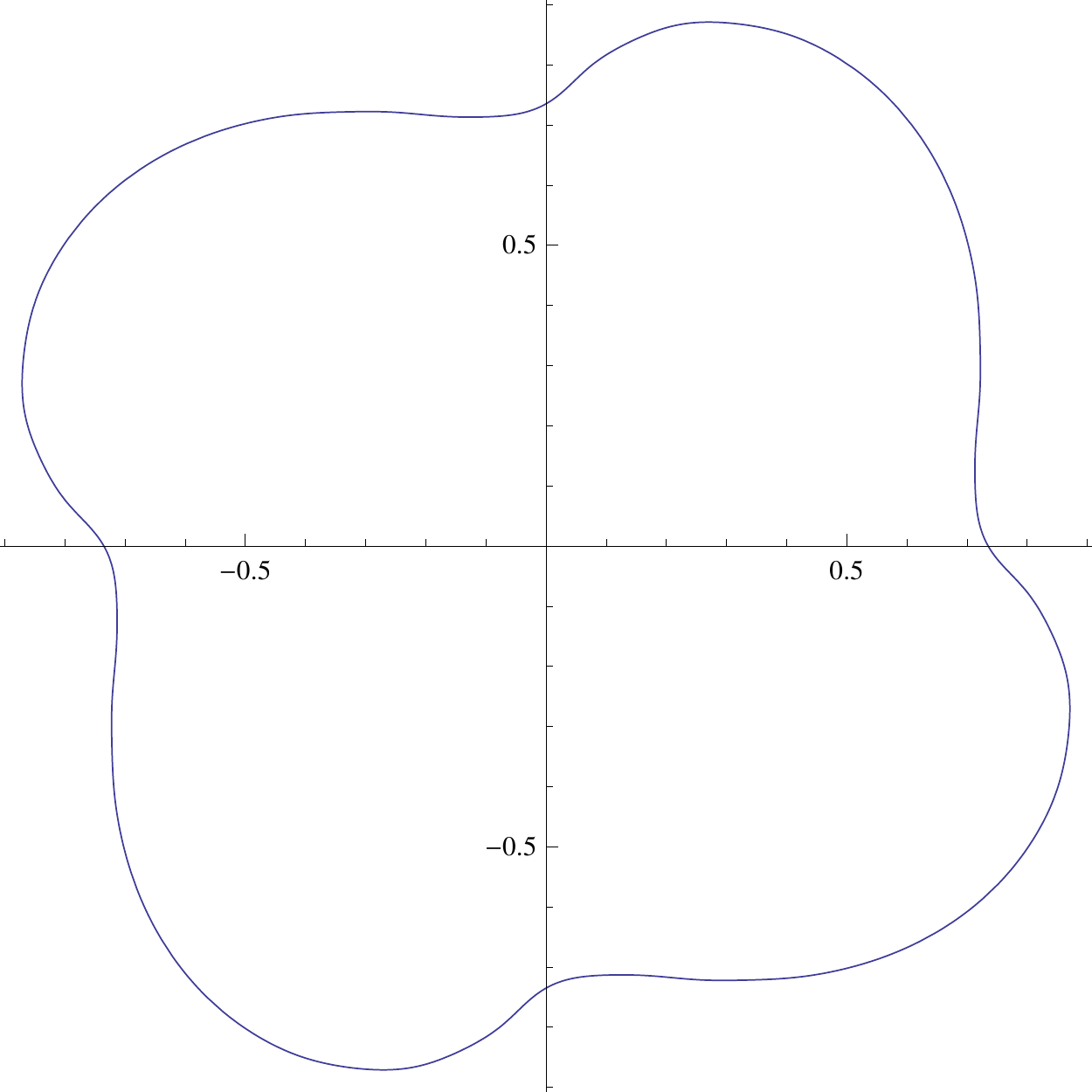}} \hspace{3mm}
\subfloat{
  \includegraphics[width=25mm]{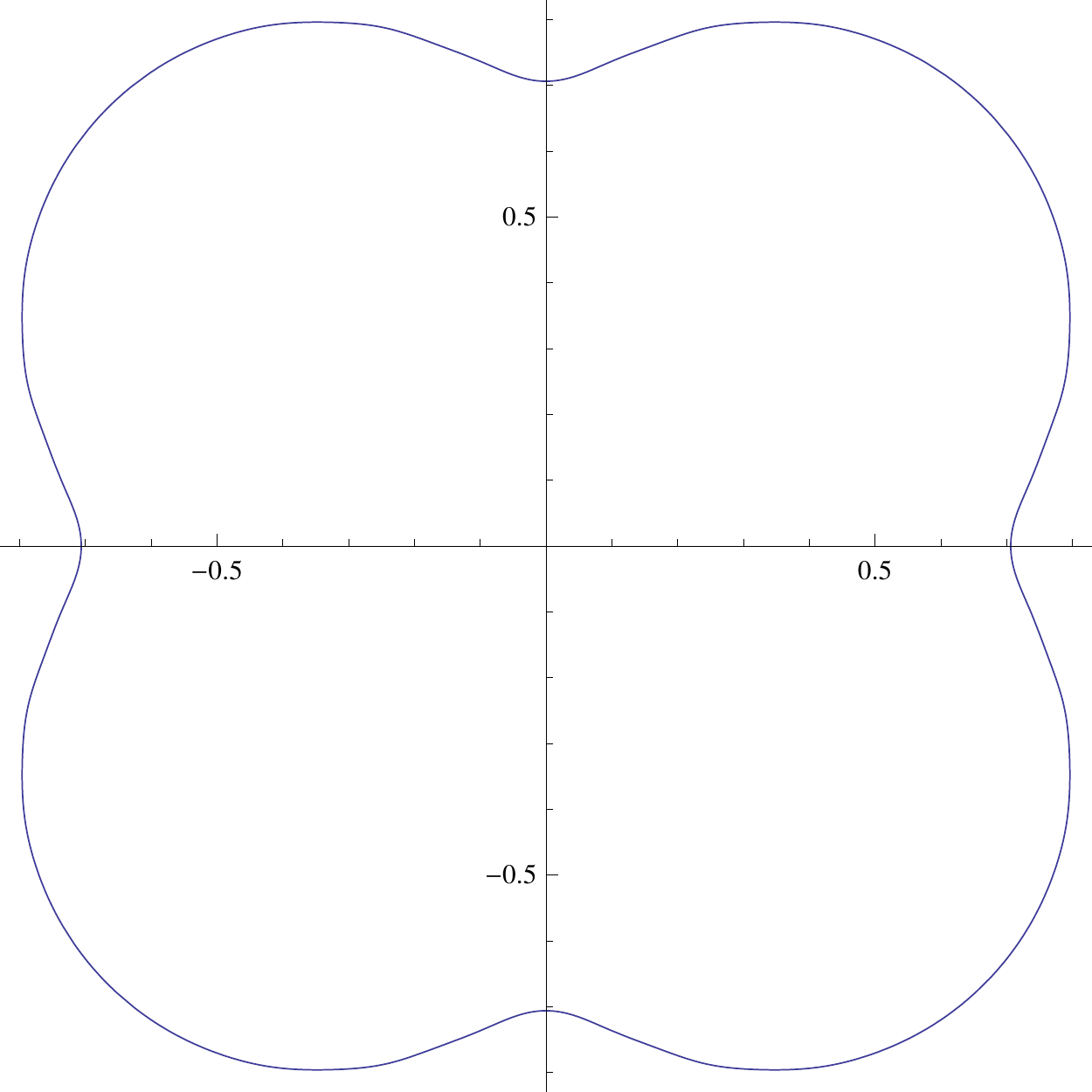}} \hspace{3mm}
\subfloat{
  \includegraphics[width=25mm]{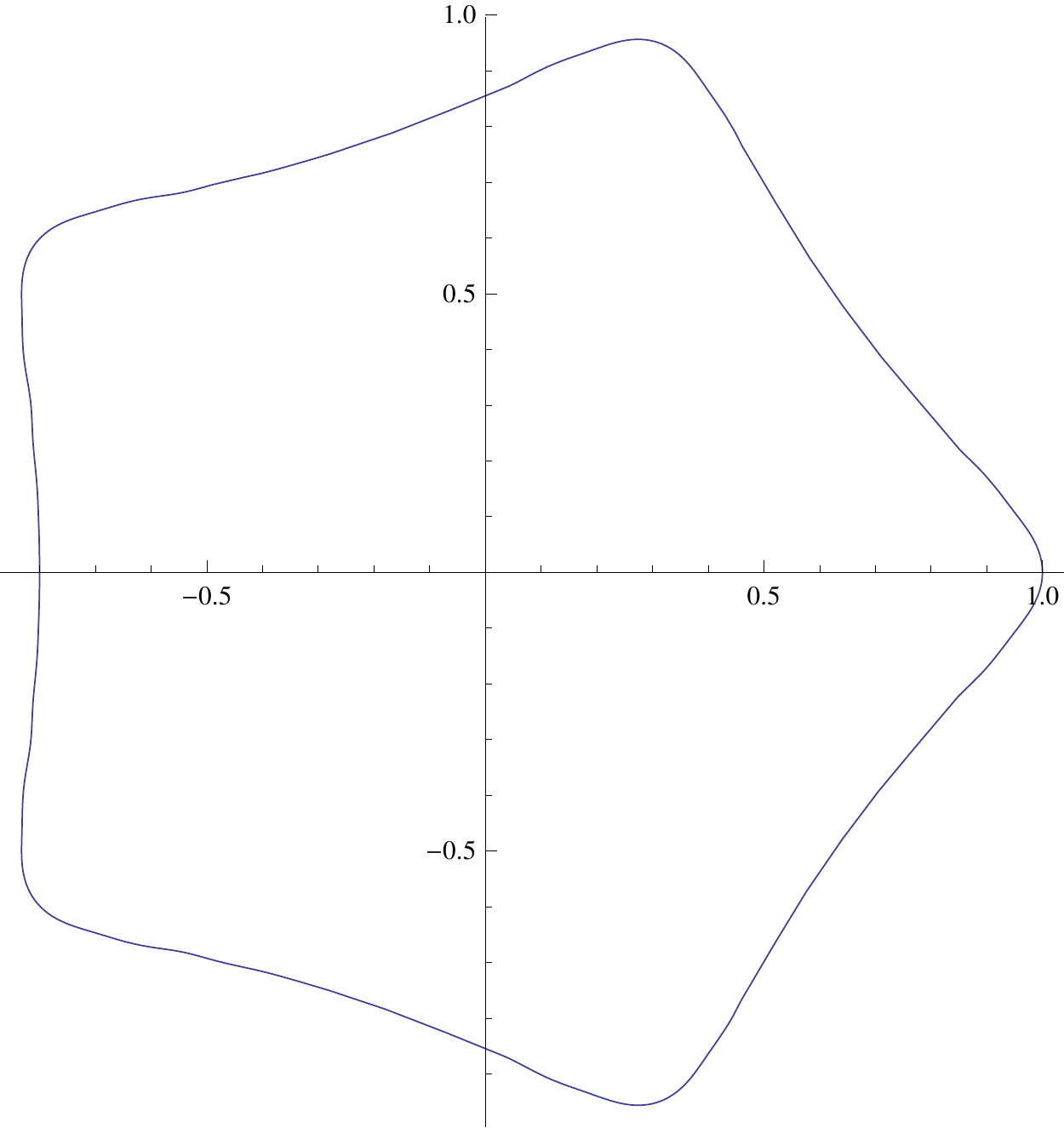}} \hspace{3mm}
\subfloat{
  \includegraphics[width=25mm]{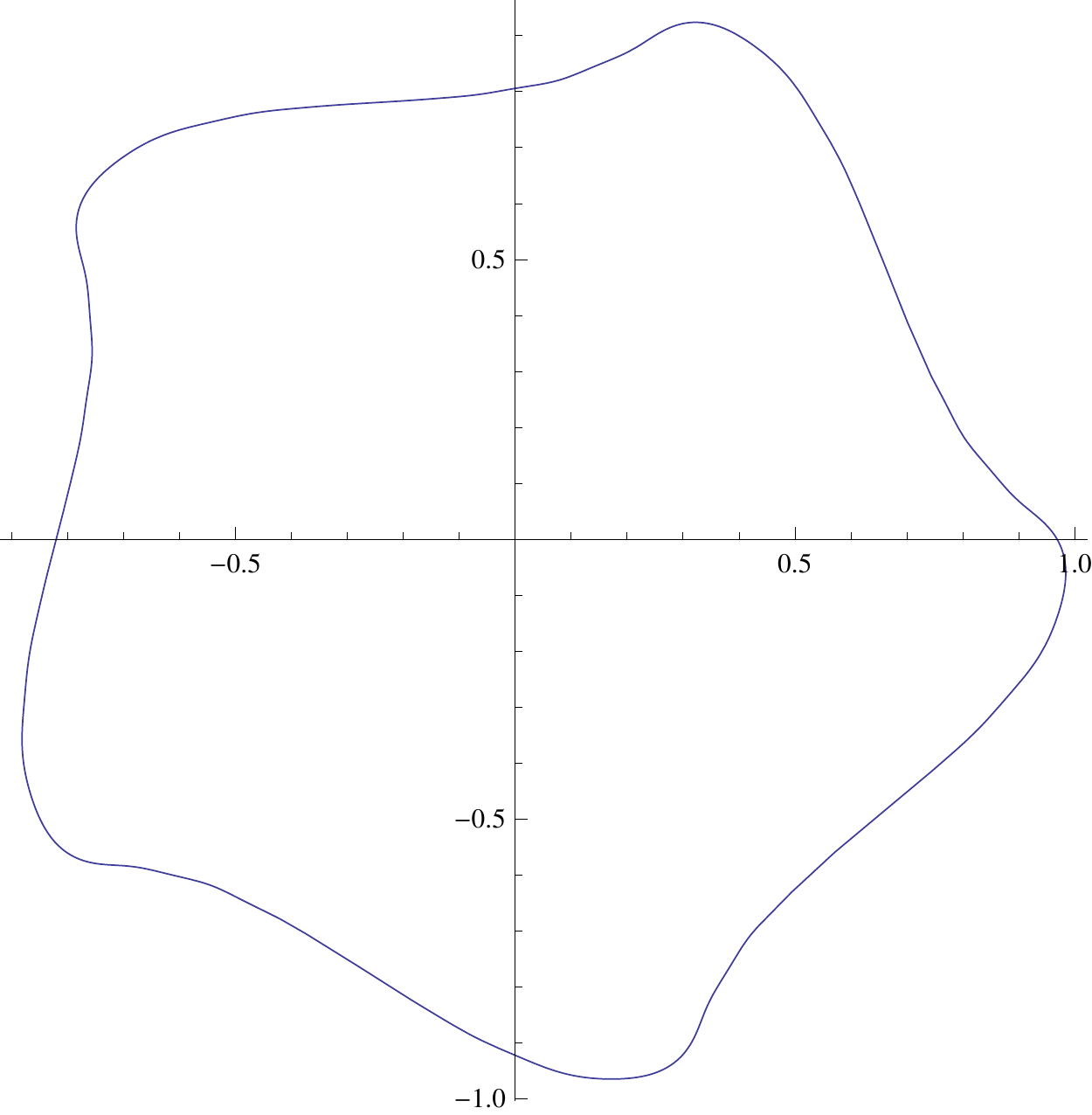}} \hspace{3mm}
\subfloat{
  \includegraphics[width=25mm]{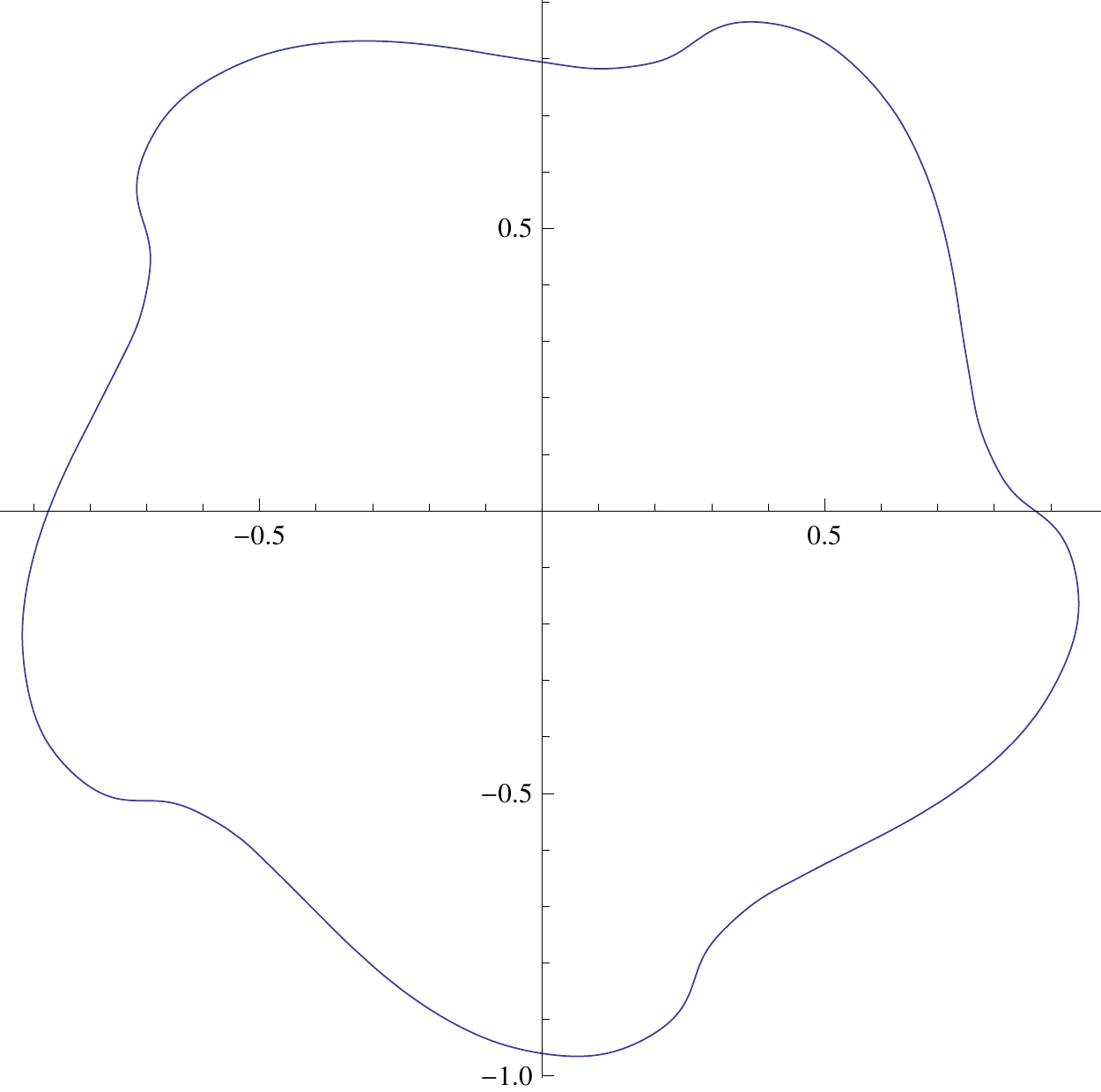}} \hspace{3mm}
\subfloat{
  \includegraphics[width=25mm]{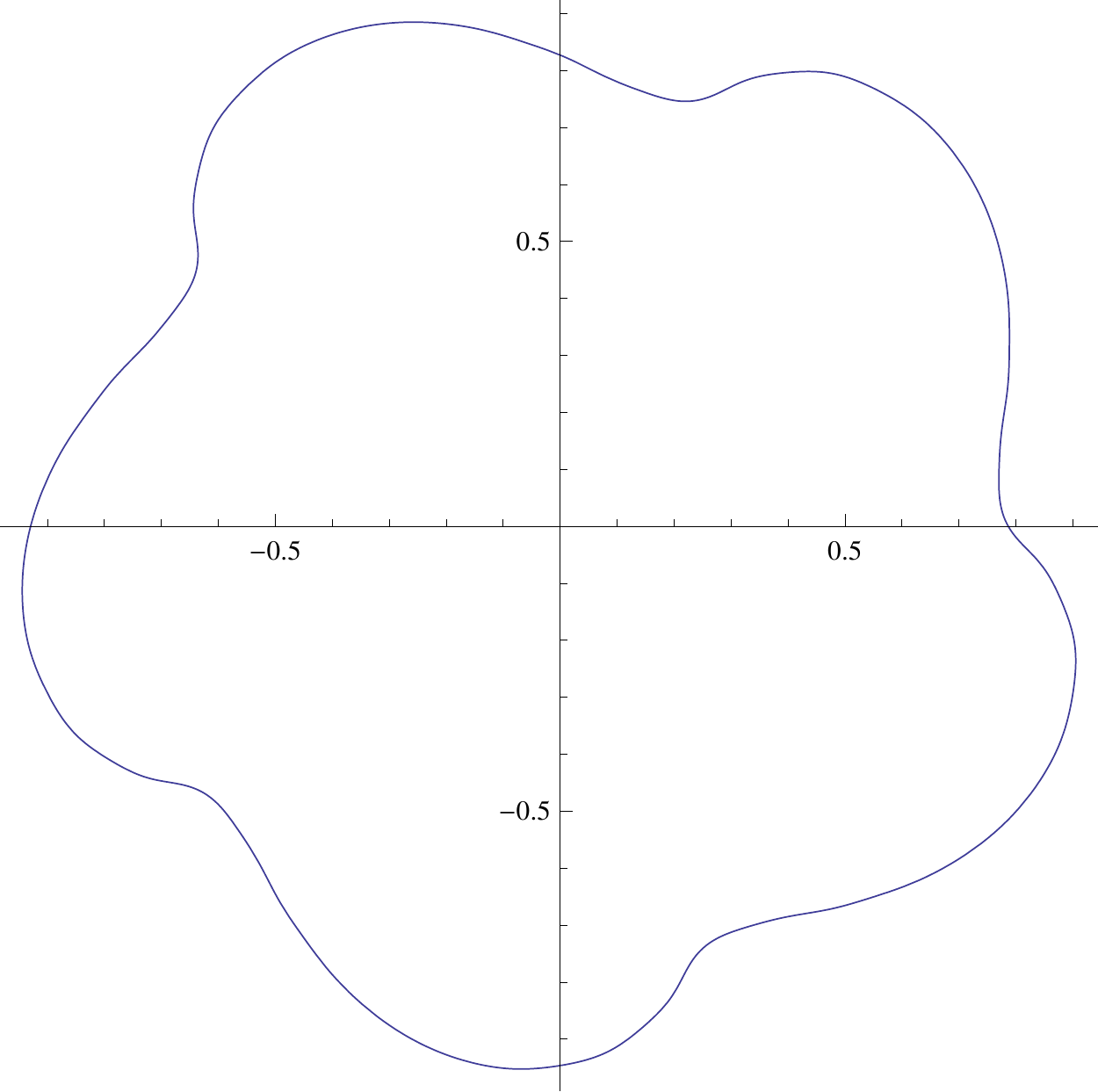}} \hspace{3mm}
\subfloat{
  \includegraphics[width=25mm]{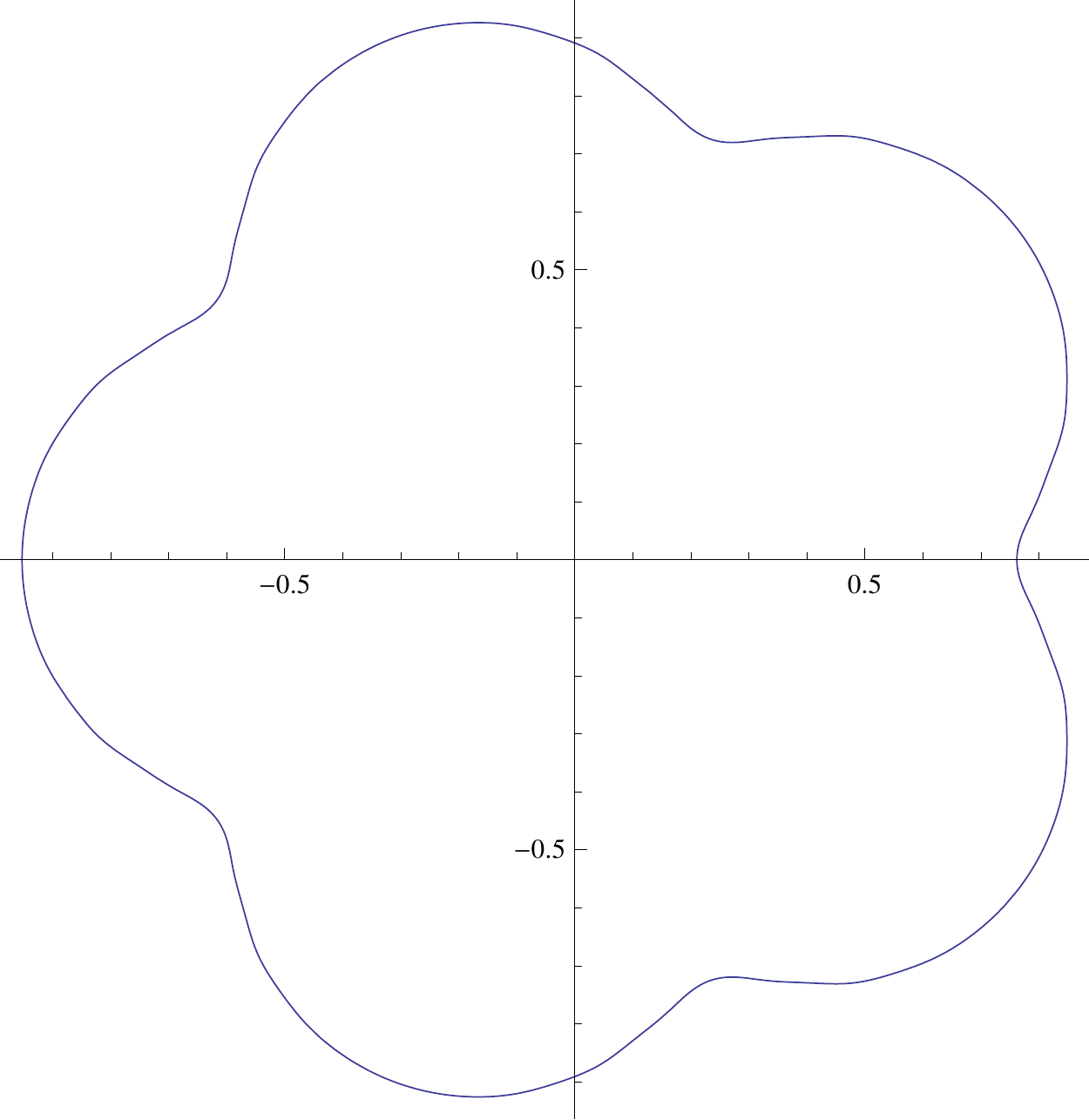}} \hspace{3mm}
\caption{\textbf{$\lambda$-deformations.}
We plot the $\lambda$-deformations for $\varphi=0,\frac{\pi}{4},\frac{\pi}{2},\frac{3\pi}{4},\pi$ for the interpolating contours between the circle and hypocycloids (ordered from left to right). We used $\epsilon=-0.35$, $\epsilon=-0.25$ and $\epsilon=-0.2$ for the $n=3,4,5$ cases respectively.
}
\label{fig:hypodef}
\end{center}
\end{figure}

\begin{figure}
\begin{center}
\subfloat[][]{
  \includegraphics[width=49mm]{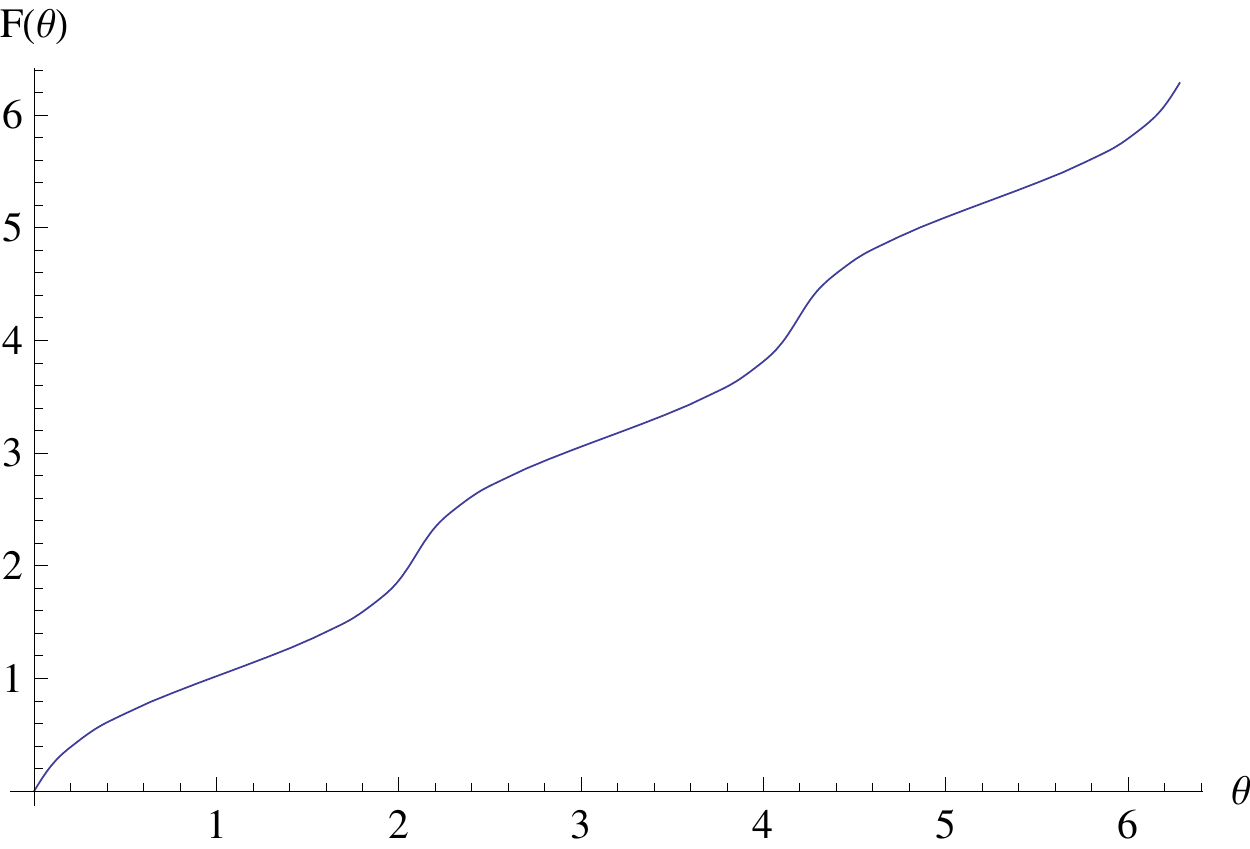}} \hspace{3mm}
\subfloat[][]{
  \includegraphics[width=49mm]{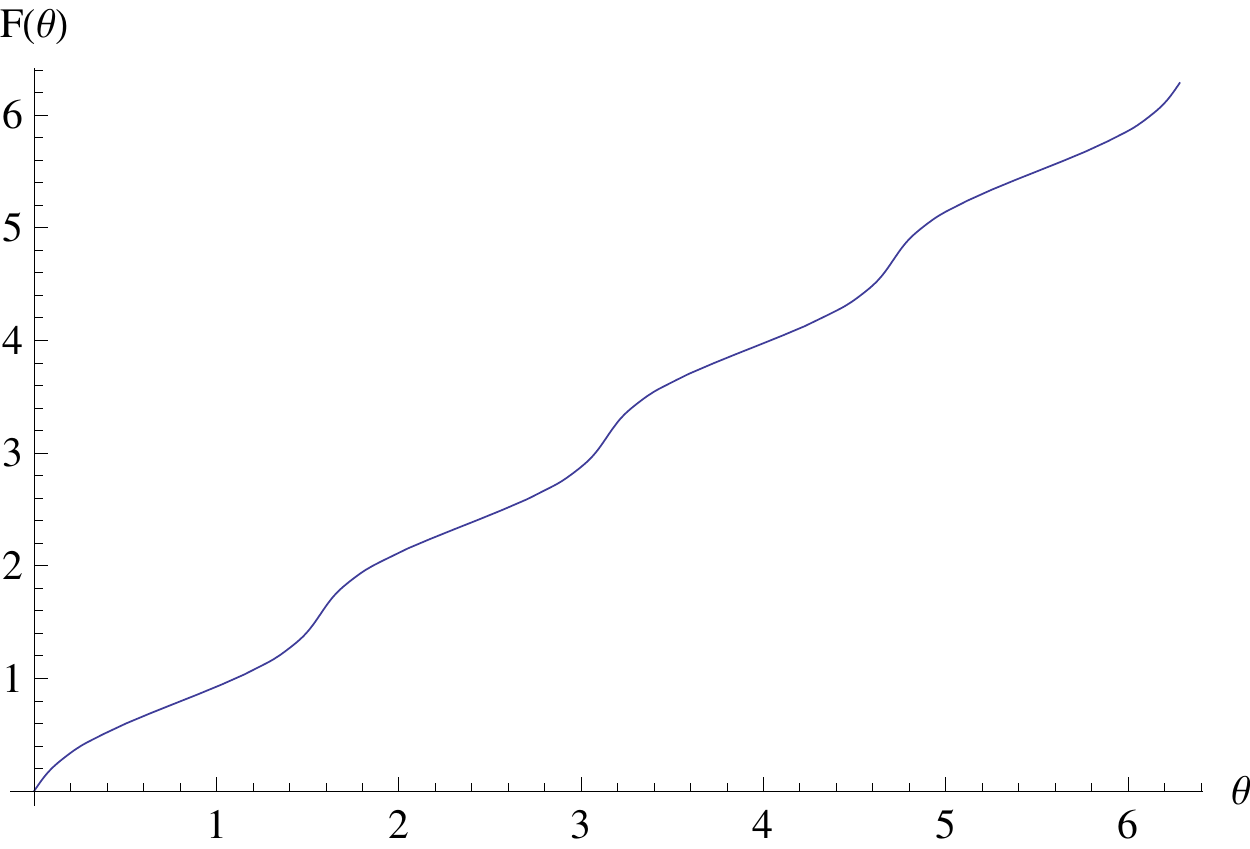}} \hspace{3mm}
\subfloat[][]{
  \includegraphics[width=49mm]{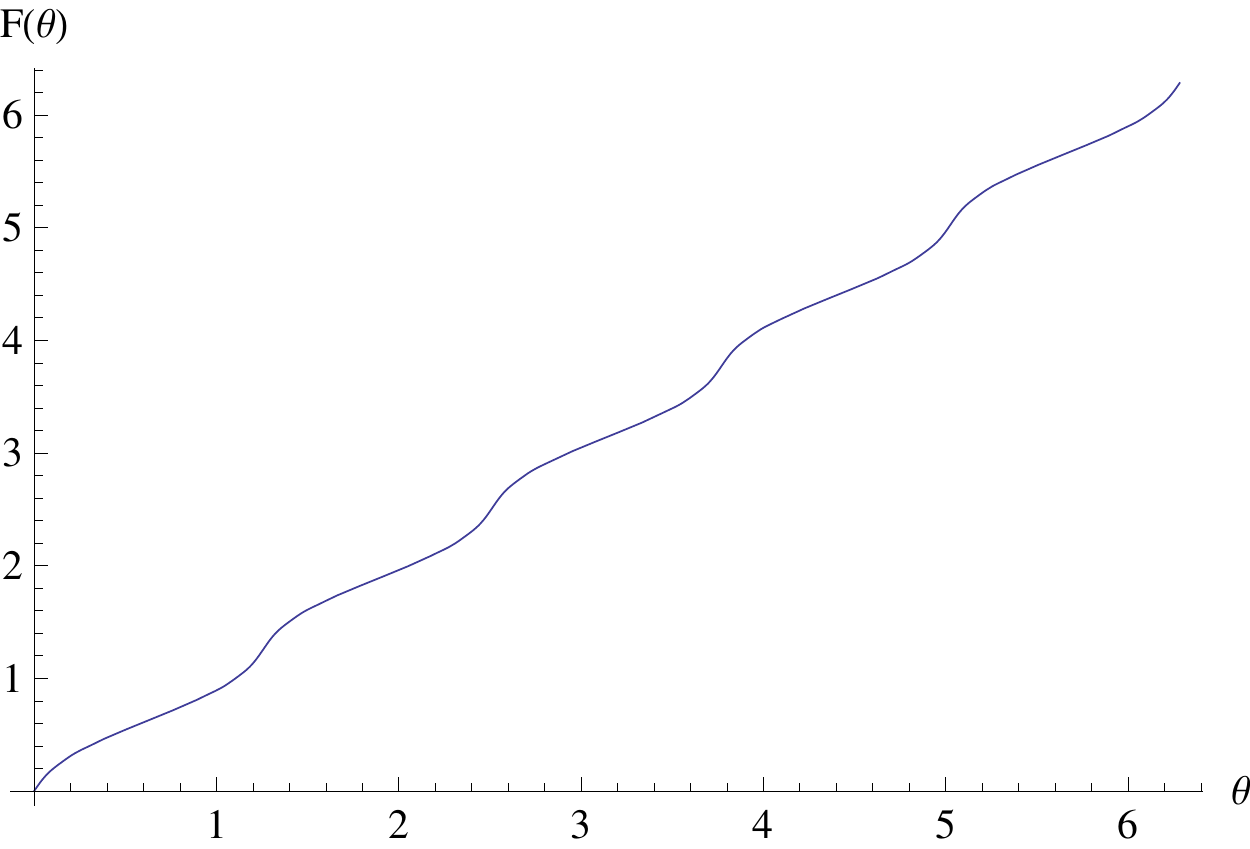}} \hspace{3mm}
\caption{\textbf{Parametrization.}
We plot the correct parametrization for the interpolating contours between the circle and hypocycloids.
(a) $n=3$, $\epsilon=-0.35$,
(b) $n=4$, $\epsilon=-0.25$,
(c) $n=5$ $\epsilon=-0.2$.
}
\label{fig:hypoparam}
\end{center}
\end{figure}

\begin{figure}
\begin{center}
\subfloat[][]{
  \includegraphics[width=49mm]{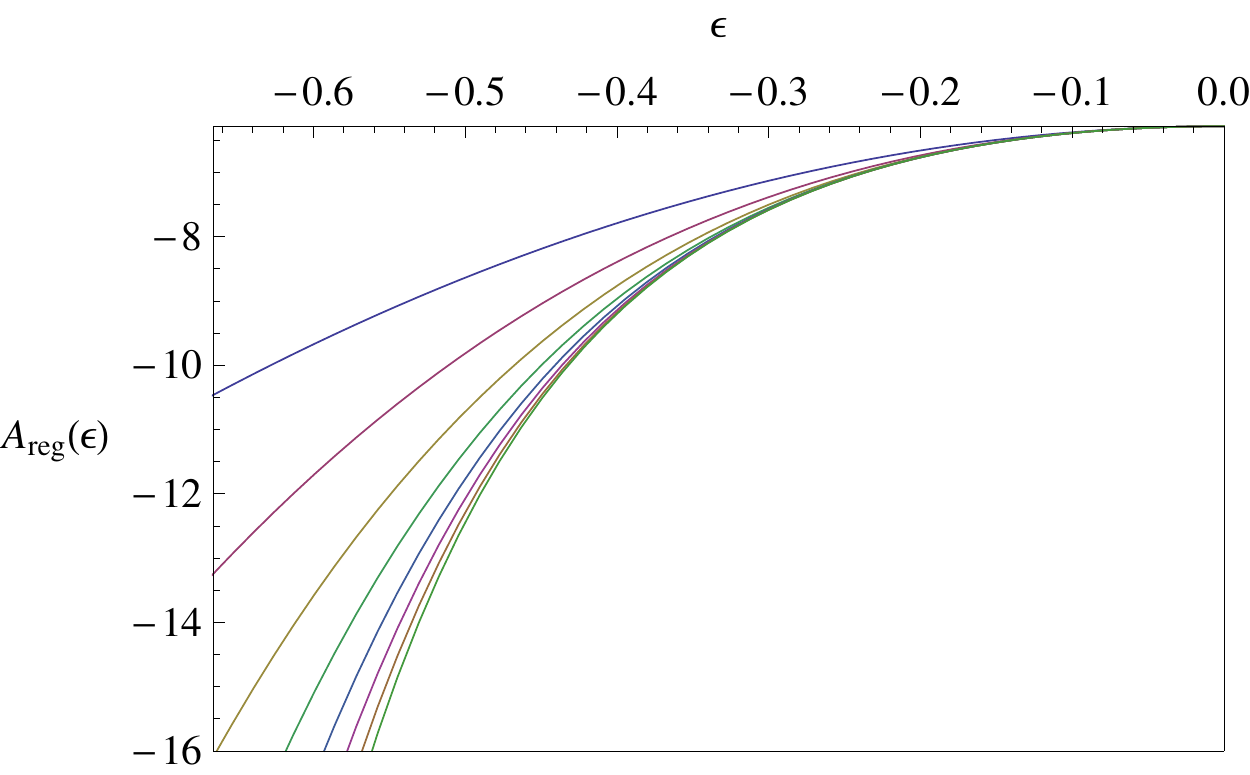}} \hspace{3mm}
\subfloat[][]{
  \includegraphics[width=49mm]{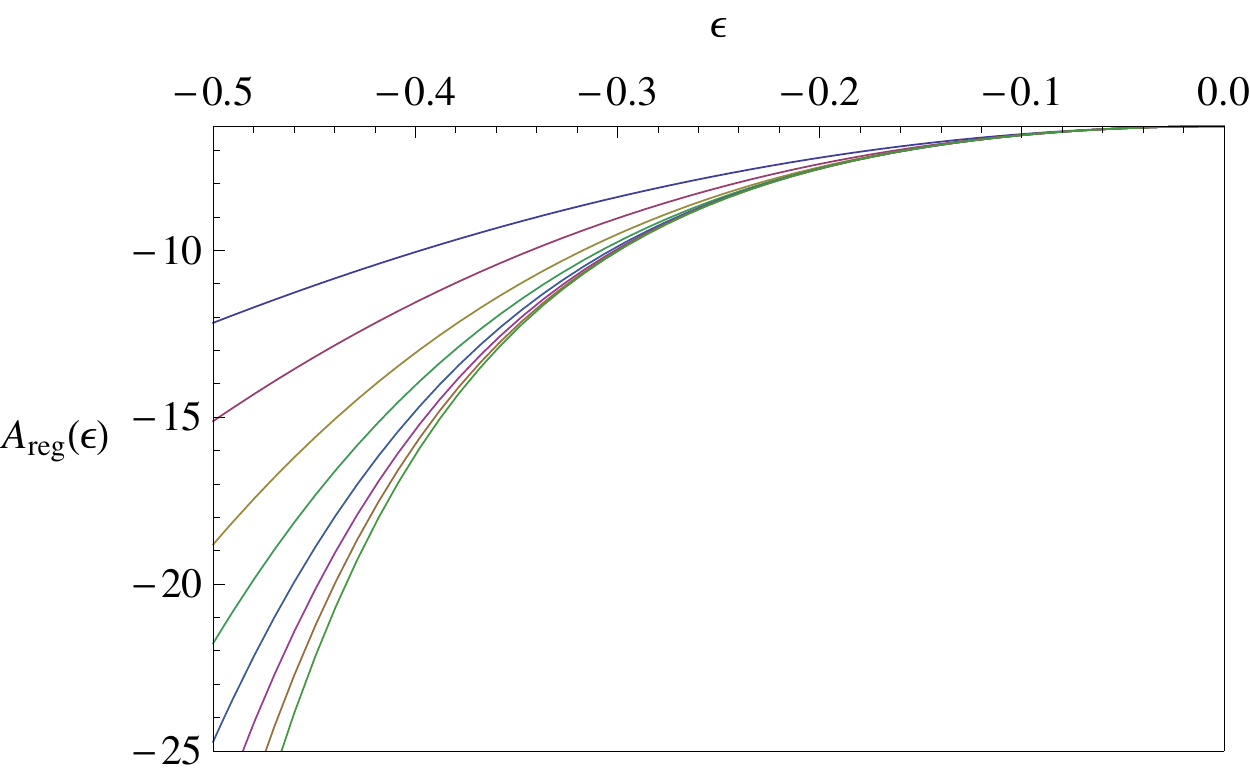}} \hspace{3mm}
\subfloat[][]{
  \includegraphics[width=49mm]{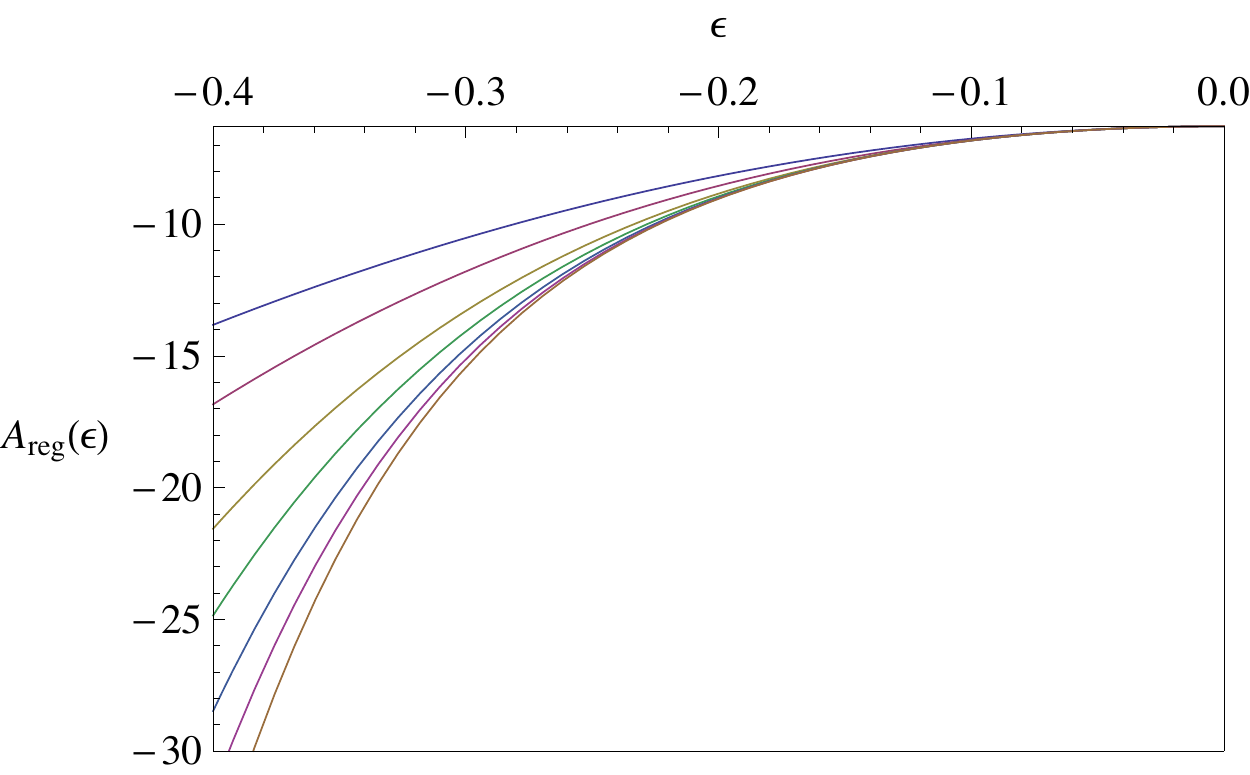}} \hspace{3mm}
\caption{\textbf{Regularized area.}
We plot the regularized area for the interpolating contours between the circle and hypocycloids.
The different lines correspond to different orders of the $\epsilon$ expansion, from $\epsilon^2$ to $\epsilon^{10}$ for $n=3,4$ and $\epsilon^{9}$ for $n=5$.
We plot for $-\frac{2}{n}<\epsilon<0$, where the cups form at $\epsilon = -\frac{2}{n}$ and the area should diverge.
(a) $n=3$,
(b) $n=4$,
(c) $n=5$.
}
\label{fig:hypoarea}
\end{center}
\end{figure}

Next, we give the regularized area of the minimal surfaces and the one loop weak coupling expectation value for the $\lambda$-deformed contours for $n=3,4,5$ (the $n=2$ case is the ellipse which we already considered).
We plot the results for the area, correct parametrization function and $\lambda$-deformations in figures \ref{fig:hypoarea}, \ref{fig:hypoparam} and \ref{fig:hypodef} respectively.

\begin{align}
A_{\text{reg}}^{n=3} =& -2 \pi -3 \pi  \epsilon ^2+3 \pi  \epsilon ^3-\frac{1293 \pi  \epsilon ^4}{280}+\frac{873 \pi  \epsilon ^5}{140}-\frac{1386017 \pi  \epsilon ^6}{156800}+\frac{1948851 \pi  \epsilon ^7}{156800}\nonumber\\
& -\frac{88672365813 \pi  \epsilon ^8}{5022617600}+\frac{31600967429 \pi  \epsilon ^9}{1255654400}-\frac{36254366537313759 \pi  \epsilon ^{10}}{1005528043520000}
+ \mathcal{O}(\epsilon^{10}),\\
A_{\text{reg}}^{n=4} =& -2\pi -\frac{15 \pi  \epsilon ^2}{2}+\frac{15 \pi  \epsilon ^3}{2}-\frac{2109 \pi  \epsilon ^4}{112}+\frac{1689 \pi  \epsilon ^5}{56}-\frac{13473123 \pi  \epsilon ^6}{224224}+\frac{24353319 \pi  \epsilon ^7}{224224}\nonumber\\
& -\frac{63592443793533 \pi  \epsilon ^8}{305249584640}+\frac{29902126613013 \pi  \epsilon ^9}{76312396160}
-\frac{29627266774400450319 \pi  \epsilon ^{10}}{39477684581823488}
+ \mathcal{O}(\epsilon^{10}),\\
A_{\text{reg}}^{n=5} =& -2\pi -15 \pi  \epsilon ^2+15 \pi  \epsilon ^3-\frac{10335 \pi  \epsilon ^4}{176}+\frac{9015 \pi  \epsilon ^5}{88}-\frac{4365315 \pi  \epsilon ^6}{15488}+\frac{9245505 \pi  \epsilon ^7}{15488}\nonumber\\
& -\frac{7582306709205 \pi  \epsilon ^8}{5075603456}+\frac{4338088174725 \pi  \epsilon ^9}{1268900864}
+ \mathcal{O}(\epsilon^{10}).
\end{align}

\begin{align}
W_{1,\varphi}^{n=3} =& -2 \pi ^2-6 \pi ^2 \epsilon ^2+6 \pi ^2 \epsilon ^3-\frac{81 \pi ^2 \epsilon ^4}{8}+\frac{57 \pi ^2 \epsilon ^5}{4}-\frac{83 \pi ^2 \epsilon ^6}{4}+\frac{237 \pi ^2 \epsilon ^7}{8}\nonumber\\
&-\frac{\pi ^2 (2133269689+3648094 \cos (2 \varphi ))\epsilon ^8}{50226176}\nonumber\\
&+\frac{\pi ^2 (764606393+3648094 \cos(2 \varphi )) \epsilon ^9}{12556544} + \mathcal{O}(\epsilon^{10}),\\
W_{1,\varphi}^{n=4} =& -2 \pi ^2-15 \pi ^2 \epsilon ^2+15 \pi ^2 \epsilon ^3-\frac{1333 \pi ^2 \epsilon ^4}{32}+\frac{1093 \pi ^2 \epsilon ^5}{16}-\frac{71179 \pi ^2 \epsilon ^6}{512}+\frac{129937 \pi ^2 \epsilon ^7}{512}\nonumber\\
&-\frac{\pi ^2 (3837945912753851+7831242589272 \cos (2 \varphi ))\epsilon ^8}{7814389366784}\nonumber\\
&+\frac{\pi ^2 (1816117657651195+7831242589272 \cos( 2 \varphi )) \epsilon ^9}{1953597341696} + \mathcal{O}(\epsilon^{10}),\\
W_{1,\varphi}^{n=5} =& -2 \pi ^2-30 \pi ^2 \epsilon ^2+30 \pi ^2 \epsilon ^3-130 \pi ^2 \epsilon ^4+230 \pi ^2 \epsilon ^5-\frac{2585 \pi ^2 \epsilon ^6}{4}+\frac{5515 \pi ^2 \epsilon ^7}{4}\nonumber\\
&-\frac{5 \pi ^2 (15805224661+34793379 \cos(2 \varphi ))\epsilon ^8}{22658944} + \mathcal{O}(\epsilon^{9}).
\end{align}

As is easily seen, in all of these examples the week coupling expansion of the $\lambda$-deformed contours is dependent of the deformation, where the dependence starts at the eighth order.

We can also find the area for general $p$, however in this case we could compute it only up to the fifth order in $\epsilon$,
\begin{align}
A_{\text{reg}} =& -2 \pi -\frac{1}{8} n \left(n^2-1\right) \pi  \epsilon ^2+\frac{1}{8} n \left(n^2-1\right) \pi  \epsilon ^3\nonumber\\
&-\frac{n \left(56+8 n-307 n^2+8 n^3+338 n^4-160 n^5+57 n^6\right) \pi  \epsilon ^4}{512 \left(4 n^2-1\right)}\nonumber\\
& +\frac{n \left(24+8 n-147 n^2+8 n^3+210 n^4-160 n^5+57 n^6\right) \pi  \epsilon ^5}{256 \left(4 n^2-1\right)}
+ \mathcal{O}(\epsilon^{6}).
\end{align}
The reason is that for higher orders the number of terms that $\alpha(z,\bar z)$ contains is proportional to $n$, which complicates the analysis.
However, for a fixed $n$ this is not a problem.

\subsection{Symmetric wavy contours}
\begin{figure}
\begin{center}
\subfloat[][]{
  \includegraphics[width=29mm]{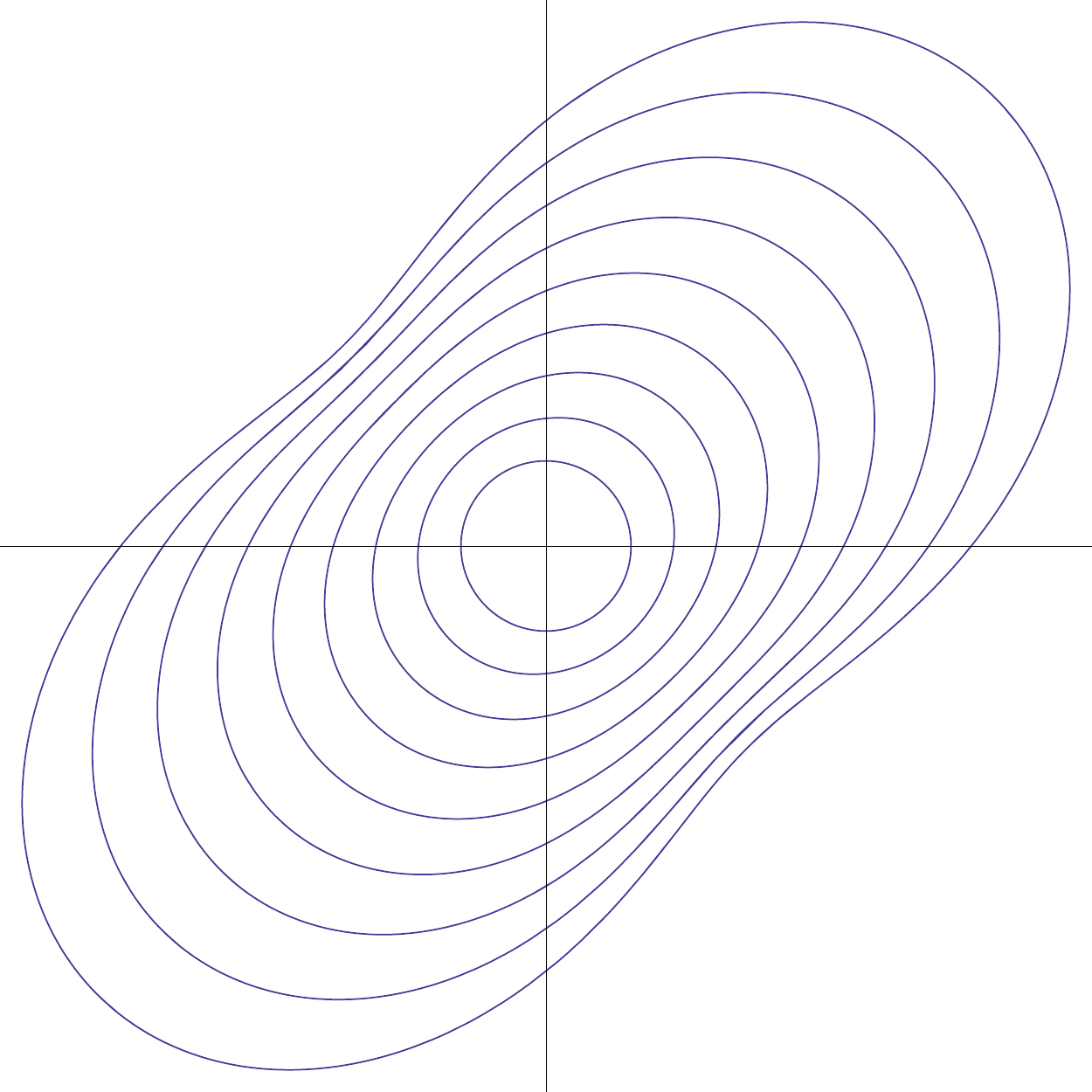}} \hspace{3mm}
\subfloat[][]{
  \includegraphics[width=29mm]{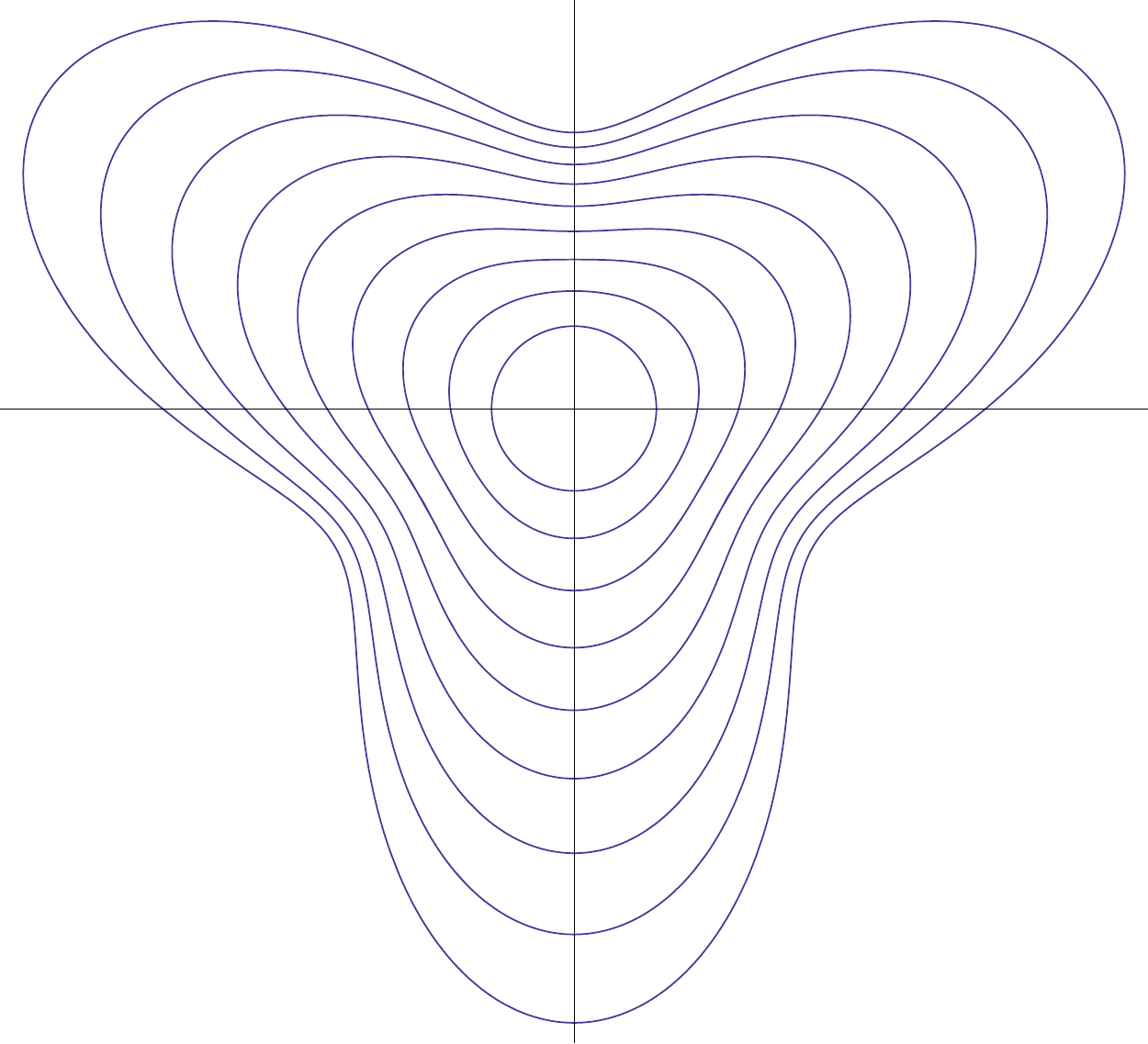}} \hspace{3mm}
\subfloat[][]{
  \includegraphics[width=29mm]{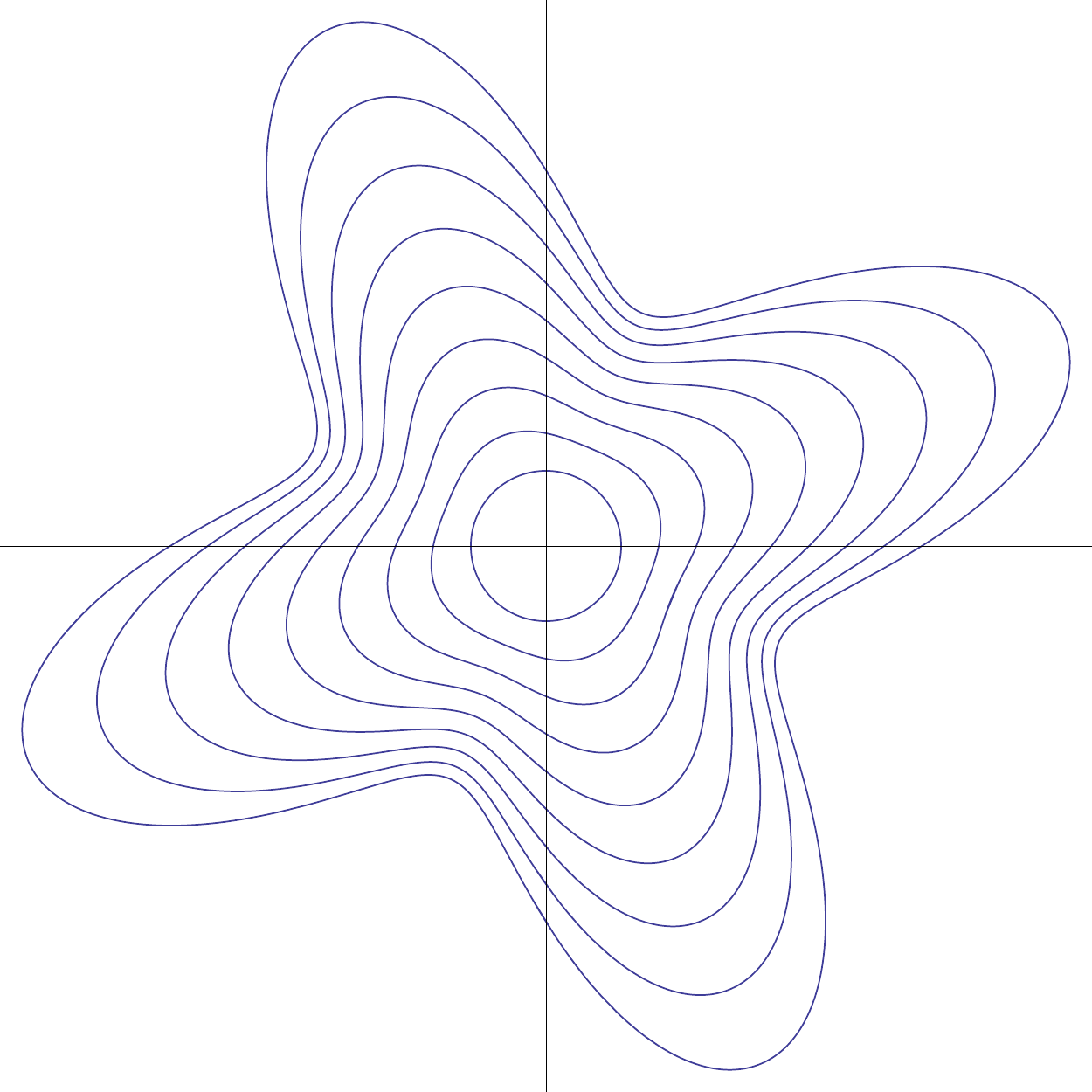}} \hspace{3mm}
\subfloat[][]{
  \includegraphics[width=29mm]{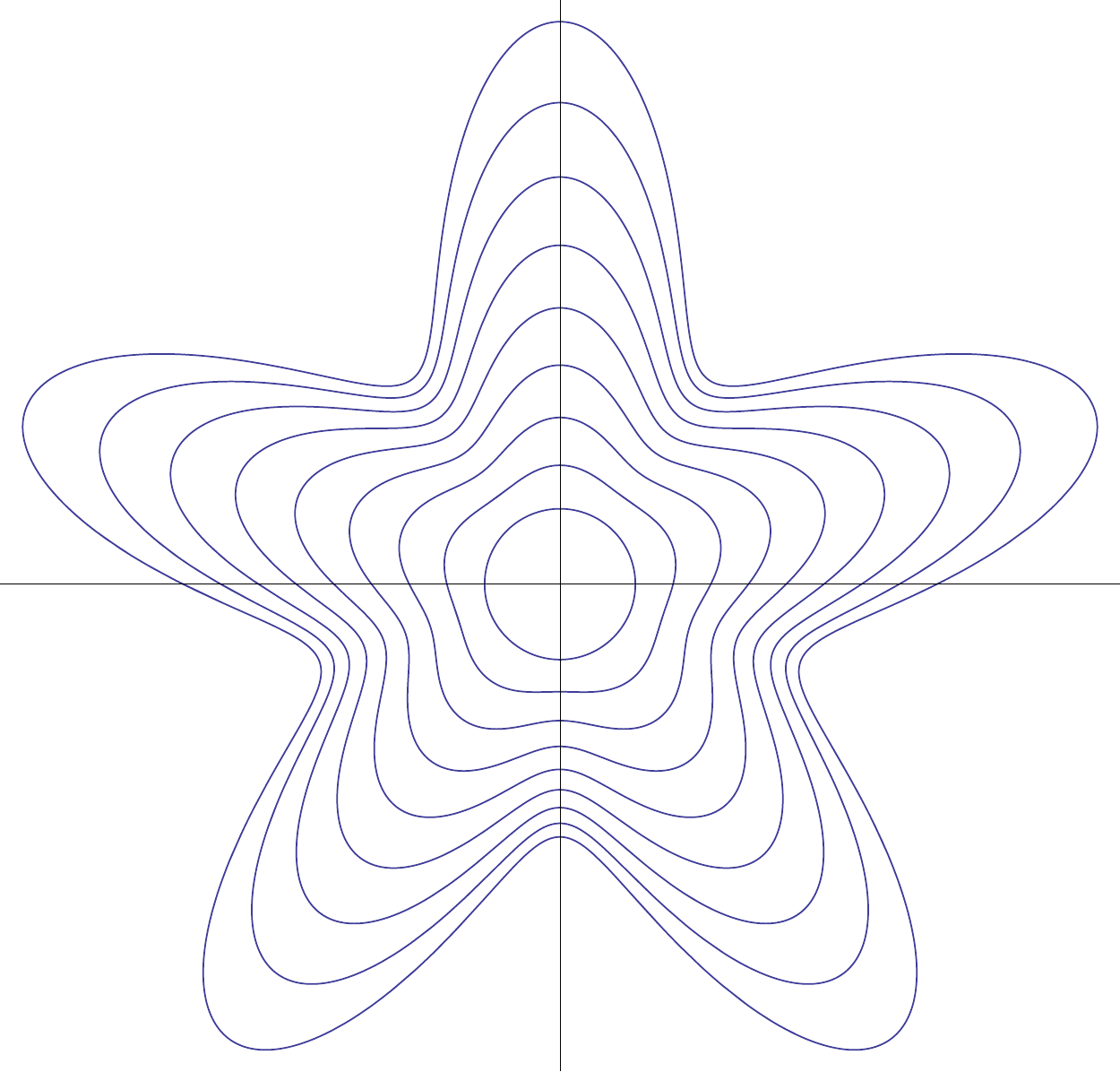}} \hspace{3mm}
\caption{\textbf{Wavy contours to circle deformation.}
Here we plot some examples of (\ref{eq:wavycont}) for $p=2,3,4,5$, for different values of $0<\epsilon$.
For clarity, the figures are scaled to avoid intersections.
}
\label{fig:wavycont}
\end{center}
\end{figure}

Another family of symmetric contours we study is given by
\begin{align}\label{eq:wavycont}
X(\theta) = e^{i F(\theta)+ \epsilon \sin(p F(\theta))},
\end{align}
see figure \ref{fig:wavycont}.

\begin{figure}
\begin{center}
\subfloat[][]{
  \includegraphics[width=49mm]{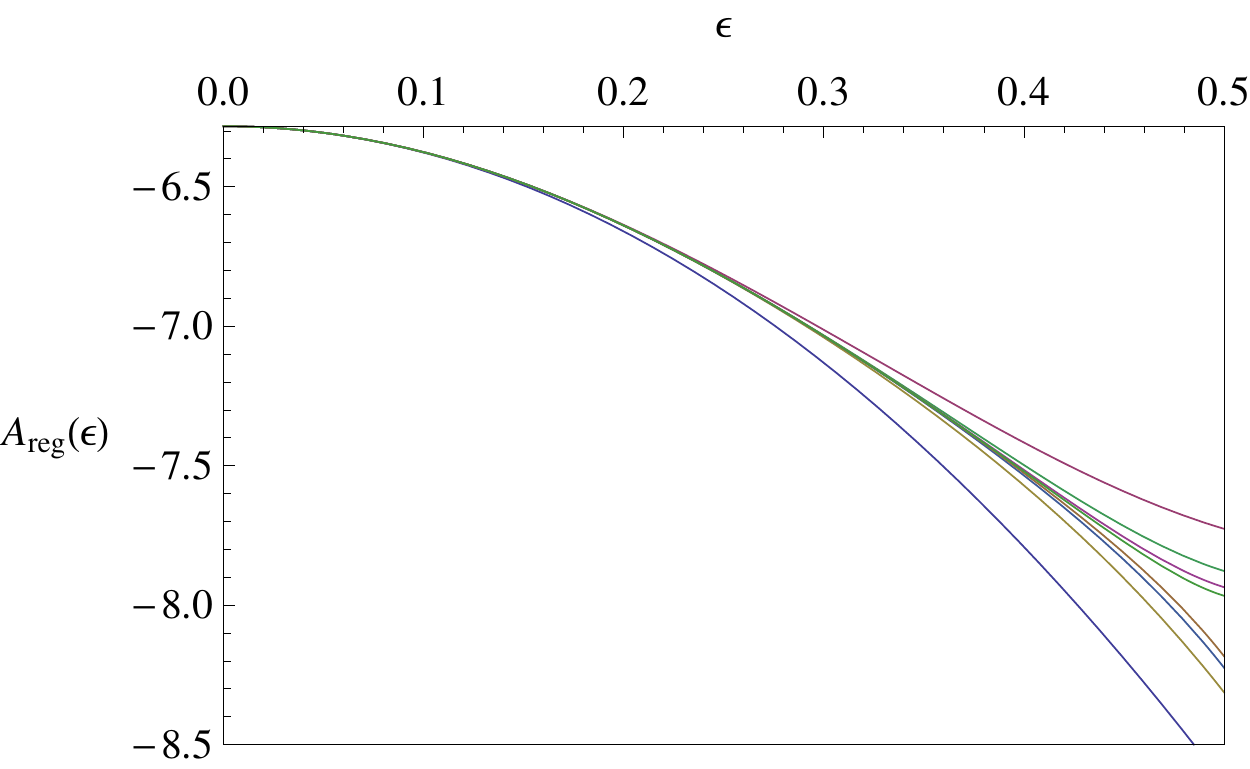}} \hspace{3mm}
\subfloat[][]{
  \includegraphics[width=49mm]{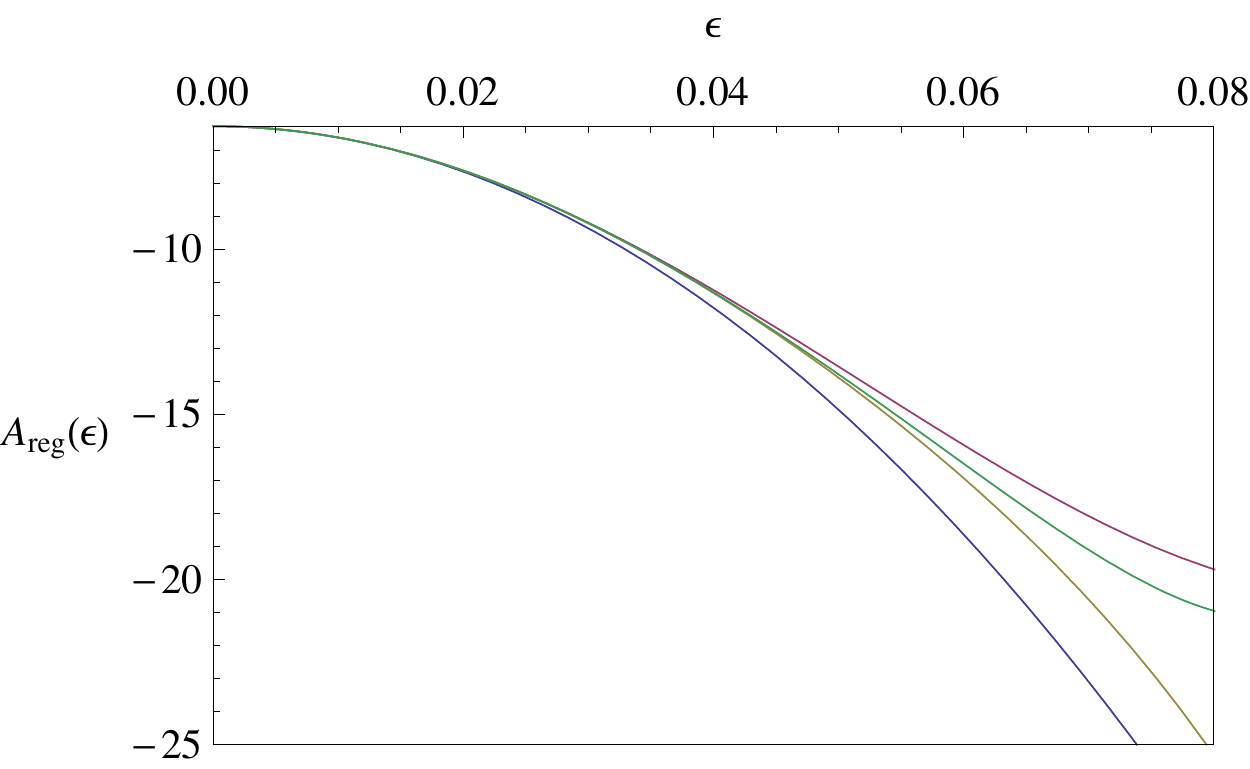}} \hspace{3mm}
\caption{\textbf{Regularized area.}
We plot the regularized area for the interpolating contours between the circle and the wavy contours.
The different lines correspond to different orders of the $\epsilon$ expansion, from $\epsilon^2$ to $\epsilon^{18}$ for $p=2$ and $\epsilon^{10}$ for $p=13$.
(a) $p=2$,
(b) $p=13$,
}
\label{fig:wavyarea}
\end{center}
\end{figure}

\begin{figure}
\begin{center}
\subfloat{
  \includegraphics[width=27mm]{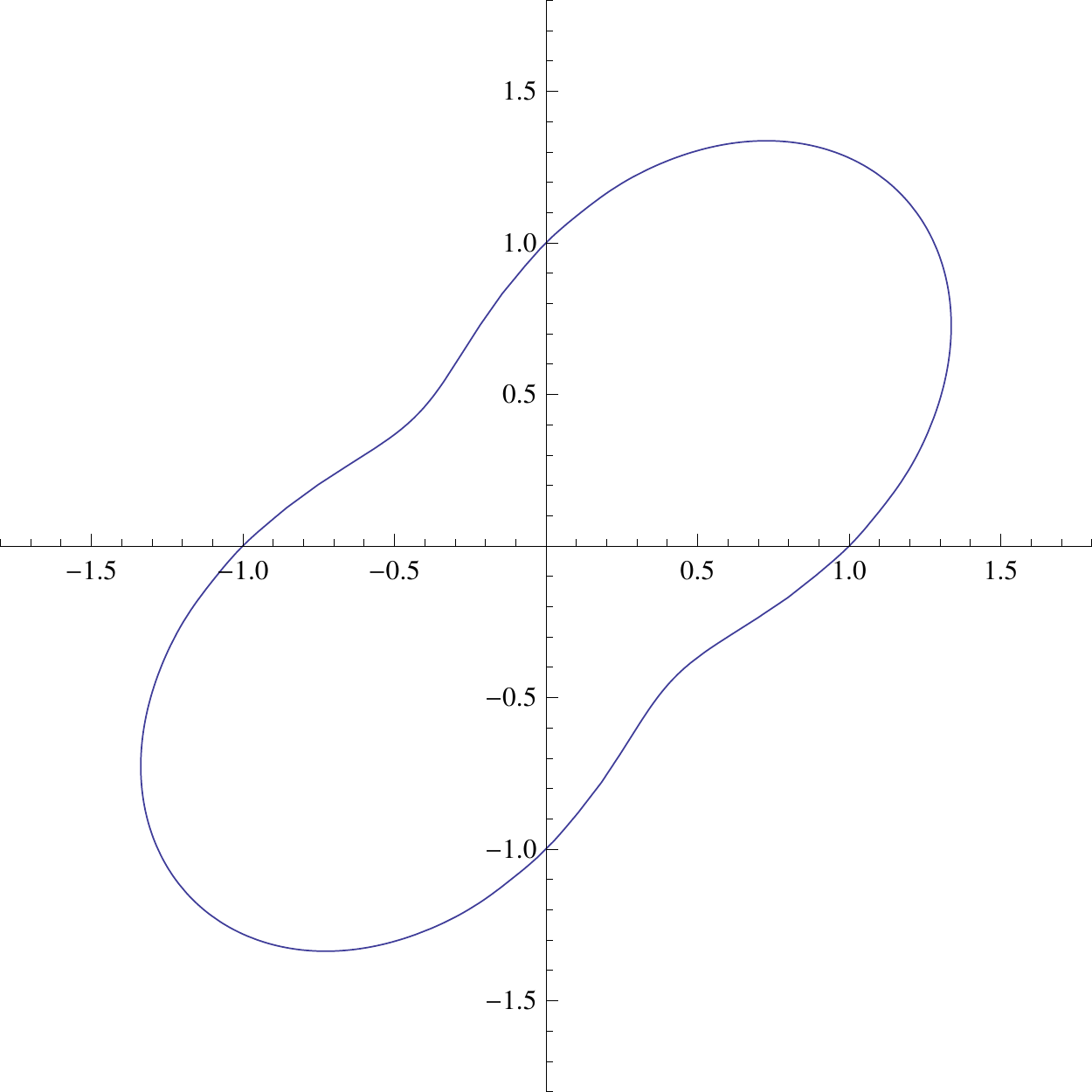}} \hspace{3mm}
\subfloat{
  \includegraphics[width=27mm]{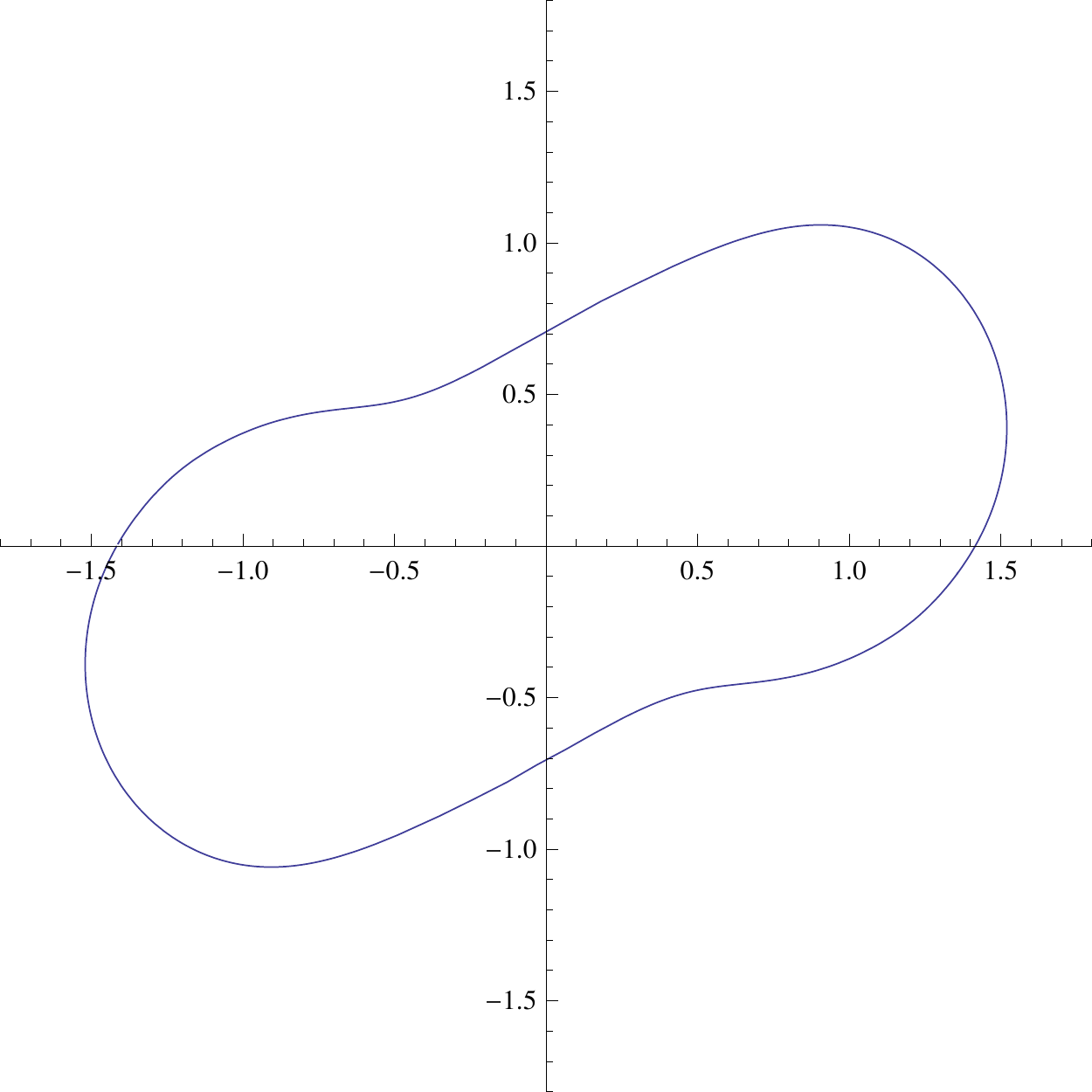}} \hspace{3mm}
\subfloat{
  \includegraphics[width=27mm]{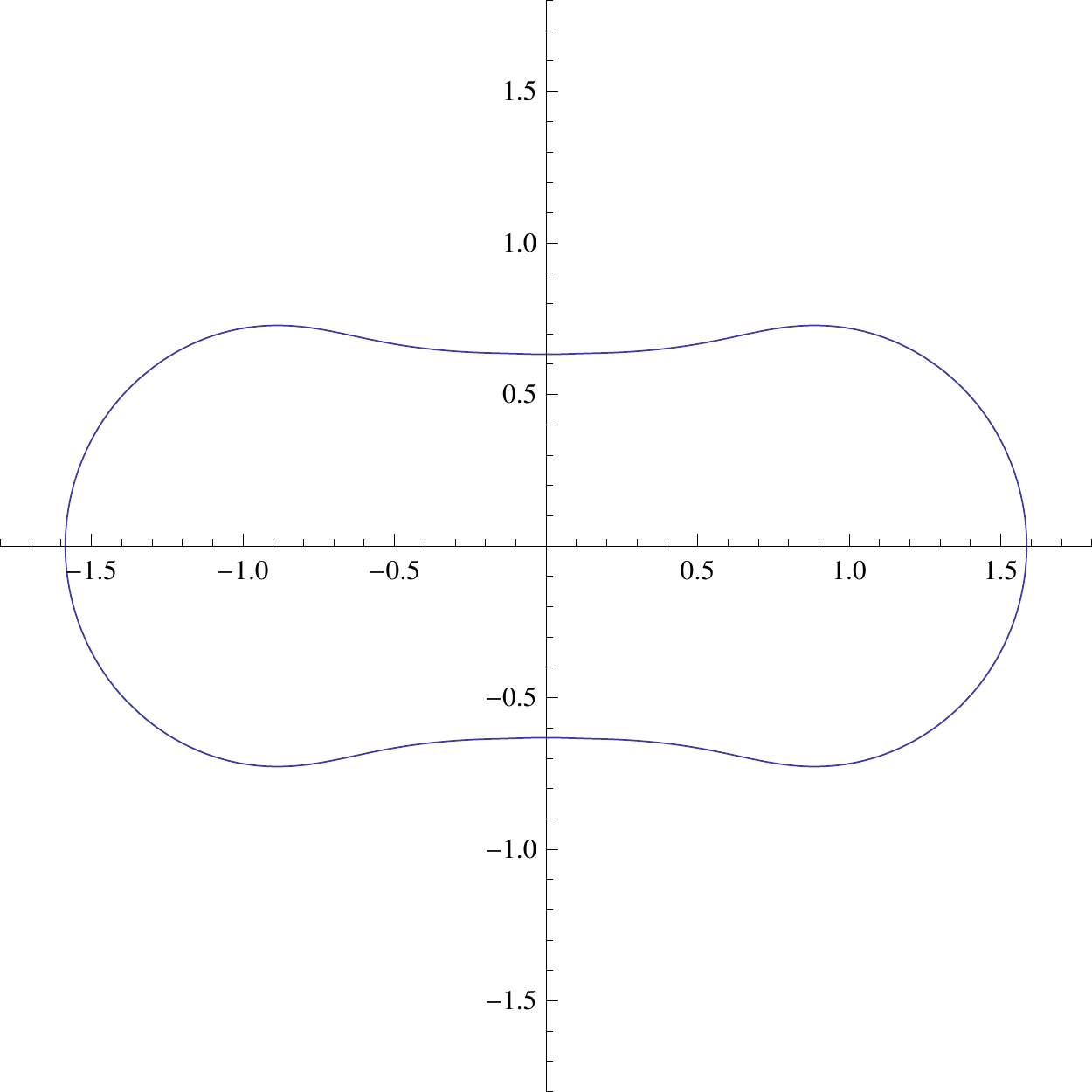}} \hspace{3mm}
\subfloat{
  \includegraphics[width=27mm]{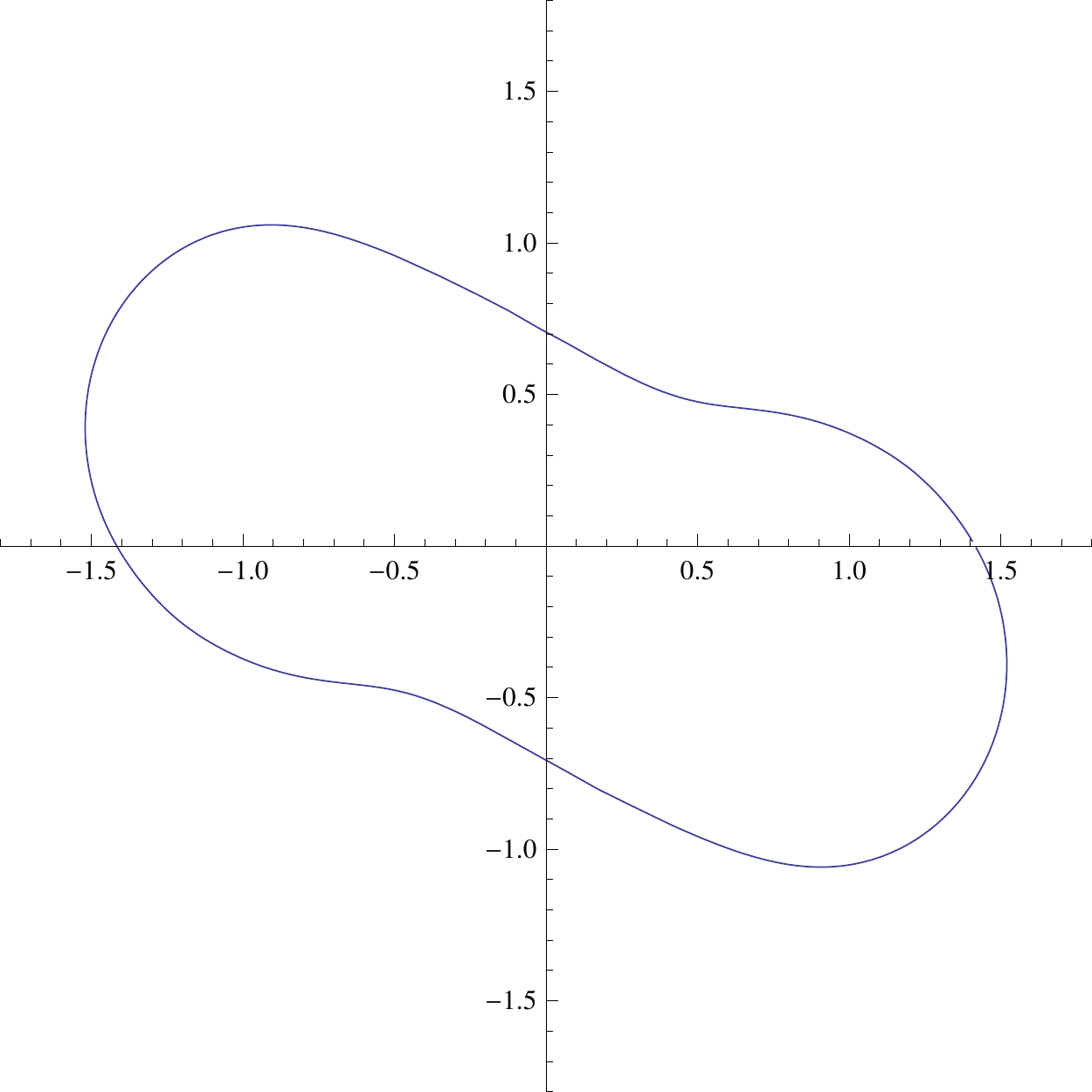}} \hspace{3mm}
\subfloat{
  \includegraphics[width=27mm]{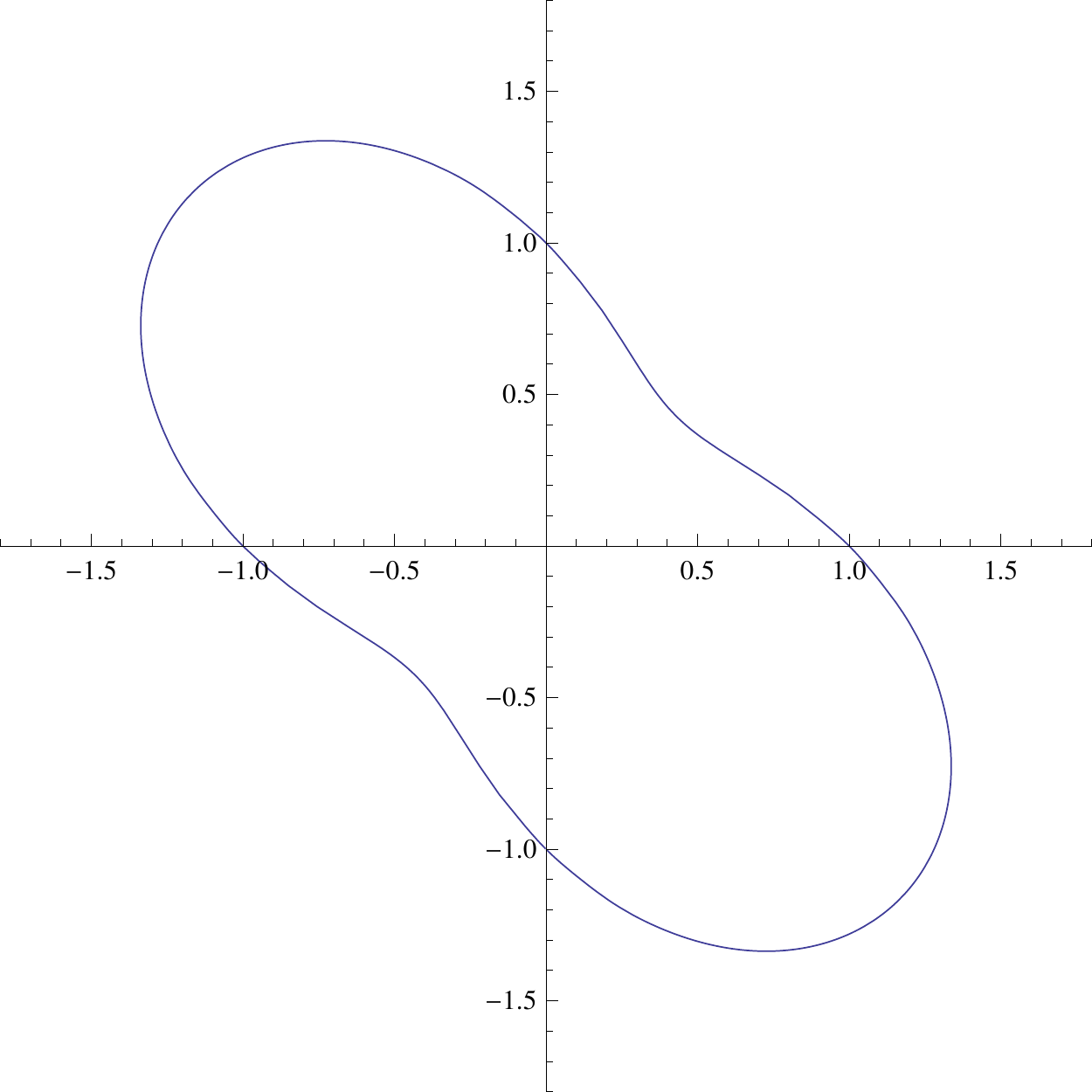}} \hspace{3mm}
\caption{\textbf{$\lambda$-deformations.}
We plot the $\lambda$-deformations for $\varphi=0,\frac{\pi}{4},\frac{\pi}{2},\frac{3\pi}{4},\pi$ for the $p=2$ case with $\epsilon=0.5$.
}
\label{fig:wavyodef}
\end{center}
\end{figure}

as in the previous example, we can study it for different specific values of $p$.
Here we consider two examples, $p=2$ where quite easily we can get to very high orders in the expansion, and see how the weak coupling integral depends on the $\lambda$-deformations in very high orders.
The second example is $p=13$, which is a bit harder, but one can still see the same dependence on $\lambda$ to the first orders in the expansion.
Next we give the results

\begin{align}
A_{\text{reg}}^{p=2} =& -2\pi -3 \pi  \epsilon ^2+\frac{93 \pi  \epsilon ^4}{20}-\frac{50143 \pi  \epsilon ^6}{4200}+\frac{510139 \pi  \epsilon ^8}{14400}-\frac{65754318359 \pi  \epsilon ^{10}}{582120000}\nonumber\\
& +\frac{1195458440855851 \pi  \epsilon ^{12}}{3178375200000}-\frac{61047851487256409 \pi  \epsilon ^{14}}{47344547250000}\nonumber\\
&
+\frac{45707069078388982419341507 \pi  \epsilon ^{16}}{10124976097716480000000}-\frac{52566325973037148254959546391187 \pi  \epsilon ^{18}}{3273637646841985463040000000}\nonumber\\
&
+ \mathcal{O}(\epsilon^{20}),\\
A_{\text{reg}}^{p=13} =& -2\pi -1092 \pi  \epsilon ^2+\frac{1660932 \pi  \epsilon ^4}{25}-\frac{3887594024353 \pi  \epsilon ^6}{570000} +\frac{679687975645852511 \pi  \epsilon ^8}{821712000}\nonumber\\
& -\frac{2652706006393624451200787779 \pi  \epsilon ^{10}}{24329522800000000}+ \mathcal{O}(\epsilon^{12}).
\end{align}

\begin{align}
W_{1,\varphi}^{p=2} =& -2 \pi ^2-6 \pi ^2 \epsilon ^2+\frac{31 \pi ^2 \epsilon ^4}{4}-\frac{985 \pi ^2 \epsilon ^6}{48}+\frac{\pi ^2 (10132841-11016 \cos(2 \varphi )) \epsilon ^8}{161280}\nonumber\\
& +\frac{\pi ^2 (-507075437357+519046632 \cos(2 \varphi )) \epsilon ^{10}}{2483712000} \nonumber\\
&-\frac{\pi ^2 (-205243491659573369+193885849040584 \cos(2 \varphi )+6619276121760 \cos(4 \varphi )) \epsilon ^{12}}{298343485440000}\nonumber\\
&+\frac{\pi ^2 \epsilon^{14}}{1628955430502400000}\bigg(-3872331127046431211507+3825980252246834960 \cos(2 \varphi )\nonumber\\
&+207824411640450672 \cos(4 \varphi )\bigg)
+\frac{\pi ^2\epsilon^{16}}{584860157767581696000000}\bigg(\nonumber\\
&-4894520980993299391859564659+5286197530092023750882320 \cos(2 \varphi )\nonumber\\
&+304417890855492833823264 \cos(4 \varphi )+4656911757360684307200 \cos(6 \varphi )\bigg)+ \mathcal{O}(\epsilon^{18}),\nonumber\\
W_{1,\varphi}^{p=13} =& -2 \pi ^2-2184 \pi ^2 \epsilon ^2+107926 \pi ^2 \epsilon ^4-\frac{275349451 \pi ^2 \epsilon ^6}{24}\nonumber\\
& +\frac{13 \pi ^2 (64529639666077152031+16900003369957344\cos (2 \varphi )) \epsilon ^8}{585469800000}\nonumber\\
& -\frac{13 \pi ^2 (17188687999825875665056111457+5500782068869693320779168 \cos(2 \varphi ))\epsilon ^{10}}{1160518237560000000}\nonumber\\
&
+ \mathcal{O}(\epsilon^{12}).
\end{align}

The $p=2$ example is quite interesting since we could compute $W_{1,\varphi}$ to higher orders than in the other examples above and see how the $\varphi$ dependence changes as we increase the order of the expansion.
A plot of the regularized area is given in figure \ref{fig:wavyarea}.
In figure \ref{fig:wavyodef} we plot the contour for different values of the phase $\lambda = e^{i \varphi}$ for $p=2$, for $p=13$ it is very hard to notice the change by the deformation for the values of $\epsilon$ for which our approximation is valid, thus we do not present the corresponding figures.

\subsection{Lima\c{c}on}
\begin{figure}
\begin{center}
\subfloat[][]{
  \includegraphics[width=30mm]{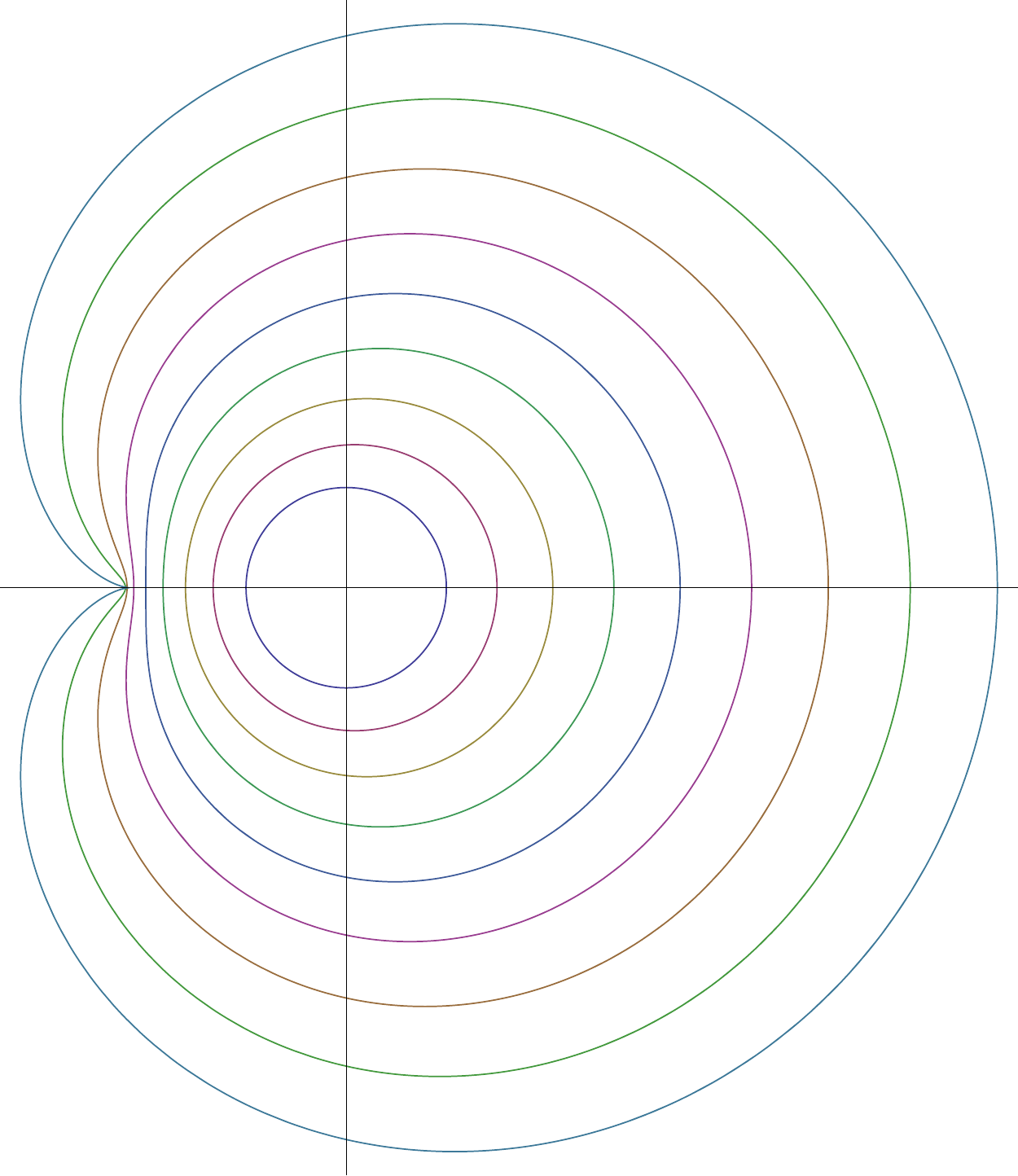}} \hspace{3mm}
\subfloat[][]{
  \includegraphics[width=60mm]{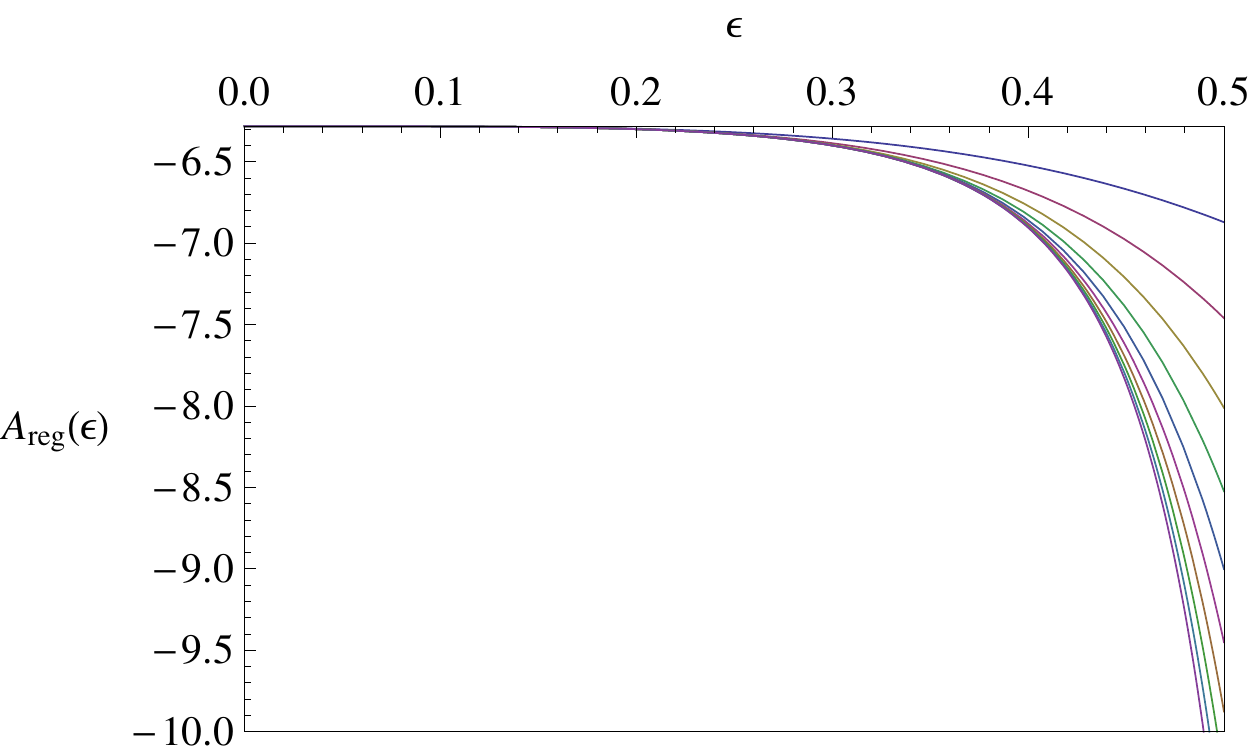}} \hspace{3mm}
\subfloat[][]{
  \includegraphics[width=50mm]{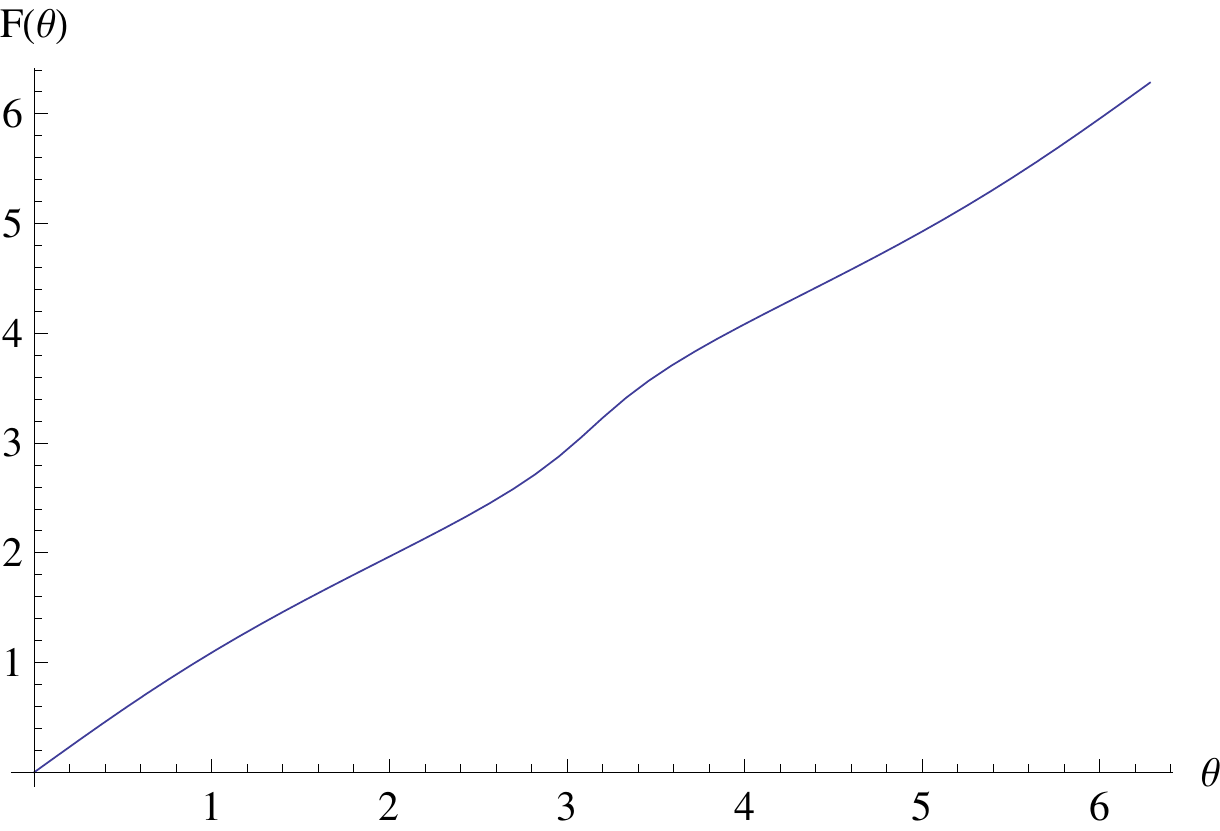}} \hspace{3mm}
\caption{\textbf{The lima\c{c}on.}
(a) The lima\c{c}on for different values of $\epsilon$ between $\epsilon=0$ (the circle) and $\epsilon = 1/2$ (the cardioid).
    For clarity, the different contours in the figure are scaled so that they do not overlap.
(b) The minimal surface area calculated using (\ref{eq:limacon_area}) by including terms up to $\epsilon^n$ terms from $n=2$ (upper blue line) to $n=22$ (lower purple line).
(c) The parametrization for $\epsilon=0.3125$, corresponding to the 6th inner contour in (a).}
\label{fig:Limacon}
\end{center}
\end{figure}

\begin{figure}
\begin{center}
\subfloat[][]{
  \includegraphics[width=25mm]{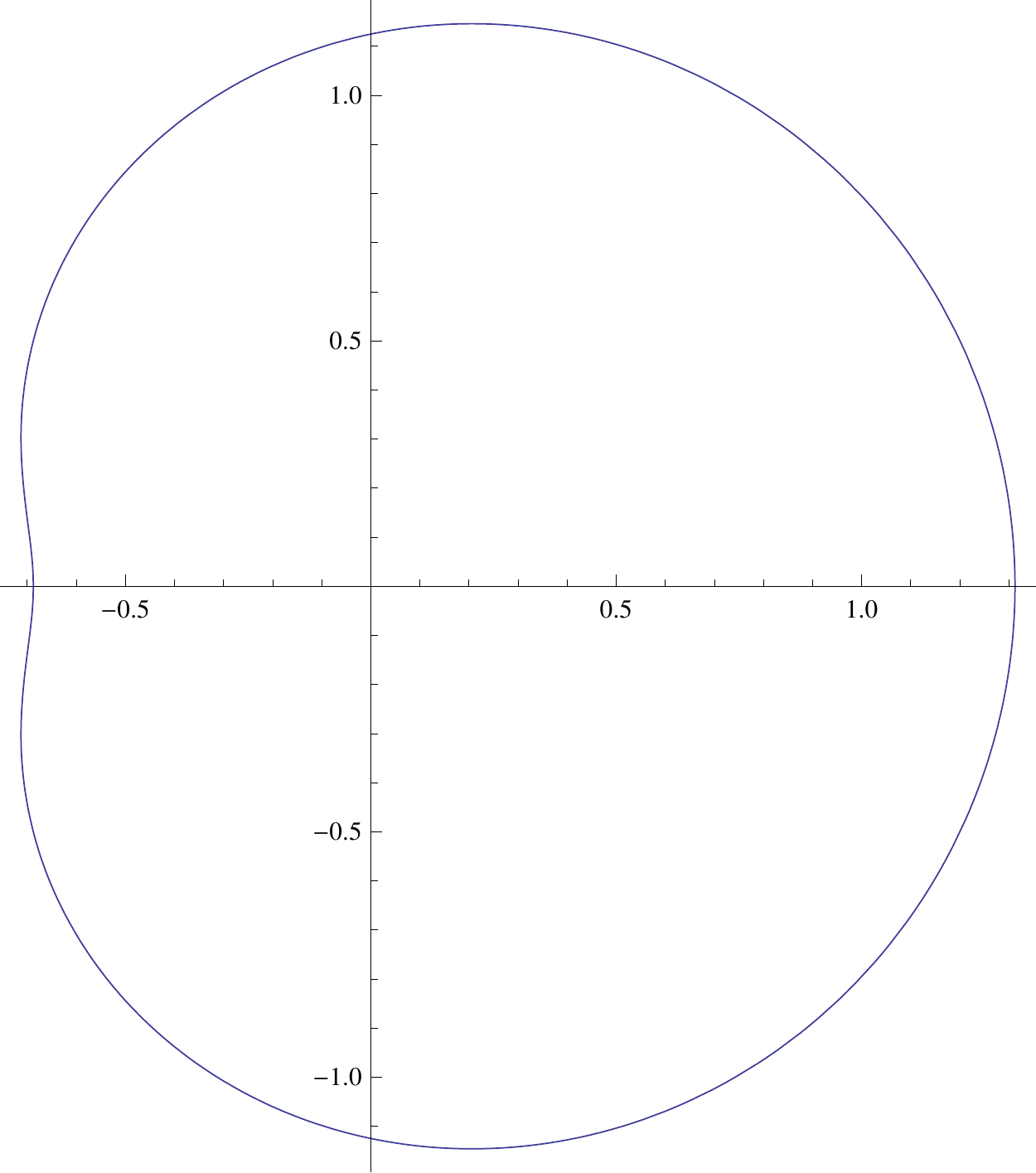}} \hspace{3mm}
\subfloat[][]{
  \includegraphics[width=25mm]{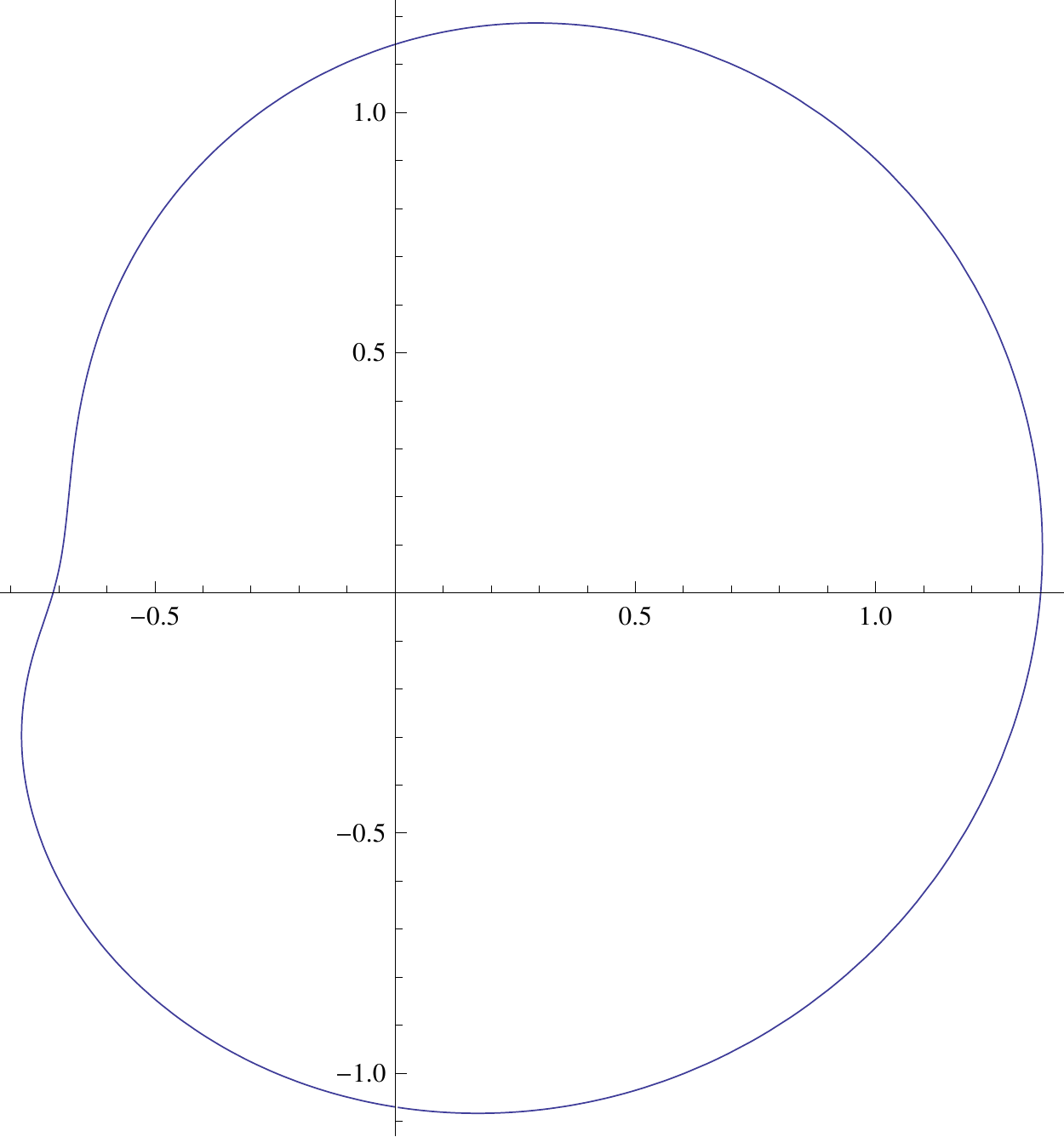}} \hspace{3mm}
\subfloat[][]{
  \includegraphics[width=25mm]{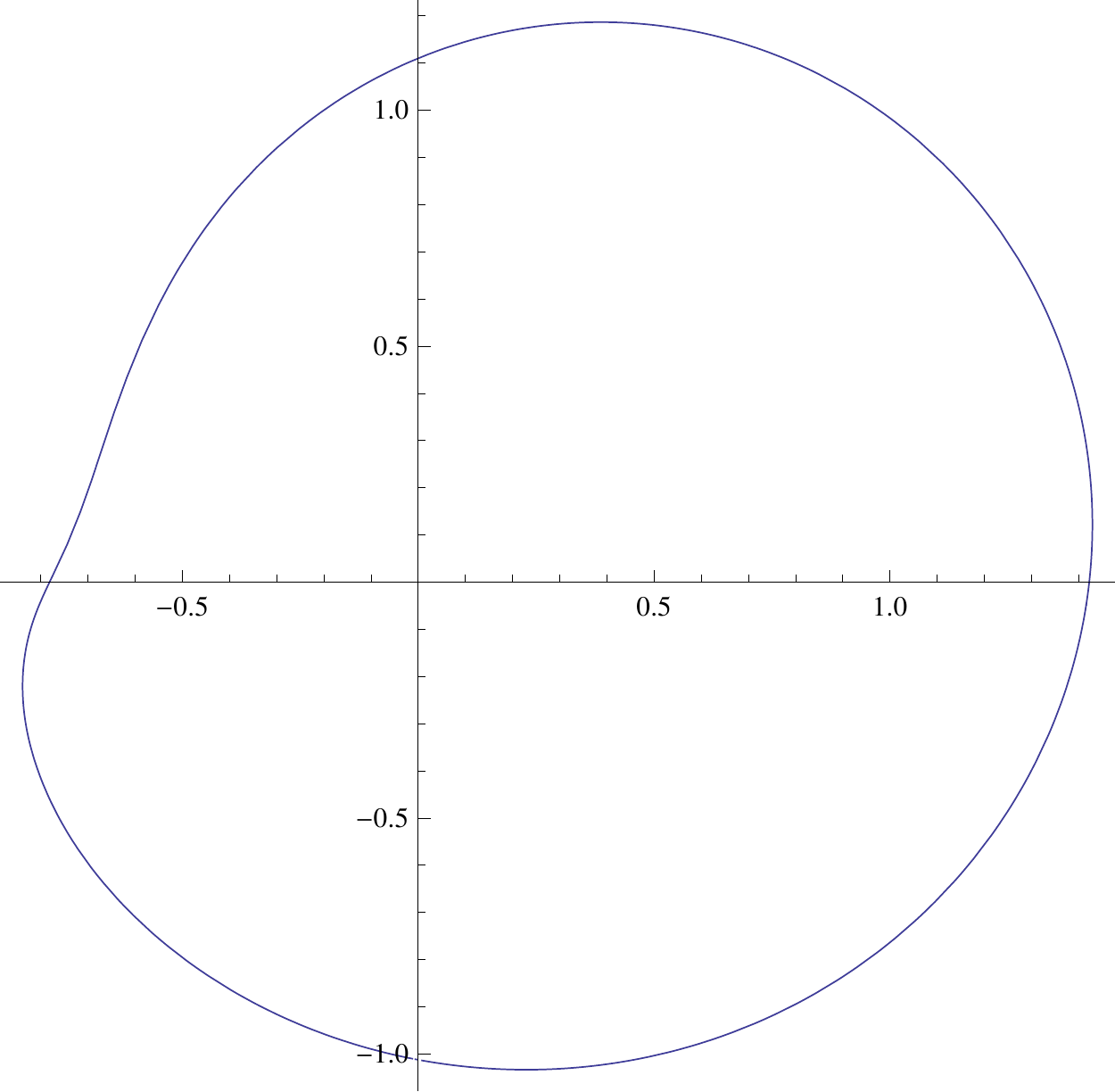}} \hspace{3mm}
\subfloat[][]{
  \includegraphics[width=25mm]{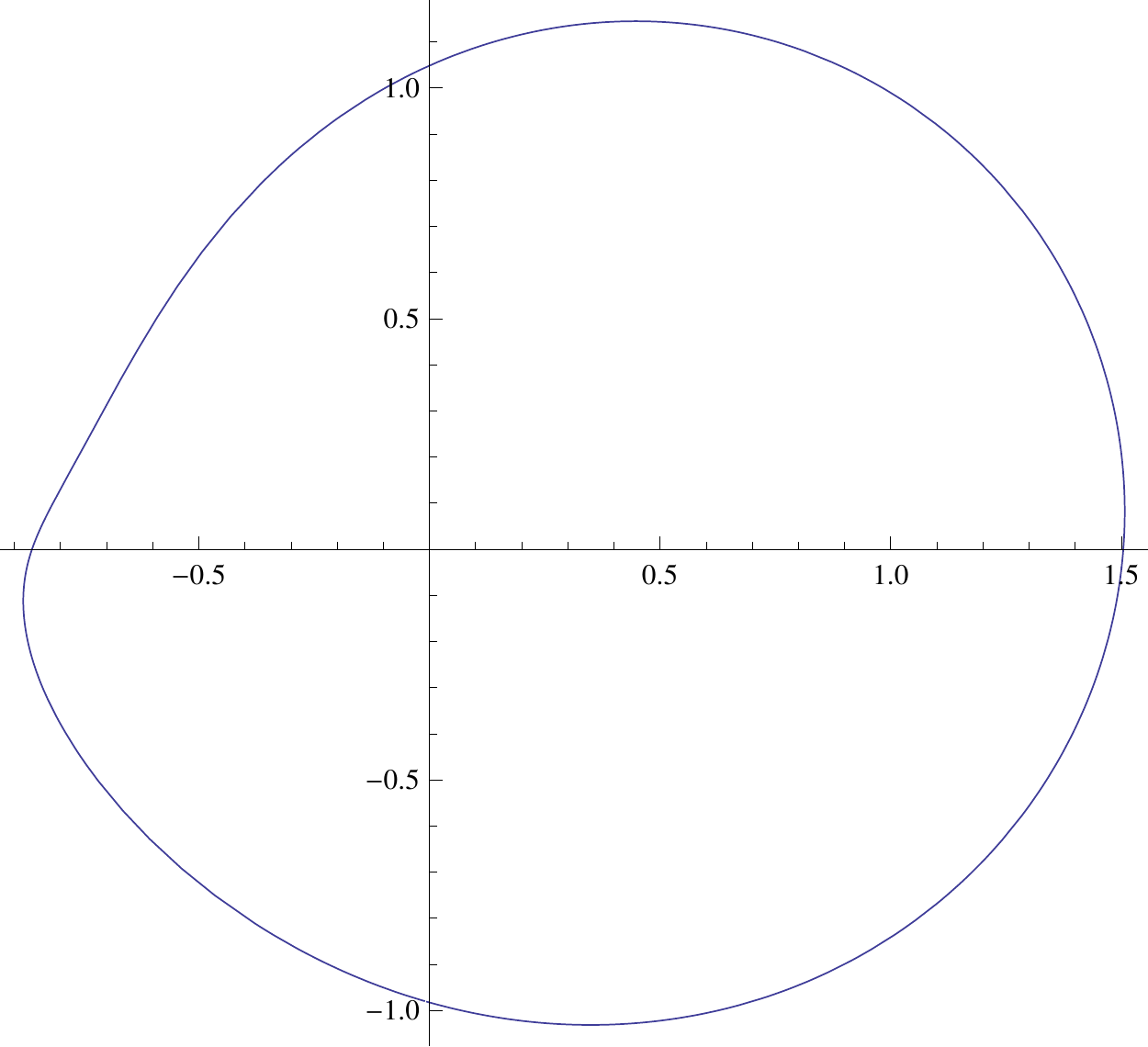}} \hspace{3mm}
\subfloat[][]{
  \includegraphics[width=25mm]{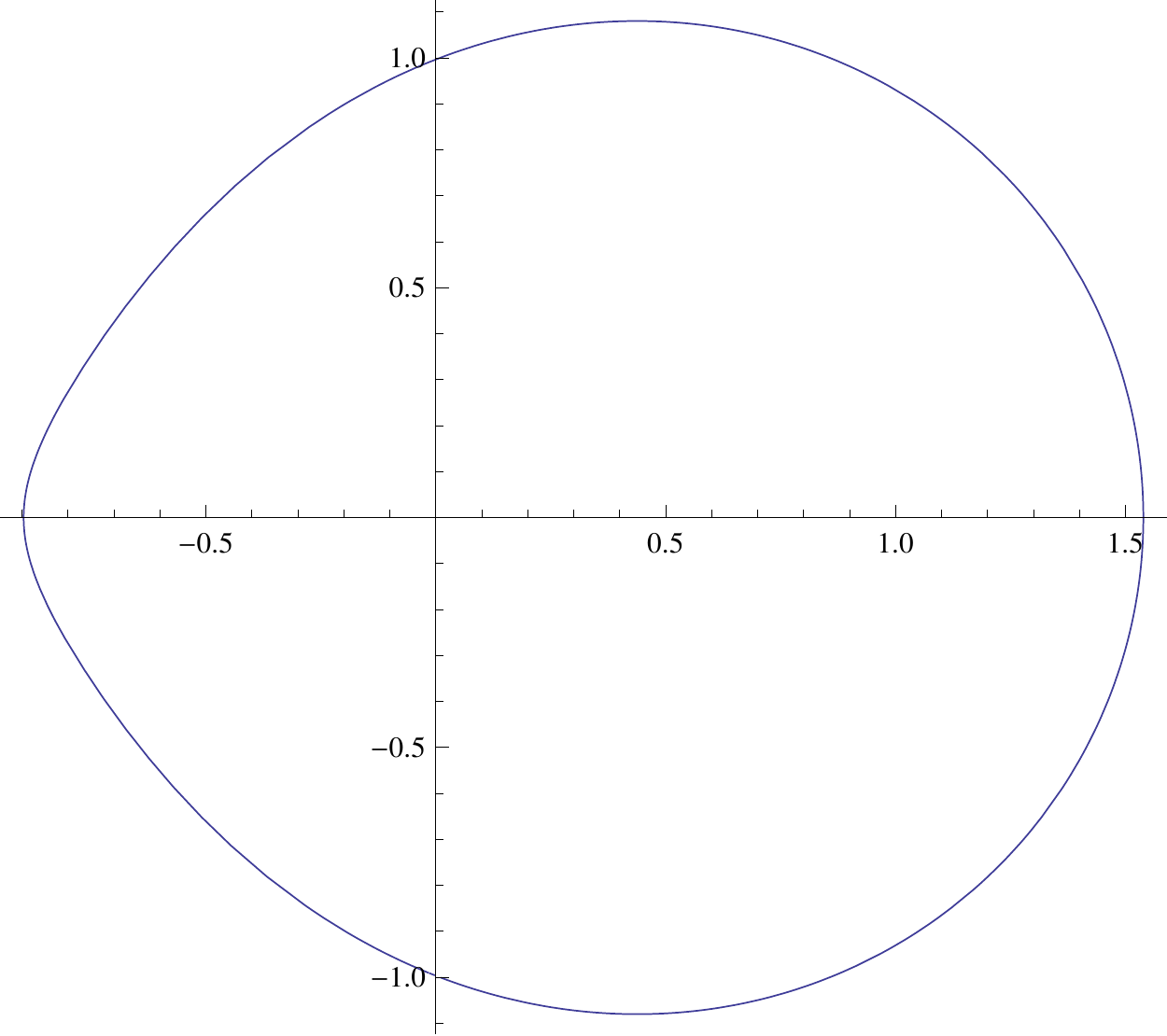}} \hspace{3mm}
\caption{\textbf{$\lambda$-deformations of the lima\c{c}on for $\epsilon = 0.3125$.}
(a) $\varphi=0$,
(b) $\varphi=\frac{\pi}{4}$,
(c) $\varphi=\frac{\pi}{2}$,
(d) $\varphi=\frac{3\pi}{4}$,
(e) $\varphi=\pi$.
}
\label{fig:Limacon_deform}
\end{center}
\end{figure}

In this case the contour is given by (see figure \ref{fig:Limacon} (a))
\begin{align}
X(\theta) = e^{i F(\theta)}+ \epsilon e^{2 i F(\theta)}.
\end{align}
As can be seen, this contour has only one reflectional symmetry.
The regularized area of the minimal surface is given by
\begin{align}\label{eq:limacon_area}
A_{\text{reg}} =& -2\pi -3 \pi  \epsilon ^4-12 \pi  \epsilon ^6-\frac{897 \pi  \epsilon ^8}{20}-\frac{834 \pi  \epsilon ^{10}}{5}-\frac{872541 \pi  \epsilon ^{12}}{1400}-\frac{820023 \pi  \epsilon ^{14}}{350}\nonumber\\
&-\frac{694490161 \pi  \epsilon ^{16}}{78400} -\frac{32988413 \pi  \epsilon ^{18}}{980}
-\frac{19389297203319 \pi  \epsilon ^{20}}{150920000}-\frac{3714023472279 \pi  \epsilon ^{22}}{7546000}
+ \mathcal{O}(\epsilon^{24}).
\end{align}
The one loop weak coupling expectation value for the $\lambda$-deformed contour is given by
\begin{align}
W_{1,\varphi} =& -2 \pi ^2-6 \pi ^2 \epsilon ^4-24 \pi ^2 \epsilon ^6-\frac{365 \pi ^2 \epsilon ^8}{4}-346 \pi ^2 \epsilon ^{10}-\frac{21063 \pi ^2 \epsilon ^{12}}{16}-\frac{20117 \pi ^2 \epsilon ^{14}}{4}\nonumber\\
& -\frac{3 \pi ^2 (115201973+2592 \cos(2 \varphi ))\epsilon ^{16}}{17920}-\frac{\pi ^2 (83104799+7776 \cos(2 \varphi ))\epsilon ^{18}}{1120}
+ \mathcal{O}(\epsilon^{20}).
\end{align}
In figure \ref{fig:Limacon} (b) and (c), we plot the regularized area and the correct parametrization function respectively.
In figure \ref{fig:Limacon_deform} we plot the contour for different values of the phase $\lambda = e^{i \varphi}$.
The relative difference between $W_{1,\varphi}$ and $W_{1,\varphi=0}$ is very small.
For example, for $\epsilon = 0.3125$ the maximal relative difference is $\frac{W_{1,\varphi=0}-W_{1,\varphi=\pi/2}}{W_{1,\varphi=0}} \simeq 8.79 \times 10^{-9}$.

\subsection{Asymmetric contour}
\begin{figure}
\begin{center}
\subfloat[][]{
  \includegraphics[width=30mm]{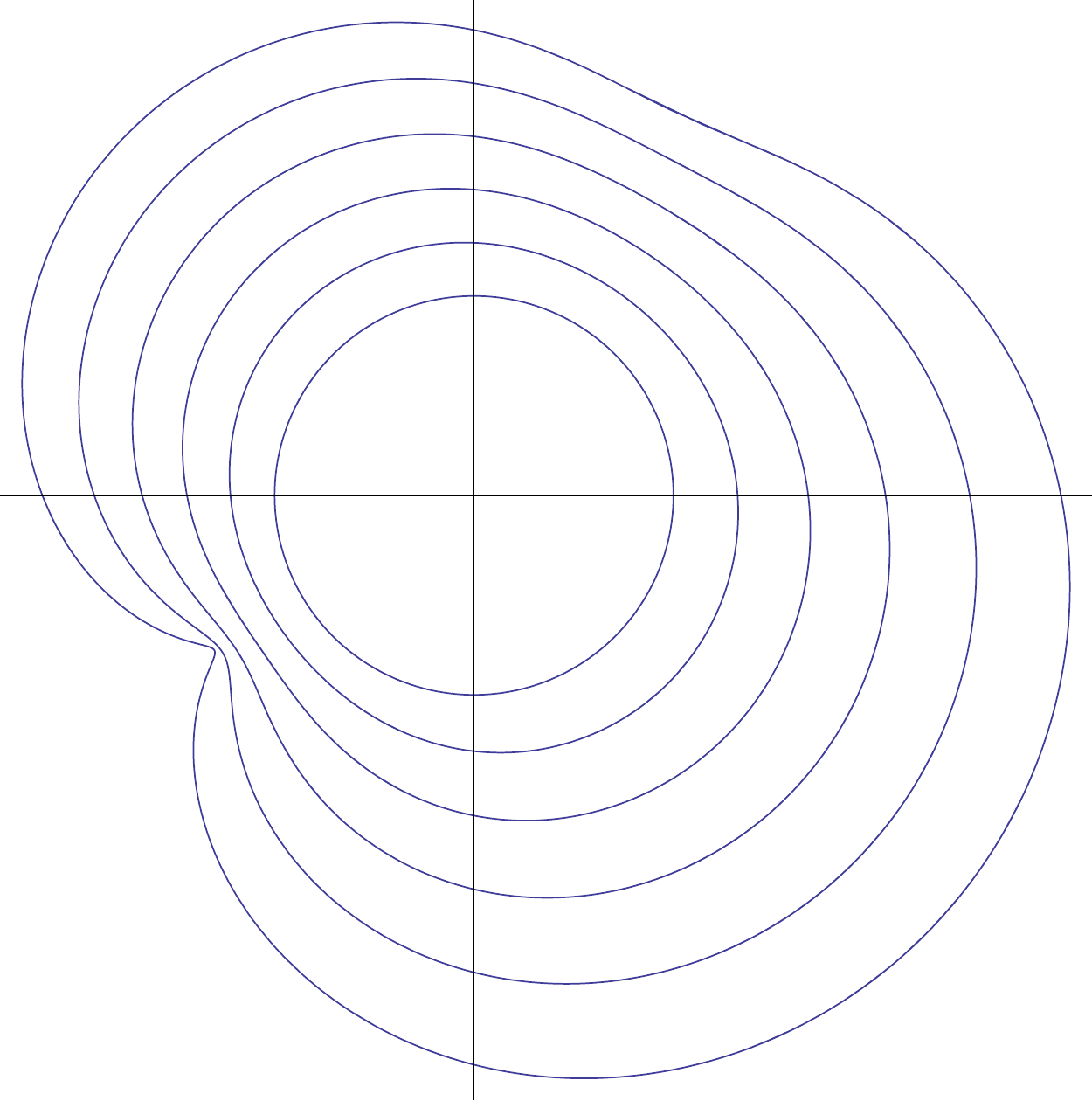}} \hspace{3mm}
\subfloat[][]{
  \includegraphics[width=60mm]{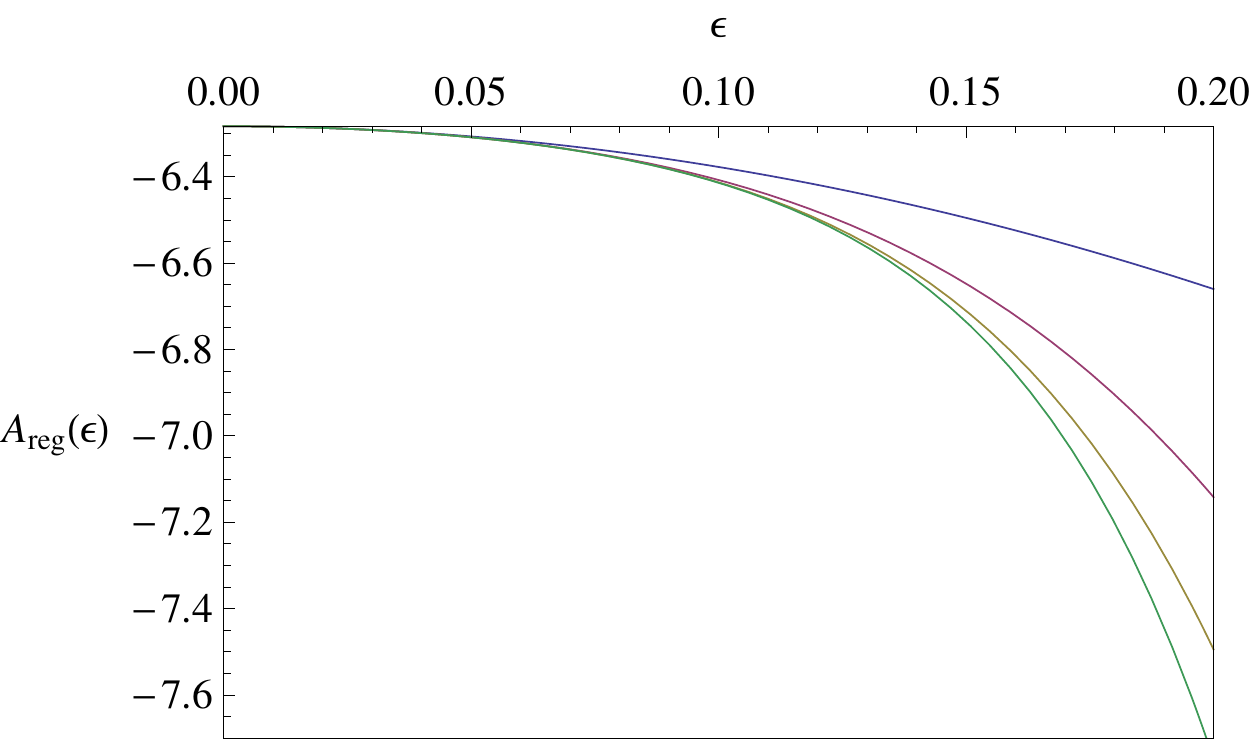}} \hspace{3mm}
\subfloat[][]{
  \includegraphics[width=50mm]{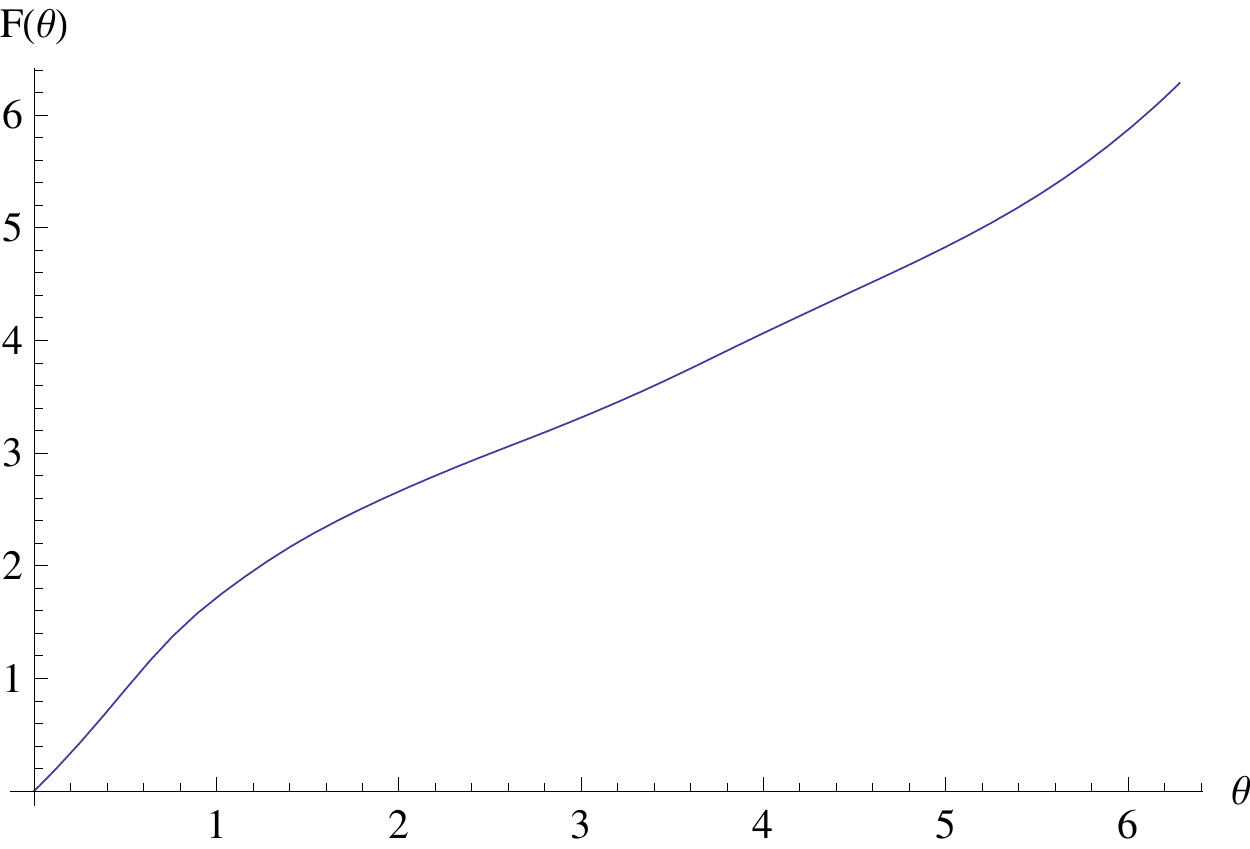}} \hspace{3mm}
\caption{\textbf{The asymmetric contour.}
(a) The asymmetric contour for different values of $\epsilon$ between $\epsilon=0$ (the circle) and $\epsilon = 0.2$.
    For clarity, the different contours in the figure are scaled so that they do not overlap.
(b) The minimal surface area calculated using (\ref{eq:asymm_area}) by including terms up to $\epsilon^n$ terms from $n=2$ (upper blue line) to $n=8$ (lower green line).
(c) The parametrization for $\epsilon=0.08$, corresponding to the 3rd inner contour in (a).}
\label{fig:asymm}
\end{center}
\end{figure}

In this case we choose the simple asymmetric contour (see figure \ref{fig:asymm} (a))
\begin{align}
X(s) = e^{i F(\theta)}+ \epsilon \left(e^{2 i F(\theta)} + i e^{3 i F(\theta)}\right).
\end{align}
The regularized area of the minimal surface is given by
\begin{align}\label{eq:asymm_area}
A_{\text{reg}} =& -2\pi -3 \pi  \epsilon ^2-\frac{1917 \pi  \epsilon ^4}{20}-\frac{350823 \pi  \epsilon ^6}{200}-\frac{2475105369 \pi  \epsilon ^8}{78400} + \mathcal{O}(\epsilon^{10}).
\end{align}
The one loop weak coupling expectation value for the $\lambda$-deformed contour is given by
\begin{align}
W_{1,\varphi} =& -2 \pi ^2-6 \pi ^2 \epsilon ^2-\frac{773 \pi ^2 \epsilon ^4}{4}-\frac{57359 \pi ^2 \epsilon ^6}{16}
-\frac{\pi ^2 (1182155647+62208 \cos(2 \varphi ))\epsilon ^8}{17920}
+ \mathcal{O}(\epsilon^{10}).
\end{align}

\begin{figure}
\begin{center}
\subfloat[][]{
  \includegraphics[width=25mm]{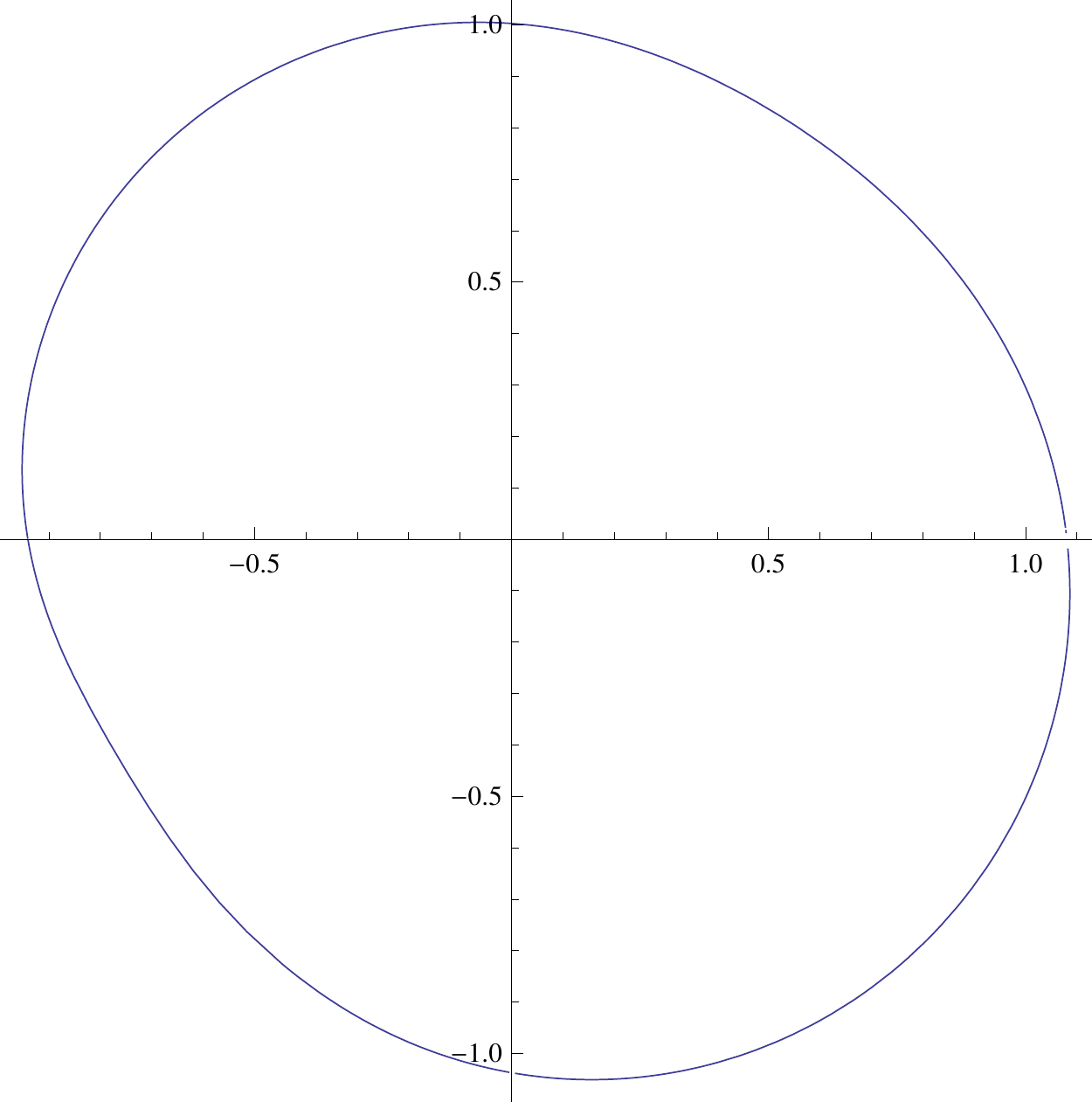}} \hspace{3mm}
\subfloat[][]{
  \includegraphics[width=25mm]{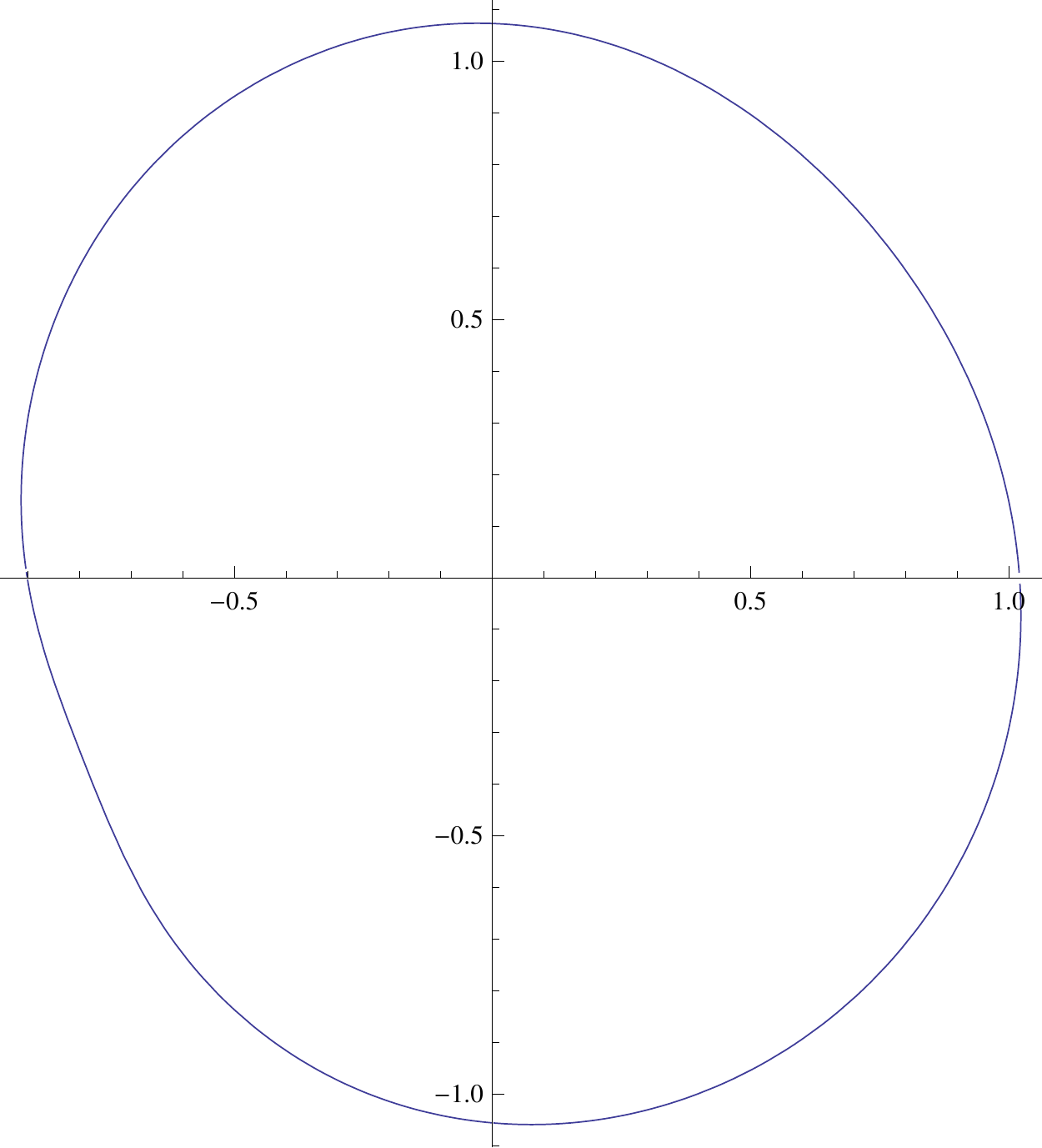}} \hspace{3mm}
\subfloat[][]{
  \includegraphics[width=25mm]{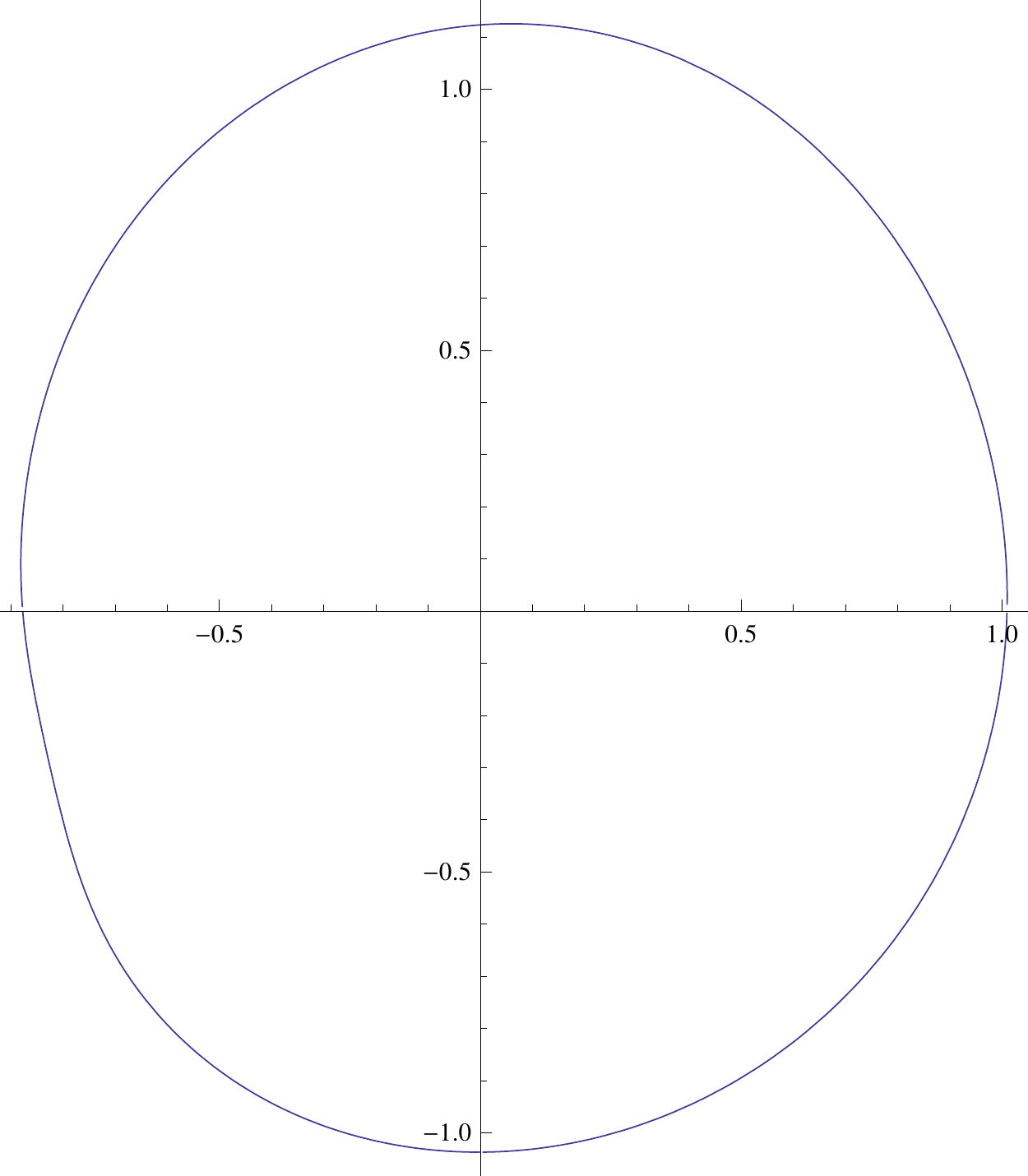}} \hspace{3mm}
\subfloat[][]{
  \includegraphics[width=25mm]{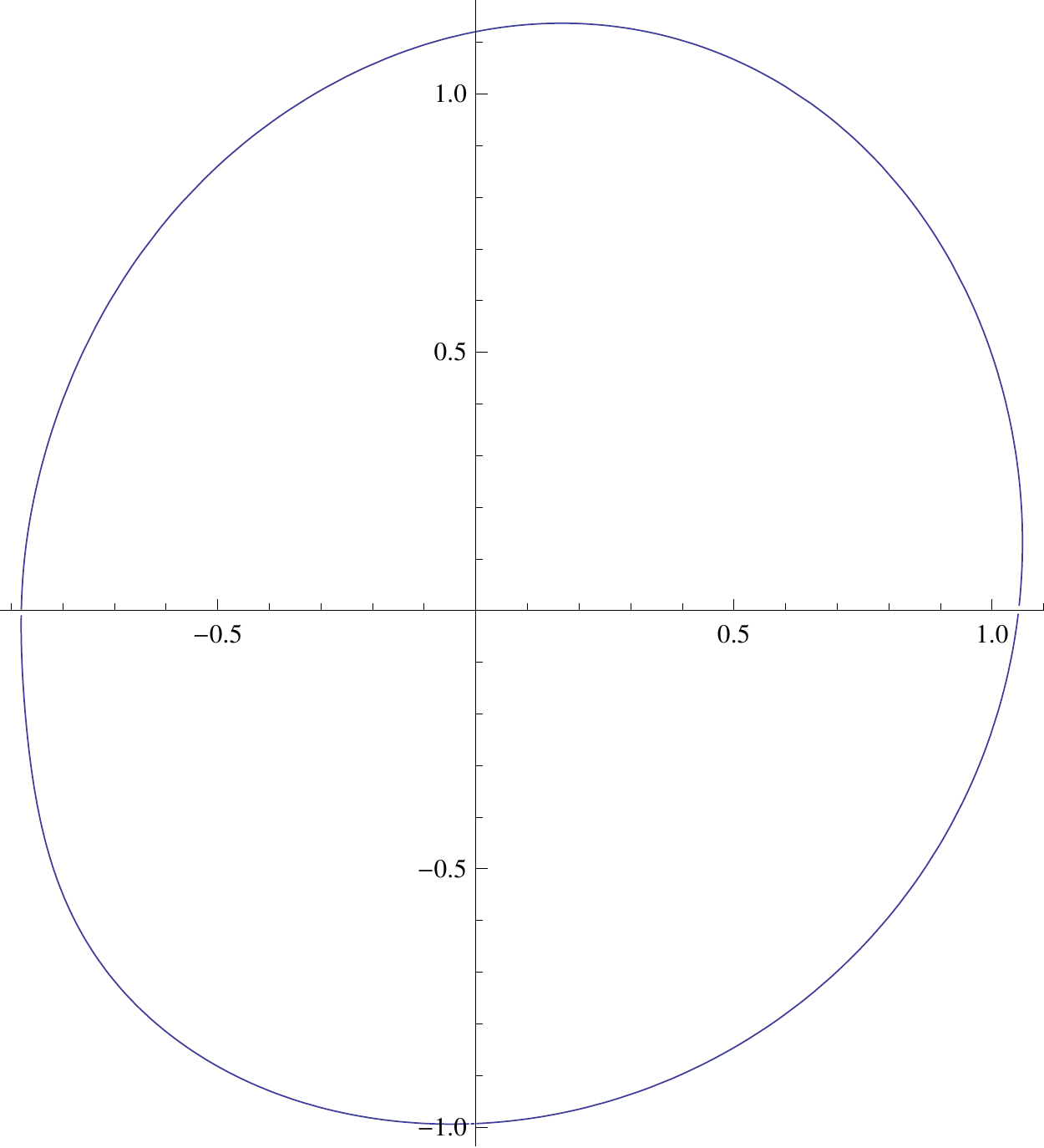}} \hspace{3mm}
\subfloat[][]{
  \includegraphics[width=25mm]{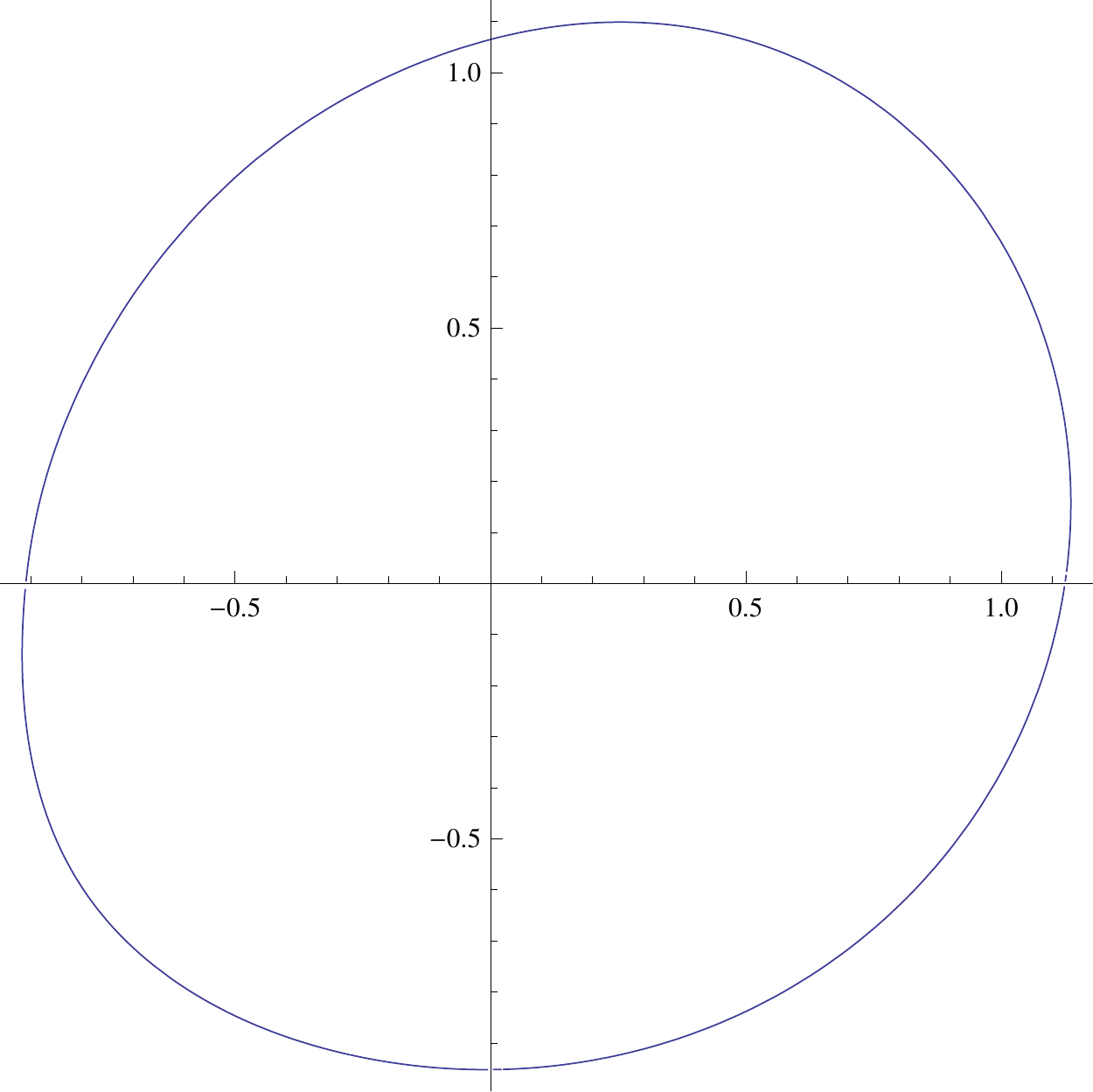}} \hspace{3mm}
\caption{\textbf{$\lambda$-deformations of the asymmetric contour for $\epsilon = 0.07$.}
(a) $\varphi=0$,
(b) $\varphi=\frac{\pi}{4}$,
(c) $\varphi=\frac{\pi}{2}$,
(d) $\varphi=\frac{3\pi}{4}$,
(e) $\varphi=\pi$.
}
\label{fig:asymm_deform}
\end{center}
\end{figure}

In figures \ref{fig:asymm} (b) and (c), we plot the regularized area and the correct parametrization function respectively.
In figure \ref{fig:asymm_deform} we plot the contour for different values of the phase $\lambda = e^{i \varphi}$.
The relative difference between $W_{1,\varphi}$ and $W_{1,\varphi=0}$ is very small.
For example, for $\epsilon = 0.07$ the maximal relative difference is $\frac{W_{1,\varphi=0}-W_{1,\varphi=\pi/2}}{W_{1,\varphi=0}} \simeq 1.97 \times 10^{-9}$.

\section{Perturbations around the infinite straight line}\label{sec:infinite_line}

In some cases it might be more convenient to map the upper half plane (UHP) to the target space Wilson loop.
For instance, this could be the case for the infinite straight line solution, which is conformally equivalent to the circle.
Furthermore, for the circular solution this is the gauge used to write the minimal surface solution in terms of theta-functions \cite{Ishizeki:2011bf}.
In this section we adapt the formulation introduced in \cite{Kruczenski:2014bla} to map the solution from the UHP instead of the unit disk, and provide an explicit example of perturbation of the straight line into a sinusoidal curve.

For the infinite straight line solution we take the worldsheet to be the UHP, where it is more convenient to use the $\sigma$ and $\tau$ coordinates.
The boundary curve is defined on the real axis, that is at $\tau = 0$.
Using the fact that $\partial_\sigma = \partial + \bar \partial$ and $\partial_\tau = i (\partial - \bar \partial)$, the generalized cosh-Gordon equation is given by
\begin{equation}
\frac{1}{4}(\p_\sigma^2 + \p_\tau^2) \alpha(\sigma,\tau) = e^{2\alpha(\sigma,\tau)}+|f(\sigma + i \tau)|^2e^{-2\alpha(\sigma,\tau)}.
\end{equation}
Similarly, the flat-connection components are given by
\begin{equation}
J_\sigma(\lambda) = \left(
      \begin{array}{cc}
        \frac{i}{2}\p_\tau \alpha & f e^{-\alpha} + \frac{1}{\lambda} e^{\alpha} \\
        \lambda e^{\alpha} -\bar f e^{-\alpha} & -\frac{i}{2}\p_\tau\alpha \\
      \end{array}
    \right),\quad
J_\tau (\lambda) = \left(
      \begin{array}{cc}
        -\frac{i}{2}\p_\sigma \alpha & i(f e^{-\alpha} - \frac{1}{\lambda} e^{\alpha}) \\
        i(\lambda e^{\alpha} +\bar f e^{-\alpha}) & \frac{i}{2}\p_\sigma \alpha \\
      \end{array}
    \right).
\end{equation}
Notice that $\alpha(\sigma,\tau)$ has to diverge at $\tau = 0$, so expanding it around that point using the EOM gives
\begin{align}
\alpha \simeq -\ln (2\tau) + \tau^2 \gamma_2(\sigma) + \tau^4 \gamma_4(\sigma) + O(\tau^5)
\end{align}
where $\gamma_4(\sigma) = \frac{1}{10}\left(16 |f(\sigma)|^2 + 2\gamma_2^2(\sigma) - \gamma_2''(\sigma)\right)$.
In order to have a single boundary curve we should require $e^{2\alpha(\sigma,\tau \neq 0)} < |\infty|$ (remember that this factor is related to the induced matric).

Defining $H(\sigma,\tau) = \frac{\psi_1(\sigma,\tau)}{\psi_2(\sigma,\tau)}$ and expanding $\p_\tau H$ close to the boundary we find $H(\sigma,0) = i \lambda$, which implies that the boundary curve is given by $\bar X = \psi_1/\tilde \psi_1$.

From the linear problem, by defining $\chi = \frac{1}{\sqrt{J_{21}^\sigma}}\psi_1$, we get $-\chi''+V(\sigma,\tau)\chi = 0$.
Close to the boundary we have
\begin{align}
V(\sigma) \simeq \frac{3}{2}\gamma_2(\sigma) + \lambda f(\sigma) - \frac{1}{\lambda} \bar f(\sigma).
\end{align}
Next, we can relate the potential to the Schwarzian derivative by $\{ \bar X(\sigma), \sigma\} = -2 V(\sigma,\tau = 0)$, so
\begin{align}
\{ X^\lambda(\sigma), \sigma\} = -3\gamma_2(\sigma) + 2 \lambda f(\sigma) - 2 \frac{1}{\lambda} \bar f(\sigma),
\end{align}
thus,
\begin{align}
\Re\{ X^\lambda(\sigma), \sigma\} & = -3\gamma_2(\sigma) \nonumber\\
\Im\{ X^\lambda(\sigma), \sigma\} & = -2i( \lambda f(\sigma) - \frac{1}{\lambda} \bar f(\sigma)) = -4 \Im (\lambda f(\sigma)).
\end{align}
Now,
\begin{align}
f(\sigma) = i \left(1+\mathcal{H}\right)\Im(f(\sigma)) = -\frac{i}{4} \left(1+\mathcal{H}\right)\Im\{ X(\sigma), \sigma\},
\end{align}
where $\mathcal{H}[f(\sigma)](\sigma') = -\frac{1}{\pi i}\mathrm{p.v.}\int_{-\infty}^{\infty}d\sigma \frac{f(\sigma)}{\sigma-\sigma'}$ represents the Hilbert transform along the line.

\subsection{Perturbation around the straight WL minimal surface}
Using the equations introduced above we can repeat the wavy analysis around the straight line and present high order results.
We start by defining the straight line solution, and continue with an example of sinusoidal perturbation\footnote{As a check, we also repeated the ellipse analysis using this parametrization for the circle, and got the same results we found in the previous section. The explicit solution looks quite different, but is still simple enough in order to carry the analysis to high orders.}.
\subsubsection{Straight Wilson line}
The simplest solution is given for the infinite straight line where $\alpha(\sigma,\tau) = -\ln (2\tau)$, $f(\sigma + i\tau) = 0$.
The Target space solution is given by
\begin{align}
\X=\left(
\begin{array}{cc}
 \frac{\kappa \left(\sigma ^2+\tau ^2\right)}{\tau } & \frac{\sigma }{\lambda  \tau } \\
 \frac{\lambda  \sigma }{\tau } & \frac{1}{\kappa \tau }
\end{array}
\right),\quad
\A = \left(
\begin{array}{cc}
 \frac{-i \sigma +\tau }{\sqrt{2\tau/\kappa }} & \frac{\sigma -i \tau }{\lambda  \sqrt{2\tau /\kappa}} \\
 -\frac{i \lambda }{\sqrt{2 \tau \kappa}} & \frac{1}{\sqrt{2 \tau \kappa}}
\end{array}
\right),
\end{align}
so the target space Poincar\'{e} coordinates are given by $\mathrm{Z} = \kappa\tau$ and $\mathrm{X} = \kappa \lambda^{-1}\sigma$ (we kept the deformation parameter which in this case rotates the line in the $\mathrm{X}-\mathrm{Y}$ plane).
\subsubsection{Sinusoidal Wilson line}
Using the equations above we add a perturbation to the straight line.
Generally, one can add a perturbabtion and solve the equations of motion and Virasoro constraints order by order.
However, the procedure is complicated and it is hard to get to high orders.
As in the previous section, the equations above simplify the analysis considerably, where one does not have to find the complete surface at each order.
Here we show how it works for the simple case of the sinusoidal line where it turns out to be possible to compute analytically very high order corrections.
The curve we would like to study is $X(\sigma) = \sigma + i \epsilon \sin\sigma$.
The procedure we use goes as follows.
We stat with a general ansatz for $\alpha(\sigma,\tau)$, $f(\sigma+i\tau)$ and the parametrization $F(\sigma)$ expanded in powers of $\epsilon$.
At each order, we first "read" $f$ from the imaginary part of the Schwarzian derivative, then we plug it in the generalized cosh-Gordon equation solving for $\alpha$ and use the boundary conditions at $\tau = 0$ and $\tau\to\infty$ to fix the integration constants. Finally we compare with the real part of the Schwarzian derivative and fix $F$.

We find that for this case the expansion is given by
\begin{align}
\alpha(\sigma,\tau) & = -\ln 2\tau^2 + \sum_{n=1}^\infty \epsilon^{2 n} \sum_{k=0}^{n-1}\cos(2 k \sigma) \alpha_{2 n, k}(\tau),\nonumber\\
f(\sigma+i \tau) & = \sum_{n=1}^\infty \epsilon^{2 n -1}\sum_{k=1}^{n} e^{i(2 k -1)(\sigma + i \tau)} f_{2 n -1,2 k -1},\nonumber\\
F(\sigma) & = \sigma + \sum_{n=1}^\infty \epsilon^{2 n}\sum_{k=1}^{n} \sin(2 k \sigma ) F_{2 n, 2 k}.
\end{align}
$\alpha_{2 n, k}(\tau)$ are simple functions involving $\tau$-exponentials times polynomials, that can be found analytically.
We can also write $\alpha(\sigma,\tau)$ in a more explicit form as
\begin{align}
\alpha(\sigma,\tau) & =
-\ln 2\tau^2 + \sum_{n=1}^{\infty}\epsilon^{2 n}\sum_{k=0}^{n-1}\cos(2 k \sigma)\frac{1}{\tau^{n-k}}\sum_{s=k}^n e^{-2 s \tau} P_{2 s -k}(\tau),
\end{align}
where $P_{n}(\tau)$ is a polynomial of degree $n$.

\begin{figure}
    \centering
    \includegraphics[trim = 0mm 0mm 0mm 0mm,clip,width = 0.6\textwidth]{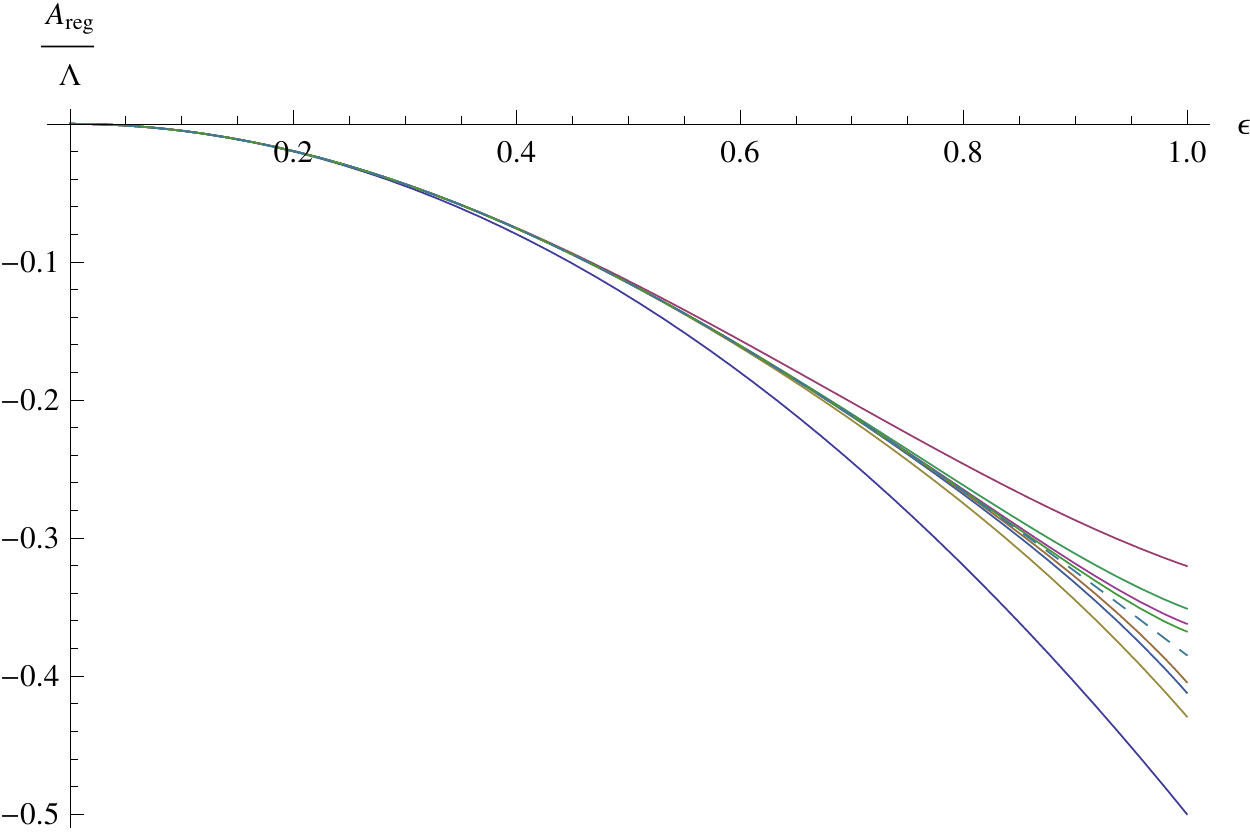}
    \caption{The area of the minimal surface ending on an sinusoidal contour parameterized by $(F(\sigma),\epsilon \sin F(\sigma))$.
    The different lines correspond to different orders in the $\epsilon$ expansion (from order $\epsilon^2$ up to order $\epsilon^{16}$), where it is seen how the lines converge. The dashed line is plotted using interpolation of the expansion coefficients up to $\epsilon^{1000}$. As one can see, the wavy line approximation (blue line) is quite good for $\epsilon \lesssim 0.4$.}
    \label{fig:sineArea}
\end{figure}

\begin{figure}
    \centering
    \includegraphics[trim = 0mm 0mm 0mm 0mm,clip,width = 0.6\textwidth]{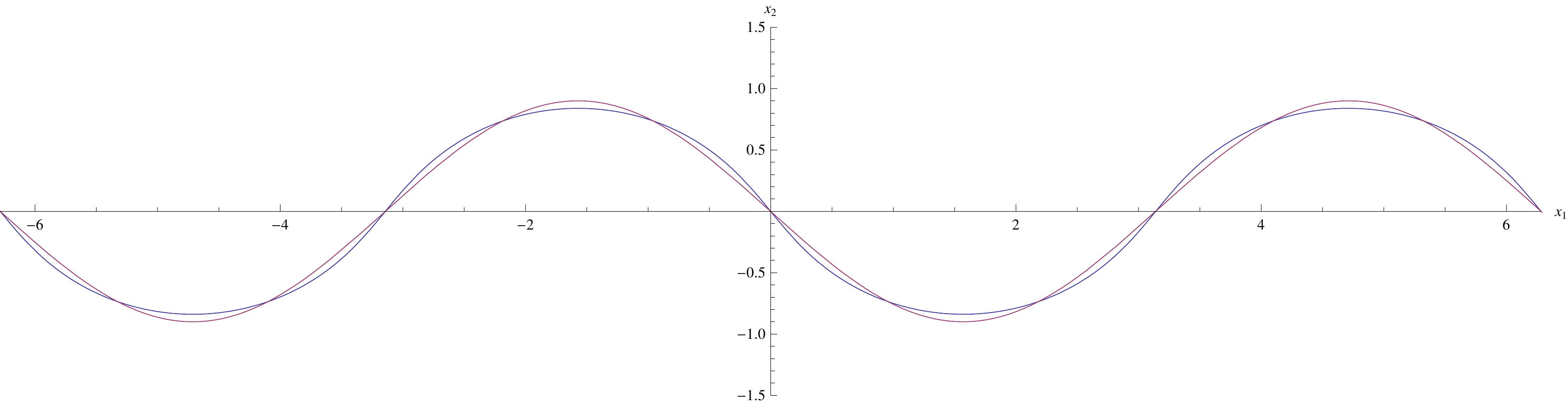}
    \caption{The blue line represents the original sinusoidal curve for up to order $\epsilon^{17}$ for $\epsilon=0.9$. The purple curve represents the resulting deformed curve with the same area, for $\lambda = e^{i\pi/2}$ (which gives the maximal deformation). In the plot we shifted the deformed line by a factor of $\varphi = \pi/2$ in order to ease the comparison between the curves.}
    \label{fig:sindeformation}
\end{figure}

After finding these functions, the area can be easily integrated to give
\begin{align}
\frac{A_{\text{reg}}}{\Lambda} = & -\frac{\epsilon ^2}{2}+\frac{23 \epsilon ^4}{128}-\frac{12059 \epsilon ^6}{110592}+\frac{1476007 \epsilon ^8}{18874368}-\frac{662835985271 \epsilon ^{10}}{10871635968000}+\frac{750989224006549 \epsilon ^{12}}{15028949562163200}\nonumber\\
&-\frac{155198331721663659127 \epsilon ^{14}}{3665727786540072960000}+\frac{132953910399497800665047 \epsilon ^{16}}{3619644351509283471360000}+ \mathcal{O}(\epsilon^{18}).
\end{align}
where $\Lambda$ is half the length of the worldsheet, $\sigma\in [-\Lambda,\Lambda]$.
This expression converge quite fast for $\epsilon < 1$.
We can interpulate a function for the coefficients in order to estimate the area close to $\epsilon = 1_{-}$.
The results are presented in figure \ref{fig:sineArea}.
We can also find the $\lambda$ deformed curve order by order, see figure \ref{fig:sindeformation}.
The deformed curve is very close to the original one, the difference shows first only in the $\epsilon^3$ correction term, while before that the deformation amounts to a translation of the curve in the $\mathrm{X}$ direction.

\section{Summary and discussion}\label{sec:discussion}

\paragraph{Minimal surfaces area expansion}
In this paper we studied minimal surface solutions in $\HH_3$ ending on various contours, by expanding perturbatively around the circular solution.
We used the formalism introduced recently in \cite{Kruczenski:2014bla} to obtain high order results for the area compared to the wavy approximation, without finding the explicit minimal surface solution.
We applied this approach to several target-space contours with different discrete symmetry properties.
We explored several explicit examples including contours which interpolate between the circle and hypocycloids (which include the ellipse), some other wavy symmetric contours, the lima\c{c}on curve and a contour with no discrete symmetries.
In the case of the ellipse which was treated in more details, we compared our results with the numerical data given in \cite{Fonda:2014cca} and found very good agreement in the range where our expansion is supposed to hold.
In all cases we found quite easily expressions for the area, far beyond the wavy approximation (at least to order $\epsilon^{8}$ and up to order $\epsilon^{22}$ in one of the examples).

Having such high order expansions, it is natural to wonder if one can express the area in a closed form or find a general expression to all orders for specific examples such as the ellipse.
In the case of the ellipse we conjectured the functional dependence to all orders in the expansion which might help in this direction.

A very interesting generalization of our analysis would be to study the problem in higher dimensional space-times such as  $\AdS_3\times \Sphere^1$ or $\AdS_{d>3}$ where very little is known compared to the $\AdS_3$.

\paragraph{$\lambda$-deformations}
As explained in \cite{Ishizeki:2011bf,Kruczenski:2014bla},
for any given boundary contour $X(\theta)$ we can define a one-parameter family of contours $X(\theta,\varphi)$ with $\varphi\in[0,2\pi]$ where $X(\theta,0) = X(\theta,2\pi) = X(\theta)$, such that the regularized area of the minimal surfaces corresponding to $X(\theta,\varphi)$ is the same.
The parameter $\varphi$ is related to the spectral parameter by $\lambda = e^{i \varphi}$, and we referred to such deformations as $\lambda$-deformations.
The origin of this symmetry of the Wilson loops at strong coupling is not clear since it is not related to the global conformal symmetry of the problem, furthermore it is not clear if the symmetry survives quantum corrections.
Moreover, given a contour $X(\theta)$, it is not known what is the shape of $X(\theta,\varphi)$ without having the analytic solution at hand.
We computed $X(\theta,\varphi)$ perturbatively in $\epsilon$ and checked the dependence of the one-loop weak coupling Wilson loop expectation value on $\varphi$.
Because of conformal symmetry one should not expect any dependence to order $\epsilon^3$,
however, at order $\epsilon^4$ there is no obvious reason for the weak coupling result to be invariant under the deformation.
For all the examples we checked, it turned out that the $\varphi$ dependence starts to show only at order $\epsilon^8$ except for the Lima\c{c}on where it starts at order $\epsilon^{16}$.
Thus, we show that the symmetry does not survive quantum corrections in general.
This fact also demonstrates the need for computing high order corrections.
These results definitely could not be found using the wavy approximation (that is to order $\epsilon^2$) and not even by going to order $\epsilon^7$.

The $\varphi$ dependence we found raises some natural questions.
Although we checked only a number of simple cases, the fact that we find a dependence only at the eighth order (or higher)
is quite suggestive, and it should be very interesting to understand if this is true for general deformations of the circle, and if this is the case what is the symmetry that protects from lower order corrections.
Another natural question is how does the Wilson loop depends on $\varphi$ at higher loop orders at weak coupling, and for the quantum corrections at strong coupling; when does the symmetry breaks and at which order in $\epsilon$?
It is also quite interesting why the dependence showed only at the 16th order for the lima\c{c}on which is a multiple of $8$ which appears in all the other examples.

\paragraph{Algebraic curve and theta-functions}

As discussed in \cite{Ishizeki:2011bf}, minimal surface solutions in $\HH_3$ correspond to Riemann surfaces of odd genus defined by an algebraic curve $y^2(\lambda) = 0$, which are invariant under $\lambda\to -1/\bar \lambda$ (in order for the solution to be real).
Non-trivial smooth solutions which end on one closed compact contour (different from the circular Wilson loop) start to appear from genus three.
Given a string solution, the algebraic curve can be found by constructing the Lax operator, which is a rational function of the spectral parameter, and taking its determinant \cite{Janik:2012ws,Cooke:2014uga}.
We applied this procedure for our examples by computing the Lax operator perturbatively in $\epsilon$.

In principle, one could imagine it is possible to deform existing cuts and add/remove cuts in order to continuously deform the target space contour until the desired contour is reached, and particularly to end up with the circular contour.
However, this is not the picture our construction yields (see also \cite{Cagnazzo:2013dqa} for analysis at leading order).
We started with the trivial curve $y^2(\lambda) = 1$ corresponding to the circular contour, and at each order we were able to construct a rational Lax operator such that $y^2(\lambda) = 1$ remains.
Although it was checked only up to a finite order in $\epsilon$ for each example,
it would be quite surprising if this result will suddenly change at higher orders.
It would be very interesting to understand why do the algebraic curves we find are trivial and whether a nontrivial algebraic curve can be extracted from our analysis.

This may be due to special properties of our starting point, the circular solution, compared to the higher genus solutions.
First we note that for the circular solution $f(z) = 0$, while the theta-functions are found when the $f(z)$ term in the generalized cosh-Gordon equation is set to $1$, after a change of coordinates followed by a gauge transformation.
However, when $f(z)$ vanishes such a transformation cannot be carried and the analysis becomes more subtle.
In contrast, by construction our expansions have a well defined limit to the circular solution.

In terms of the Riemann surface, the circular solution is a limiting case of the genus one solution where the length of the two cuts shrinks to zero, namely, it is a genus zero solution with two poles.
The limit of adding a pair of new cuts on top of this solution is not obvious in terms of the theta function solution and it should be interesting to see if such a procedure can generate new minimal surface solutions.

Moreover, the worldsheet picture of the contour (i.e. zeros of the theta functions) is very different compared to the known higher genus solutions.
For the circle, the worldsheet contour is a finite piece of an infinite straight line which is periodic in the target space.
On the other hand, in the case of higher genus solutions the worldsheet contour is a closed loop which is mapped to a closed contour in target space. Thus, it is not clear if and how the worldsheet straight line can be continuously deformed to such closed loops.
Notice that this is different from the genus one case which corresponds to two concentric circles contour in target space.
There, it is possible to make the target space contours wavy by adding two cuts \cite{Kruczenski:2013bsa} or more and the straight lines on the worldsheet which are mapped to the two concentric circles become wavy.
Therefore, it would be interesting to check if our analysis can be repeated for these solutions, and if it yields an algebraic curve which is consistent with the known exact results.
In this case there are several complications, first the equations involve the Jacobi elliptic functions which are much harder to solve.
Moreover, now one has to take care of two independent contours and allow for $f(z)$ to be meromorphic inside the unit disk since the solution is defined in the annulus.
For example in the case of two concentric circles $f(z) \propto z^{-2}$, although if one maps the solution from the UHP (or more precisely from the finite strip) $f(z)$ is a constant as expected since the solution coincides with the theta-functions solution (see e.g. \cite{Zarembo:1999bu,Drukker:2005cu,Dekel:2013kwa}).

\section*{Acknowledgement}
I would like to thank M. Cooke, N. Drukker, T. Klose, E. Tonni and K. Zarembo for valuable discussions.
I also thank M. Cooke, N. Drukker and K. Zarembo for their comments on the manuscript.
This work is supported by the ERC advanced grant No 341222.

\appendix

\section{Higher order general analysis}\label{app:general_higher_orders}
In this appendix we give the general solution for $\alpha(z,\bar z)$ to second order and $f(z)$ to third order.
These expressions are much more complicated than the lower terms in the expansion, since conformal symmetry is less restrictive.
For this purpose we have to solve the generalized cosh-Gordon equation to order $\epsilon^2$
\begin{align}
4(1-r^2)^2 |f_1(z)|^2 + 8\frac{\alpha_2(r,\theta)}{(1-r^2)^2}-\frac{1}{r^2}\left(r^2\p_r^2 +r\p_r + \p_\theta^2\right)\alpha_2(r,\theta) = 0.
\end{align}
We can solve this equation by expanding $f_1(z) = \sum_{n=0}^{\infty} f_{1,n} z^n$, where the coefficients $f_{1,n}$ are given by
\begin{align}\label{eq:f1G1relation}
f_{1,n} = -\frac{1}{2}(n+1)(n+2)(n+3)G^r_{1,n+2},
\end{align}
where $G^r_{1,n}$ is a coefficient defined by $G^r_1(s) = \sum_{n\in \mathbb{Z}} G^r_{1,n} e^{i n s}$.
Plugging the $f_1(z)$ expansion, we find the general solution
\begin{align}
\alpha_2(r,\theta) = & \left(r^2-1\right)\sum_{n=0}^{\infty}\sum_{q=0}^{\infty} e^{i \theta (n-q)}f_{1,n} f_{1,q} \Bigg(\frac{r^{4+n+q} }{(3+n) (3+q)}
-\frac{(4 (3+q)+n (4+q)) r^{2+n+q}}{(2+n) (3+n) (2+q) (3+q)}\nonumber\\
&+\frac{2(n+q+4)}{(1+n) (2+n) (3+n) (1+q) (2+q) (3+q)}\sum_{k=0}^{\text{Min}(n,q)}(q+n+1-2k)r^{n+q-2k}\Bigg).
\end{align}
From here we extract $\beta_2(\theta)$ which is needed in order to find the parametrization, and is given by expanding around $r = 1$,
\begin{align}
\beta_2(\theta) = &
\sum_{n=0}^\infty \sum_{q=0}^\infty f_{1,n} \bar f_{1,q} e^{i(n-q)\theta}
\begin{cases}
    \frac{q-2(1+n)}{3(1+n)(2+n)(3+n)},& n\geq q\\
    \frac{n-2(1+q)}{3(1+q)(2+q)(3+q)},& q\geq n\\
\end{cases}.
\end{align}
Notice that when $n=q$ the expressions coincide.
Let us further define
\begin{align}
B^+(\theta) = &
\sum_{n=0}^\infty \sum_{q=0}^\infty f_{1,n} \bar f_{1,q} e^{i(n-q)\theta}\frac{q-2(1+n)}{3(1+n)(2+n)(3+n)},\nonumber\\
B^-(\theta) = &
\sum_{n=0}^\infty \sum_{q=0}^\infty f_{1,n} \bar f_{1,q} e^{i(n-q)\theta}\frac{n-2(1+q)}{3(1+q)(2+q)(3+q)}.
\end{align}
We can use these expressions and the Hilbert transform in order to express $\beta_2(\theta)$ without using the conditional expression, in order to finally express it in terms of the perturbation,
\begin{align}
\beta_2(\theta) = \frac{1}{2}\left(B^+ + B^-\right)  + \frac{1}{2}\mathcal{H}\left(B^+ - B^-\right).
\end{align}
Notice that there is no double counting of the zero modes since the Hilbert transform kills them.
Next we insert $f_{1,n}$ in terms of the $G^r_{1,n}$'s using the relation (\ref{eq:f1G1relation}) which gives 
\begin{align}
B^+(\theta) = &
\frac{1}{12}\sum_{n=0}^\infty \sum_{q=0}^\infty G^r_{1,n+2} e^{i(n+2)\theta} \bar G^r_{1,q+2} e^{-i(q+2)\theta}(q-2(1+n))(1+q)(2+q)(3+q),\nonumber\\
B^-(\theta) = &
\frac{1}{12}\sum_{n=0}^\infty \sum_{q=0}^\infty G^r_{1,n+2} e^{i(n+2)\theta} \bar G^r_{1,q+2} e^{-i(q+2)\theta}(n-2(1+q))(1+n)(2+n)(3+n).
\end{align}
Since $B^-(\theta) = \bar B^+(\theta)$ we shall concentrate only on $B^+(\theta)$.
These expressions are related to the derivatives of $G$ as follows,
\begin{align}
B^+(\theta) = &
\frac{1}{12}\left(2 \p G \mathcal{L}_3 \bar G+G\mathcal{L}_3 \p \bar G\right),
\end{align}
where here $G = G_1^r$ with the zeroth and first modes removed.
The relation with the Schwarzian derivative implies
\begin{align}
\beta_2(\theta)
=
-\frac{1}{12}\mathcal{L}_3\left(G^i_2+F_2-\frac{1}{2} \p (G^i_1)^2 + \frac{1}{4} \p (G^r_1)^2 \right)
+\frac{1}{24}\left(2 \p G_1^r \mathcal{L}_3 G_1^r + G_1^r \mathcal{L}_3 \p G_1^r\right).
\end{align}
Defining $F_2 = K - \left(G^i_2-\frac{1}{2} \p (G^i_1)^2 + \frac{1}{4} \p (G^r_1)^2 \right)$ (where $K = K(\theta)$) we get
\begin{align}
\mathcal{L}_3 K = -12 \beta_2(\theta)
+\frac{1}{2}\left(2 \p G_1^r \mathcal{L}_3 G_1^r + G_1^r \mathcal{L}_3 \p G_1^r\right).
\end{align}
After some manipulations we find
\begin{align}
K & =
\frac{1}{2}\mathcal{L}_3 ^{-1}\left(\mathcal{H}\left[
2\p G_1^r \mathcal{L}_3 H_1^r
+G_1^r \mathcal{L}_3 \p H_1^r
\right]\right),
\end{align}
where $H_1^r(\theta) \equiv \mathcal{H}[G_1^r(\theta')](\theta)$, i.e. $G_1^r = g_1^r(\theta) + \bar g_1^r(\theta)$ so $H_1^r = g_1^r(\theta) - \bar g_1^r(\theta)$, with $g_1^r(\theta) = \sum_{n=1}^{\infty} g_{1,n}^r e^{i n \theta}$.
Notice that $\mathcal{L}_3 ^{-1} e^{i n \theta} = -i\frac{e^{i n \theta}}{n(1-n^2)}$, so for our periodic functions
\begin{align}
\mathcal{L}_3 ^{-1} f(\theta) & = e^{i \theta}\int d\theta' e^{-2 i \theta'}\int d\theta'' e^{i \theta''} \int d\theta''' f(\theta''')\nonumber\\
& = \int d\theta f(\theta) - \frac{1}{2}\left(e^{- i \theta}\int d\theta e^{i \theta}f(\theta) + e^{i \theta}\int d\theta e^{- i \theta}f(\theta)\right).
\end{align}
Notice also that $\mathcal{L}_3$ and its inverse commute with $\mathcal{H}$.
Let us now define
\begin{align}
\Im\{X(\theta),\theta\} = \sum_{n=0}^{\infty} \epsilon^n \mathcal{L}_3 S_n(\theta),
\end{align}
and plug $F_1(\theta)$ and $F_2(\theta)$ in $\Im\{X(\theta),\theta\}$, which to third order gives
\begin{align}
\mathcal{L}_3 S_3(\theta) = & -\bigg[
\mathcal{L}_3\left(G_3^r - G_1^i G_2^{r'} - G_2^i G_1^{r'}-\frac{1}{8}G_1^{r'}\p (G_1^r)^2 + \frac{1}{2}((G_1^i)^2 G_1^{r'})\right)
\nonumber\\
+2 K^{'} \mathcal{L}_3 G_1^{r} & +K \mathcal{L}_3 G_1^{r'}
-\frac{1}{4}\left((G_1^{r'})^2 -  2 G_1^r G_1^{r''}\right) \mathcal{L}_3 G_1^{r}
-\frac{1}{2}\left((G_1^{r'})^3 + 3 G_1^r G_1^{r'} G_1^{r''}\right)
\bigg]
+\mathcal{O}(\epsilon^4),
\end{align}
now $f_3$ follows from (\ref{eq:f_intermsof_ImXs}).

\subsection{Alternative derivation for the $\alpha_2(z,\bar z)$ solution}
We notice there exist an explicit particular solution for $\alpha_2(z,\bar z)$ in terms of the perturbation
\begin{align}
\alpha_2^{(p)}(z,\bar z) = W \bar W - \frac{z \bar{\tilde{G}}_1^r W + \bar z \tilde{G}_1^r \bar W}{(1 - z \bar z )},\quad
\text{where}\quad
W(z,\bar z) = \frac{1}{2}\left(z(z \bar z -1)\tilde{G}_1^{r''}-2\tilde{G}_1^{r'}\right),
\end{align}
where $\tilde G(z)$ is related to $f(z)$ and is defined in (\ref{eq:defOfGtilde}).
The homogenous solution which should be added (the one which is regular at $r=0$, as $\alpha_2^{(p)}$ is) is given by
\begin{align}
\alpha_2^{(h)}(z,\bar z) = \frac{1+z \bar z}{1-z \bar z}\left(T(z) + \bar T(\bar z)\right) + z \p T(z) + \bar z \bar \p \bar T(\bar z),
\end{align}
where $T(z)$ is an arbitrary holomorphic function and $\bar T(\bar z)$ is its complex conjugate.
Expanding the particular solution around $r=1$ yields
\begin{align}
\alpha_2^{(p)} \simeq \frac{z \bar{\tilde{ G_1^r}} \p \tilde G_1^r + \bar z {\tilde{ G_1^r}} \bar\p \bar{\tilde {G}}_1^r}{2(1-r)}\bigg|_{r\to 1}=\frac{1}{2(1-r)}\sum_{n,q} (n +q) e^{i(n-q)\theta}\tilde G_n \bar{\tilde{G}}_q.
\end{align}
On the other hand,
\begin{align}
\alpha_2^{(h)} \simeq \frac{T +\bar T }{1-r}\bigg|_{r\to 1}=\frac{1}{1-r}\sum_{k} \left(e^{i k\theta} T_k + e^{- i k\theta} \bar T_k\right),
\end{align}
which fixes $T$ in terms of $\tilde G$ using the boundary conditions.

\bibliographystyle{nb}
\bibliography{refs}

\begin{thebibliography}{10}
\ifx\href\asklfhas\newcommand{\href}[2]{#2}\fi
\ifx\arxivref\asklfhas\newcommand{\arxivref}[1]{\href{http://arxiv.org/abs/#1}{#1}}\fi
\ifx\doiref\asklfhas\newcommand{\doiref}[2]{\href{http://dx.doi.org/#1}{#2}}\fi
\raggedright
\small
\parskip 0pt

\bibitem{Maldacena:1997re}
J.~M.~Maldacena,
\textit{``{The large N limit of superconformal field theories and
  supergravity}''},
\textsf{Adv.~Theor.~Math.~Phys.~2,~231~(1998)},
\texttt{\arxivref{hep-th/9711200}}.
%
\bibitem{Beisert:2010jr}
N.~Beisert, C.~Ahn, L.~F.~Alday, Z.~Bajnok, J.~M.~Drummond et~al.,
\textit{``{Review of AdS/CFT Integrability: An Overview}''},
\textsf{\doiref{10.1007/s11005-011-0529-2}{Lett.Math.Phys.~99,~3~(2012)}},
\texttt{\arxivref{1012.3982}}.
%
\bibitem{Maldacena:1998im}
J.~M.~Maldacena,
\textit{``{Wilson loops in large N field theories}''},
\textsf{\doiref{10.1103/PhysRevLett.80.4859}{Phys.Rev.Lett.~80,~4859~(1998)}},
\texttt{\arxivref{hep-th/9803002}}.
%
\bibitem{Rey:1998ik}
S.-J.~Rey and J.-T.~Yee,
\textit{``{Macroscopic strings as heavy quarks in large N gauge theory and
  anti-de Sitter supergravity}''},
\textsf{\doiref{10.1007/s100520100799}{Eur.Phys.J.~C22,~379~(2001)}},
\texttt{\arxivref{hep-th/9803001}}.
%
\bibitem{Pohlmeyer:1975nb}
K.~Pohlmeyer,
\textit{``{Integrable Hamiltonian Systems and Interactions Through Quadratic
  Constraints}''},
\textsf{\doiref{10.1007/BF01609119}{Commun.~Math.~Phys.~46,~207~(1976)}}.
%
\bibitem{DeVega:1992xc}
H.~De~Vega and N.~G.~Sanchez,
\textit{``{Exact integrability of strings in D-Dimensional De Sitter
  space-time}''},
\textsf{\doiref{10.1103/PhysRevD.47.3394}{Phys.Rev.~D47,~3394~(1993)}}.
%
\bibitem{babich1993}
M.~Babich and A.~Bobenko,
\textit{``Willmore tori with umbilic lines and minimal surfaces in hyperbolic
  space''},
\textsf{\doiref{10.1215/S0012-7094-93-07207-9}{Duke~Mathematical~Journal~72,~151~(1993)}},
\href{http://dx.doi.org/10.1215/S0012-7094-93-07207-9}{\texttt{http://dx.doi.org/10.1215/S0012-7094-93-07207-9}}.
%
\bibitem{Ishizeki:2011bf}
R.~Ishizeki, M.~Kruczenski and S.~Ziama,
\textit{``{Notes on Euclidean Wilson loops and Riemann Theta functions}''},
\textsf{\doiref{10.1103/PhysRevD.85.106004}{Phys.Rev.~D85,~106004~(2012)}},
\texttt{\arxivref{1104.3567}}.
%
\bibitem{Kruczenski:2013bsa}
M.~Kruczenski and S.~Ziama,
\textit{``{Wilson loops and Riemann theta functions II}''},
\textsf{\doiref{10.1007/JHEP05(2014)037}{JHEP~1405,~037~(2014)}},
\texttt{\arxivref{1311.4950}}.
%
\bibitem{Cooke:2014uga}
M.~Cooke and N.~Drukker,
\textit{``{From algebraic curve to minimal surface and back}''},
\texttt{\arxivref{1410.5436}}.
%
\bibitem{Kruczenski:2014bla}
M.~Kruczenski,
\textit{``{Wilson loops and minimal area surfaces in hyperbolic space}''},
\texttt{\arxivref{1406.4945}}.
%
\bibitem{Toledo:2014koa}
J.~C.~Toledo,
\textit{``{Smooth Wilson loops from the continuum limit of null polygons}''},
\texttt{\arxivref{1410.5896}}.
%
\bibitem{Alday:2010vh}
L.~F.~Alday, J.~Maldacena, A.~Sever and P.~Vieira,
\textit{``{Y-system for Scattering Amplitudes}''},
\textsf{\doiref{10.1088/1751-8113/43/48/485401}{J.Phys.~A43,~485401~(2010)}},
\texttt{\arxivref{1002.2459}}.
%
\bibitem{Semenoff:2004qr}
G.~W.~Semenoff and D.~Young,
\textit{``{Wavy Wilson line and AdS / CFT}''},
\textsf{\doiref{10.1142/S0217751X0502077X}{Int.J.Mod.Phys.~A20,~2833~(2005)}},
\texttt{\arxivref{hep-th/0405288}}.
%
\bibitem{Polyakov:2000ti}
A.~M.~Polyakov and V.~S.~Rychkov,
\textit{``{Gauge field strings duality and the loop equation}''},
\textsf{\doiref{10.1016/S0550-3213(00)00183-8}{Nucl.Phys.~B581,~116~(2000)}},
\texttt{\arxivref{hep-th/0002106}}.
%
\bibitem{Galakhov:2008ax}
D.~Galakhov, H.~Itoyama, A.~Mironov and A.~Morozov,
\textit{``{Deviation from Alday-Maldacena Duality For Wavy Circle}''},
\textsf{\doiref{10.1016/j.nuclphysb.2009.06.009}{Nucl.Phys.~B823,~289~(2009)}},
\texttt{\arxivref{0812.4702}}.
%
\bibitem{Fonda:2014cca}
P.~Fonda, L.~Giomi, A.~Salvio and E.~Tonni,
\textit{``{On shape dependence of holographic mutual information in AdS4}''},
\texttt{\arxivref{1411.3608}}.
%
\bibitem{Ryu:2006bv}
S.~Ryu and T.~Takayanagi,
\textit{``{Holographic derivation of entanglement entropy from AdS/CFT}''},
\textsf{\doiref{10.1103/PhysRevLett.96.181602}{Phys.Rev.Lett.~96,~181602~(2006)}},
\texttt{\arxivref{hep-th/0603001}}.
%
\bibitem{Dekel:2013dy}
A.~Dekel,
\textit{``{Algebraic Curves for Factorized String Solutions}''},
\texttt{\arxivref{1302.0555}}.
%
\bibitem{Cagnazzo:2013dqa}
A.~Cagnazzo,
\textit{``{Integrability and Wilson loops: the wavy line contour}''},
\texttt{\arxivref{1312.6891}}.
%
\bibitem{Janik:2012ws}
R.~A.~Janik and P.~Laskos-Grabowski,
\textit{``{Surprises in the AdS algebraic curve constructions: Wilson loops and
  correlation functions}''},
\textsf{\doiref{10.1016/j.nuclphysb.2012.03.018}{Nucl.Phys.~B861,~361~(2012)}},
\texttt{\arxivref{1203.4246}}.
%
\bibitem{Zarembo:1999bu}
K.~Zarembo,
\textit{``{Wilson loop correlator in the AdS / CFT correspondence}''},
\textsf{\doiref{10.1016/S0370-2693(99)00717-0}{Phys.Lett.~B459,~527~(1999)}},
\texttt{\arxivref{hep-th/9904149}}.
%
\bibitem{Drukker:2005cu}
N.~Drukker and B.~Fiol,
\textit{``{On the integrability of Wilson loops in AdS(5) x S**5: Some periodic
  ansatze}''},
\textsf{\doiref{10.1088/1126-6708/2006/01/056}{JHEP~0601,~056~(2006)}},
\texttt{\arxivref{hep-th/0506058}}.
%
\bibitem{Dekel:2013kwa}
A.~Dekel and T.~Klose,
\textit{``{Correlation Function of Circular Wilson Loops at Strong
  Coupling}''},
\textsf{\doiref{10.1007/JHEP11(2013)117}{JHEP~1311,~117~(2013)}},
\texttt{\arxivref{1309.3203}}.
%
\end{thebibliography}

\end{document}